\documentclass[a4paper,11pt]{article}
\pdfoutput=1 % if your are submitting a pdflatex (i.e. if you have
\usepackage{jheppub} % for details on the use of the package, please
\usepackage[T1]{fontenc} % if needed
\usepackage{bm,bbm}
\usepackage{graphics}
\usepackage{subcaption}
\usepackage{amssymb}
\usepackage{hyperref}
\usepackage{slashed}
\usepackage{mathrsfs}
\usepackage{esint}
\usepackage[normalem]{ulem}
\usepackage{longtable}
\usepackage{cancel}
\hypersetup{
} 
\usepackage{xspace}
\usepackage[separate-uncertainty]{siunitx}%,multi-part-units = single]{siunitx}
\usepackage{physics}
\usepackage{mathtools}
\usepackage{graphicx}
\usepackage{verbatim}
\usepackage{multirow}

\newcommand{\msb}{{\overline{\mathrm{MS}}}}

\newcommand{\sherpa}{\textsc{Sherpa}\xspace}
\newcommand{\herwig}{\textsc{Herwig}\xspace}
\newcommand{\pythia}{\textsc{Pythia}\xspace}

\newcommand{\nnll}{N\textsuperscript{2}LL\xspace}

\newcommand{\lqcd}{\Lambda_\mathrm{QCD}}
\newcommand{\mMC}{m_t^\mathrm{MC}}
\newcommand{\tmin}[1]{#1_\mathrm{min}}
\newcommand{\taumin}{\tmin{\tau}}
\newcommand{\mhat}{\hat{m}}
\newcommand{\shat}{\hat{s}}
\newcommand{\nflav}[1]{^{(#1)}}
\newcommand{\CFalpha}[1]{C_F\frac{\alpha_s(#1)}{4\pi}}
\newcommand{\CFalphafive}[1]{C_F\frac{\alpha^{(5)}_s(#1)}{4\pi}}
\newcommand{\CFalphasix}[1]{C_F\frac{\alpha^{(6)}_s(#1)}{4\pi}}
\newcommand{\plusFunc}[3]{\mathcal{L}_{#1}^{#2}\left(#3\right)}
\newcommand{\msbar}{\overline{\mathrm{MS}}}
\newcommand{\alphaflavpi}[2]{\Biggl[\frac{\alpha_s\nflav{#1}(#2)}{4 \pi}\Biggr]}
\newcommand{\mpole}{m_t^\mathrm{pole}}
\newcommand{\gap}{\bar{\delta}}

\newcommand{\alphaem}{\alpha_\mathrm{em}}
\newcommand{\quark}{\mathrm{qu}}
\newcommand{\gluon}{\mathrm{gl}}
\DeclareMathOperator{\li2}{Li_2}

\newcommand{\python}{\textsc{Python}\xspace}
\newcommand{\nbar}{{\bar{n}}}
\newcommand{\alphapi}[1]{\left[\frac{\alpha_s(#1)}{4 \pi}\right]}
\definecolor{purple}{rgb}{0.57, 0.36, 0.51}
\definecolor{ocorr}{RGB}{0,200,0}
\newcommand{\zq}{z_\mathrm{q}}
\newcommand{\zg}{z_\mathrm{g}}

\usepackage{xcolor}

\title{\boldmath
Top Quark Mass Calibration for Monte Carlo Event Generators - An Update
}

\preprint{
\begin{flushright}
UWThPh-2023-16\\
DESY-23-127\\
\end{flushright}
}

\author[a]{Bahman Dehnadi,}
\author[b]{Andr\'e H. Hoang,}
\author[b]{Oliver L. Jin,}
\author[c]{and Vicent Mateu}

\affiliation[a]{Deutsches Elektronen-Synchrotron DESY, Notkestr. 85, 22607 Hamburg, Germany}
\affiliation[b]{Faculty of Physics, University of Vienna, Boltzmanngasse 5, A-1090 Wien, Austria}
\affiliation[c]{Departamento de F\'isica Fundamental e IUFFyM, Universidad de Salamanca,\\E-37008 Salamanca, Spain}

\emailAdd{bahman.dehnadi@desy.de}
\emailAdd{andre.hoang@univie.ac.at}
\emailAdd{oliver.jin.uni@gmail.com}
\emailAdd{vmateu@usal.es}

\abstract{
We generalize and update our former top quark mass calibration framework for Monte Carlo (MC) event generators based on the $e^+e^-$ hadron-level 2-jettiness $\tau_2$ distribution in the resonance region for boosted $t\bar t$ production, that was used to relate the \pythia~8.205 top mass parameter $m_t^\mathrm{MC}$ to the MSR mass $m_t^\mathrm{MSR}(R)$ and the pole mass $m_t^\mathrm{pole}$. The current most precise direct top mass measurements specifically determine $m_t^\mathrm{MC}$. The updated framework includes the addition of the shape variables sum of jet masses $\tau_s$ and modified jet mass $\tau_m$, and the treatment of two more gap subtraction schemes to remove the ${\cal O}(\Lambda_{\rm QCD})$ renormalon related to large-angle soft radiation. These generalizations entail implementing a more versatile shape-function fit procedure and accounting for a certain type of $(m_t/Q)^2$ power corrections to achieve gap-scheme and observable independent results. The theoretical description employs boosted heavy-quark effective theory (bHQET) at next-to-next-to-logarithmic order (N$^2$LL), matched to soft-collinear effective theory (SCET) at N$^2$LL and full QCD at next-to-leading order (NLO), and includes the dominant top width effects. Furthermore, the software framework has been modernized to use standard file and event record formats. We update the top mass calibration results by applying the new framework to \pythia~8.305, \herwig~7.2 and \sherpa~2.2.11. Even though the hadron-level resonance positions produced by the three generators differ significantly for the same top mass parameter $\mMC$ value, the calibration shows that these differences arise from the hadronization modeling. Indeed, we find that $m_t^{\rm MC}$ agrees with $m_t^{\rm MSR}(1\,\mbox{GeV})$ within $200$\,MeV for the three generators and differs from the pole mass by $350$ to $600$\,MeV.
}

\begin{document}
\maketitle
\flushbottom

\section{Introduction}
The top quark mass $m_t$ is one of most important parameters of the Standard Model (SM). Due to its large size, it plays an important role in many quantitative and conceptual aspects of the SM~\cite{Cabibbo:1979ay,Alekhin:2012py,Buttazzo:2013uya,Branchina:2013jra,Branchina:2014usa,Baak:2014ora}. Its value also becomes increasingly important as an input in constraining the potential effects of physics beyond the SM~\cite{Andreassen:2014gha}. The most precise determinations of this parameter are  based on so called ``direct measurements'' where kinematical observables depending on the momenta of the top decay products (jets and/or charged leptons) in $t\bar{t}$ events are measured and  compared to the corresponding predictions obtained from Monte Carlo (MC) event-generator simulations. Even though these MC event generators (MCs) are based on first principles, due to conceptual as well as practical limitations (and to gain generality), their main ingredients ---\,parton shower and hadronization models\,--- use approximations. Modeling assumptions in the hadronization process lead to a large set of free parameters which
partly affect the parton showering description (e.g.\ the shower cut parameter). These parameters are fixed by tuning the MCs to standard observables in $e^+e^-$ facilities and also the large hadron collider (LHC) to achieve an optimal reproduction of experimental measurements. Even though an adequate data description can be achieved, the physical meaning of the MCs inherent QCD parameters including the top quark mass $\mMC$, which is determined in direct measurements, becomes partly uncontrolled. 

The current particle data group (PDG) world average for direct measurements reads $\mMC=(172.69\,\pm\, 0.30)$\,GeV~\cite{ParticleDataGroup:2020ssz} and uses, among others, the respective combinations by CMS  $\mMC=(172.44\,\pm\,0.48)$\,GeV~\cite{CMS:2015lbj}, ATLAS $\mMC=(172.69\,\pm\,0.48)$\,GeV~\cite{ATLAS:2018fwq} and Tevatron  $\mMC=(174.30\,\pm\,0.65)$\,GeV~\cite{CDF:2016vzt}. Recently, there has been a very precise direct measurement not yet included in the world average $\mMC=\SI{171.77(37)}{\GeV}$ from CMS~\cite{CMS:2023ebf}. Future projections for the HL-LHC indicate that uncertainties as small as $\SI{200}{\MeV}$ for individual measurements may eventually be reached~\cite{Azzi:2019yne}. The basis of the direct measurements are reconstructed observables defined on the top quark decay product momenta, highly sensitive to the top quark mass, based on the idealization of considering the top quark as a physical particle. The approximation of on-shell top quarks with a factorized decay is also the foundation of state-of-the-art MCs. These observables are, however, strongly affected by soft gluon radiation as well as non-perturbative effects, where currently no consistent theoretical predictions based on systematic analytic methods exist. The direct top mass measurements are therefore solely based on MCs, and even though they have reached a high level of sophistication concerning the treatment of top quark decay products, the result for $\mMC$ must be interpreted with some care when used as an input for theoretical predictions~\cite{Azzi:2019yne,Hoang:2020iah,Schwienhorst:2022yqu}.
  
At this time, a number of first-principle insights have been obtained concerning the theoretical interpretation of the top quark MC mass parameter $\mMC$, which is, as a matter of principle, tied to the precision and implementation of the parton shower. The latter is the essential perturbative component of the MCs. At the purely partonic level, it can be shown for the coherent branching parton shower algorithm and inclusive shape variables (where coherent branching is NLL precise), that $\mMC$ differs from the pole mass by a term proportional to $Q_0\times\alpha_s(Q_0^2)$, where $Q_0$ is the transverse-momentum shower cut~\cite{Hoang:2018zrp}. It has been suggested that a similar relation applies to any parton shower~\cite{Hoang:2008xm,Hoang:2014oea}, and evidence supporting this view has  been provided in Ref.~\cite{Baumeister:2020mpm} by numerical analyses for the dipole shower. However, an analytic proof  for the dipole shower, comparable to that of coherent branching in Ref.~\cite{Hoang:2018zrp}, is still missing. Conceptually, the shower cut $Q_0$ acts like an infrared factorization scale that can be controlled by a renormalization group equation that is linear rather than logarithmic~\cite{Hoang:2018zrp}. Physically, the shower cut $Q_0$ is also a resolution scale, below which real and virtual soft radiation are unresolved and cancel. It is therefore reasonable to associate $\mMC$ with a low-scale short distance mass such as the MSR mass \mbox{$m_t^{\rm MSR}(R=Q_0)$}~\cite{Hoang:2008yj,Hoang:2017suc,Hoang:2020iah} where the scale $R$ acts as an IR resolution scale for self-energy corrections as well. Using the MSR mass also avoids the appearance of the pole-mass renormalon which would add an additional uncertainty between $\SI{110}{MeV}$~\cite{Beneke:2016cbu} and $\SI{250}{\MeV}$~\cite{Hoang:2017btd}. However, in practical MCs, where the shower cut is treated as a tuning parameter, the meaning of $\mMC$ may also be influenced by details of the hadronization models~\cite{Hoang:2020iah}. This latter source of uncertainty has not yet been investigated quantitatively up to now, as it is non-trivial to disentangle their effects from the dynamics of the parton showers. The insights just described have been obtained in the context of $e^+e^-$ collisions. They should in principle also apply for hadron colliders, but initial-state radiation processes such as multi parton interactions and underlying event, for which no systematic theoretical description exists at this time, make concrete quantitative statements on the precise theoretical interpretation of $\mMC$ more difficult. It was stated in Ref.~\cite{Hoang:2020iah} that for the time being one may identify $\mMC$ with the MSR mass at the scale $R=1.3$\,GeV with an uncertainty of $0.5$\,GeV. This quantification should be scrutinized through explicit phenomenological analyses. 

Alternatively to the conceptual insights just mentioned, a number of studies to numerically relate $\mMC$ to the top quark mass in a well-defined renormalization scheme have been carried out. In Ref.~\cite{Kieseler:2015jzh} a simultaneous measurement of  $\mMC$ and the inclusive $t\bar t$ cross section at the LHC was suggested, intended for a $\mMC$-independent measurement of the top quark mass from fixed-order cross section theoretical calculations. The method also yielded a quantification of the relation between $\mMC$ and the pole and $\overline{\rm MS}$ masses with an uncertainty of $2$\,GeV which, however, depends on the set of parton distribution functions employed for the analysis. A more precise direct calibration method was developed in Ref.~\cite{Butenschoen:2016lpz}, where hadron-level N$^2$LL resummed and NLO matched theoretical predictions for the 2-jettiness distribution in the highly top-mass sensitive resonance region for boosted top production in $e^+e^-$ annihilation were fitted to Pythia 8.205~\cite{Sjostrand:2014zea} pseudo-data samples. The theoretical factorization framework to determine the 2-jettiness distribution was developed in Refs.~\cite{Fleming:2007qr,Fleming:2007xt} and is based on soft-collinear effective theory (SCET)~\cite{Bauer:2000ew,Bauer:2001ct,Bauer:2001yt} and boosted heavy-quark effective theory~\cite{Fleming:2007qr,Fleming:2007xt}. Since the 2-jettiness distribution is an inclusive event-shape closely related to thrust, apart from a systematic resummation of soft, collinear and ultra-collinear QCD corrections, also a first-principle parametrization of the hadronization effects was employed. This yields a systematic hadron-level prediction depending on QCD parameters, such as the top mass (in any renormalization scheme) and the strong coupling, as well as the parameters of a non-perturbative shape function which was originally developed for inclusive $B$-meson decays in the endpoint region~\cite{Ligeti:2008ac}. Furthermore, using a low-scale short-distance mass such as the MSR mass $m_t^{\rm MSR}(R)$ and the gap subtraction formalism~\cite{Hoang:2007vb}, all ${\cal O}(\Lambda_{\rm QCD})$ renormalon effects, which arise from ultra-collinear and large-angle soft radiation, can be removed systematically while at the same time avoiding the appearance of large logarithms. All these ingredients were combined to obtain a hadron-level cross section for the 2-jettiness distribution at N$^2$LL$+$NLO in Ref.~\cite{Dehnadi:2016snl}\footnote{Recently, the N$^3$LL resummation of the leading-power cross section for boosted top pair production in the resonance region has been achieved in Ref.~\cite{Bachu:2020nqn}.}. These theoretical predictions were used in the calibration analysis of Ref.~\cite{Butenschoen:2016lpz} and the following numerical relations were found: $\mMC = m_t^{\rm pole} + (0.57 \pm 0.29)$\,GeV and $\mMC = m_t^{\rm MSR}(1\,\mbox{GeV}) + (0.18 \pm 0.23)$\,GeV. A similar analysis in the context of the LHC was performed by the ATLAS collaboration in Ref.~\cite{ATL-PHYS-PUB-2021-034} using soft-drop groomed~\cite{Krohn:2009th} boosted top jet mass distributions, based on the NLL hadron-level theoretical description developed in Refs.~\cite{Hoang:2017kmk,Hoang:2019ceu}, which are compatible with the $e^+e^-$ calibration results, but have much larger uncertainties.

In this article, an update and a generalization of the calibration analysis of Ref.~\cite{Butenschoen:2016lpz} is presented. The work is improved in several aspects: (i) In order to study observable independence, in addition to  the 2-jettiness $\tau_2$ distribution two additional shape variables, namely the sum of jet masses $\tau_s$ and the modified jet mass $\tau_m$, are considered. The conceptual subtlety is that these three shape variables are affected differently by
\begin{equation}
\mhat_t^2 \equiv \frac{m_t^2}{Q^2}\,,
\end{equation}
(massive) power corrections which can be larger than the precision achieved in Ref.~\cite{Butenschoen:2016lpz}. We study these power corrections and provide a well-motivated prescription to tame them. (ii) To test the dependence on the gap subtraction scheme (to treat ${\cal O}(\Lambda_{\rm QCD})$ renormalons stemming from large-angle soft radiation), we implement and study two additional gap subtraction schemes, one of which was already employed in Ref.~\cite{Bachu:2020nqn}. To deal with these two additional gap schemes we improve significantly the flexibility of the shape-function fit parameters. (iii) While the calibration analysis in Ref.~\cite{Butenschoen:2016lpz} was solely for \pythia~8.205, here we also calibrate $\mMC$ for \herwig 7.2.1~\cite{Bellm:2015jjp} and \sherpa 2.2.11~\cite{Sherpa:2019gpd} (and we update the calibration for \pythia~8.305~\cite{Bierlich:2022pfr}).
(iv) In contrast to the custom-written calibration software framework used in~\cite{Butenschoen:2016lpz}, here we employ \textsc{Rivet}~\cite{Bierlich:2019rhm} for the observables, paired with in-house analysis tools to convert event-by-event kinematic information into histograms in the \textsc{yoda} format, such that the workflow now works with all major MCs that support \textsc{Rivet} directly or the event record format \textsc{HepMC}~\cite{Buckley:2019xhk}. (v) Finally, we also present details for all theoretical ingredients that were employed in the original calibration analysis~\cite{Butenschoen:2016lpz}, but not displayed there due to lack of space.

Within the theoretical uncertainties of our theoretical N$^2$LL$\,+\,$NLO description we find observable and gap-scheme independence for the $\mMC$ calibration, and reconfirm the numerical results obtained in the original analysis of Ref.~\cite{Butenschoen:2016lpz}. The probably most interesting outcome is that, while the hadron-level distributions for the three shape variables differ considerably between \pythia, \herwig and \sherpa for the same $\mMC$ input value, the calibration results for the relation of this parameter to the MSR mass are compatible within uncertainties of about $200$\,MeV. It turns out that the bulk of the differences observed for the hadron-level cross sections is associated to different modeling of hadronization effects among the three MCs. 

The content of this article is as follows:
In Sec.~\ref{sec:observables} we introduce the three shape variables used in our calibration analysis and show the corresponding predictions for the cross section using \pythia, \herwig and \sherpa for boosted top production in $e^+e^-$ annihilation. These MC pseudo-data are used as the input for the top quark mass calibrations carried out in the subsequent sections. In Sec.~\ref{sec:bhqet} a detailed description of the N$^2$LL$+$NLO differential cross section for the shape variables in the resonance region used for the calibration analysis is provided. Here we also discuss the generalizations concerning the gap subtraction schemes and the $\mhat_t^2$ power corrections that were not considered in Ref.~\cite{Butenschoen:2016lpz}. The fit procedure, data processing and our approach to determine uncertainties are explained in Sec.~\ref{sec:fitImprovements}. Section~\ref{sec:consistencyresults} focuses on a first application of the updated calibration framework, namely reproducing the results given in Ref.~\cite{Butenschoen:2016lpz}, which were based on the original calibration setup. Here we also introduce the graphical representation of the calibration results used in the following sections of the article. In Sec.~\ref{sec:shapefctfits} we discuss the generalizations of the calibration framework needed to reliably carry out fits in the two additional gap subtraction schemes. Since performing these is in general quite costly and cumbersome, we introduce a  minimal modification of the scale setting procedure that translates into a faster $\chi^2$ minimization that we also use in the final calibration analysis. The role of $\mhat_t^2$ power corrections and the necessity to partially account for them within the singular bHQET cross section to achieve observable-independent calibration fits are discussed in Sec.~\ref{sec:pcanalysis}. In Sec.~\ref{sec:finalresults} we present the final results and Sec.~\ref{sec:conclusions} contains our conclusions. We added four appendices showing the NLO fixed-order QCD results for the three shape distributions needed for the matching calculations and providing some basic formulae concerning the renormalization-group evolution factors, the three gap subtraction schemes and the definition of distributions. In Appendix~\ref{sec:MCsettings} we provide the relevant entries for the input files we used to generate the \pythia~8.305, \herwig~7.2 and \sherpa~2.2.11 shape distributions.

\section{Shape Observables}
\label{sec:observables}

In the calibration analyses carried out in this article we consider three $e^+e^-$ inclusive event shapes. They are equivalent in the dijet limit concerning the dominant singular QCD effects, but differ at ${\cal O}(m_t^2/Q^2)$, which constitute the most relevant subleading power corrections to the factorized and resummed treatment of the singular contributions.

The first observable is {\it 2-jettiness}  $\tau_2$ defined as \cite{Stewart:2009yx}
\begin{equation}
\label{eq:tau2def}
\tau_2 = \frac{1}{Q} \min_{\vec{n}_t}\sum_i(E_i - |\vec{n}_t \cdot \vec{p}_i|)\,,
\end{equation}
where the sum runs over all final-state particles with momenta $\vec{p}_i$. The maximum defines the thrust axis $\vec{n}_t$ and $Q$ is the center of mass energy. If the masses of the final-state particles are neglected $\tau_2$ agrees with thrust \cite{Farhi:1977sg}.  Since the event shapes are computed with the momenta of the top-quark decay products (which can be considered as light) $\tau_2$ is numerically close to thrust for unstable top-pair production. The $\tau_2$ distribution has a distinguished peak at its lower endpoint region that is very sensitive to the top mass, which we call the resonance region. For $Q\gg m_t$ this region  is dominated by dijet-like events where the top quarks are boosted and decay inside narrow back-to-back cones. This kinematic situation is the basis for the factorized treatment of the peak region, where the dominant large-angle soft QCD dynamics is analogous to that of $e^+e^-$ thrust at LEP energies. For a stable top quark the lower endpoint is
\begin{equation}
\label{eq:tau2min}
\tau_{2,{\rm min}} = 1-\sqrt{1-4\mhat_t^2} = 2 \mhat_t^2 + 2 \mhat_t^4 + {\cal O}(\mhat_t^6)\,,
\end{equation}
illustrating the strong top mass sensitivity. In fact, at tree-level and for stable tops, the distribution is proportional to a Dirac delta function peaking at $\tau_{2,{\rm min}}$. The expression for $\tau_{2,{\rm min}}$ also shows the importance of ${\cal O}(\mhat_t^2)$ power corrections since the $\mhat_t^4$ term in the expanded expression corresponds to a shift in the top quark mass of $2$ to $5$\,GeV for $Q$ in the range of $700$ to $1400$\,GeV. It is quite obvious that, at the level of precision of our calibration analysis, besides the power corrections in $\tau_{2,{\rm min}}$ shown above (which can be accounted for in a trivial manner), also other more subtle sources of $\mhat_t^2$ power corrections need to be considered. By construction, apart from a broadening due to the finite top-quark width, 2-jettiness is insensitive to the details of the decay products dynamics as long as the final-state kinematics does not affect the direction of $\vec{n}_t$. For $Q\gg m_t$ the out-of-hemisphere decays are $\mhat_t^2$-suppressed, but (for unpolarized electron-positron beams) the top quarks, in their rest frame, decay to a good approximation isotropically such that this effect only modifies the overall normalization and not the resonance peak location~\cite{Fleming:2007qr}. This class of power corrections is therefore not considered in our theoretical description. Thus, in the resonance (or peak) region, top quarks are so boosted that their decay products end up in the same hemisphere. Hence, the leading-order finite-width effects are fully accounted for convolving the distribution with a Breit-Wigner function. The peak region is therefore specified by the following condition:
\begin{equation}
\tau-\tau_{2,\rm min} \sim \frac{m_t\Gamma_t}{Q^2}=\mhat_t^2 \frac{\Gamma_t}{m_t}\,,
\end{equation}
where $\Gamma_t\approx 1.4\,{\rm GeV}\ll m_t$ is the top quark width. The peak location is, however, also strongly affected by perturbative and non-perturbative QCD corrections.

The second observable we consider is the {\it sum of jet masses} (sJM) $\tau_s$, also referred to as the hemisphere mass sum. The plane perpendicular to the thrust axis $\vec{n}_t$ defines the top and antitop hemispheres, called $a$ and $b$. This plane is used to define the normalized (squared) invariant masses
\begin{equation}
\rho_{a,b} = \frac{1}{Q^2} \Biggl(\, \sum_{ i \in a,b } p_i^\mu\Biggr)^{\!\!2}\,,
\end{equation} 
where the sum runs over all final-state particles in either hemisphere $a$ or $b$. The sum of jet masses is therefore defined as
\begin{equation}
\label{eq:tausdef}
\tau_s = \rho_a+\rho_b\,.
\end{equation}
For a stable top quark its lower endpoint is
\begin{equation}
\label{eq:tausmin}
\tau_{s,{\rm min}} = 2\mhat_t^2 \,,
\end{equation}
and the differential distribution shows the same features as $2$-jettiness. If all \mbox{$\mhat_t^2$-suppressed} power corrections are neglected, $\tau_2$ and $\tau_s$ are equivalent in the lower endpoint region, so that the dominant singular QCD effects are equivalent as well. However, as we shall show in the course of our analysis, the $\mhat_t^2$ power corrections in the measurement function related to (perturbative as well as non-perturbative) large-angle soft radiation are particularly sizable compared to 2-jettiness (for which they are absent). This is discussed in detail in Sec.~\ref{sec:absorbconcept}.

The third observable we consider is called {\it modified jet mass} (mJM) $\tau_m$, and defined from sJM by
\begin{equation}
\label{eq:taummdef}
\tau_m = \tau_s + \frac{1}{2} \tau_s^2\,,
\end{equation}
so that 
\begin{equation}
\label{eq:taummin}
\tau_{m,{\rm min}} =2\mhat_t^2+2\mhat_t^4\,.
\end{equation}
It has the important feature that the previously mentioned $\mhat_t^2$ power corrections to the large-angle soft radiation effects are absent as is also the case for 2-jettiness. We use the modified jet mass variable $\tau_m$ as an important diagnostic tool for our treatment of power corrections. In fact, as we shall show, in contrast to sJM, 2-jettiness is the observable least sensitive to $\mhat_t^2$ power corrections in our implementation to account for them. 

Note that in the context of having massive particles in the final state, different schemes exist specifying precisely how the energies and momenta of the final-state particles enter the shape-variable definition. The scheme we have adopted for the three shape variables $\tau_2$, $\tau_s$ and $\tau_m$ has been called ``massive scheme'' in Ref.~\cite{Salam:2001bd} and ensures that the leading non-perturbative correction (encoded quantitatively in the moment $\Omega_1$, see Sec.~\ref{sec:facttheorem}) is universal with respect to the effects of non-zero hadron masses~\cite{Salam:2001bd,Mateu:2012nk}. When the ``massive scheme'' is used for stable heavy quarks, the sensitivity to their mass is increased as compared to other choices~\cite{Bris:2020uyb,Lepenik:2019jjk}.

\begin{figure}[t]
	\makebox[\textwidth]{\includegraphics[width=\textwidth]{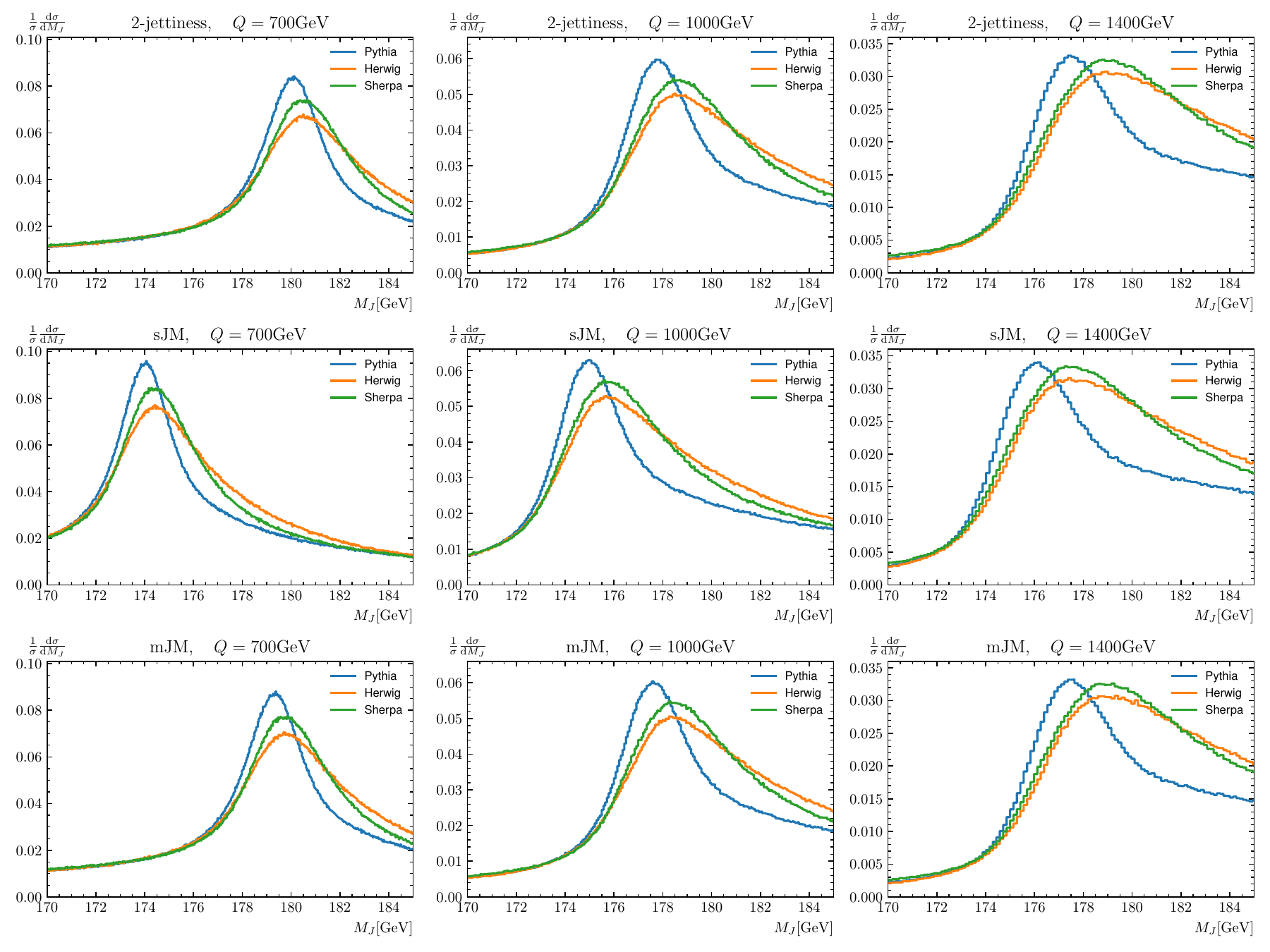}}
	\caption{Comparison of MC 2-jettiness, sJM and mJM event-shape distributions, as a function of the jet mass variable $M_J=Q \sqrt{\tau/2}$, between \pythia (blue), \herwig (orange) and \sherpa (green) at different c.m.\ energies $Q$ and for the input MC top mass \mbox{$m_t^{\rm MC}=173$\,GeV}, see Sec.~\ref{sec:dataprocessing} for details. For each distribution $10^7$ events were produced and clustered in $\tau$ bins of size $5\times 10^{-5}$.}
	\label{fig:MCdistributions}
\end{figure}

The three event-shape distributions in the peak region generated by \pythia~8.305~\cite{Bierlich:2022pfr}, \sherpa 2.2.11~\cite{Sherpa:2019gpd} and \herwig 7.2~\cite{Bellm:2015jjp} (using their standard settings) for $m_t^{\rm MC}=173$\,GeV and boosted-top pair production at center of mass (c.m.)~energies $Q=700$, $1000$ and $1400$\,GeV are displayed in Fig.~\ref{fig:MCdistributions} as a function of the jet mass variable $M_J=Q\sqrt{\tau/2}$, where $\tau$ stands for $\tau_2$, $\tau_s$ and $\tau_m$. The scaling of $M_J$ with respect to $\tau$ visualizes directly the top mass sensitivity of the three shape variables since $M_J$ would be equal to the input top mass at tree-level for $\Gamma_t=0$ and neglecting $\mhat_t$ power corrections. The differences in the peak positions between the shape variables and for different $Q$ values visualizes the sizable impact of the $\mhat_t^2$ power corrections. In addition, the shift of the peak positions to values much larger than $173$\,GeV is due to collinear and soft radiation, and in particular non-perturbative effects which are $Q$-dependent as well. It is also conspicuous that there are considerable differences in the shape and the peak locations generated by the three MC event generators. While \pythia predicts a quite narrow and distinct peak shape, \herwig and \sherpa yield a broader resonance region with \herwig showing the widest peak distribution. Furthermore, the peak positions for \herwig and \sherpa are located at significantly larger $M_J$ values. One of the most interesting conceptual aspects of the  analysis presented in this article is showing how all these differences affect the result for the $m_t^{\rm MC}$ calibration, since the theoretical framework must be capable of disentangling the perturbative radiation and the non-perturbative effects at the observable hadron level in order to provide reliable results for the top quark mass. For the framework presented here, it is essential that the calibration fits involve MC pseudo data from different $Q$ values.

We finally note that in principle also the $C$-parameter~\cite{Parisi:1978eg,Donoghue:1979vi}, in the modified version introduced in Refs.~\cite{Gardi:2003iv,Lepenik:2019jjk}, could be a good candidate as a top-mass sensitive shape variable for the calibration. The singular QCD effects are closely related to the ones for the thrust-like shape variables above (see Refs.~\cite{Hoang:2014wka,Hoang:2015hka}). However, as was shown by a thorough N$^2$LL$\,+\,$NLO analysis in Ref.~\cite{Preisser:2018yfv}, the $C$-parameter is highly sensitive to the way in which top-quark decay products are emitted, which causes a considerable broadening of the distribution in the resonance region that depends on the dynamics of the decay process and flattens the peak distribution in a way which cannot be accounted for with the Breit-Wigner smearing. This effect strongly reduces the top quark mass sensitivity and is so sizable that the $C$-parameter is not suitable for top mass calibration at the intended precision. 

\section{Resummed Cross Section at N$\mathbf{{}^2}$LL+NLO with Power Corrections}
\label{sec:bhqet}

\subsection{Factorization Formula in the Peak Region for the Singular Cross Section}
\label{sec:facttheorem}
A factorization theorem that resums large QCD logarithms in the resonance region of the 2-jettiness $\tau_2$ distribution for $e^+e^-\rightarrow t\bar{t}+X$ was derived in Refs.~\cite{Fleming:2007qr, Fleming:2007xt} using a sequence of effective field theories (EFTs). The factorization formula  also applies to the sum of jet masses $\tau_s$ and the modified jet mass $\tau_m$ distributions in the resonance region. In the following subsection, for the convenience of the reader, we briefly review the basic theoretical ingredients at N$^2$LL order precision, which have already been discussed at N$^3$LL in Ref.~\cite{Bachu:2020nqn}. Here we use the same notations as in Ref.~\cite{Bachu:2020nqn}, and generically refer to the shape variable as $\tau$.

The factorization formula in the resonance region is derived in two steps~\cite{Fleming:2007qr, Fleming:2007xt}. The first one is matching QCD to SCET in order to integrate out fluctuations at the production scale $Q$, leading to an expansion in $\tau\sim\mhat_t^2\ll1$, and resums logarithms of combinations of $\tau$ and $\mhat_t^2$. At leading power, the three shape variables $\tau_2$, $\tau_s$ and $\tau_m$ are equivalent. The resulting factorization formula exhibits the separation of large-angle soft and collinear dynamics known from massless quark event shapes (with $\lambda\sim\sqrt{\tau}\sim\mhat_t\ll 1$ the SCET power counting parameter) and is valid in the tail region of the distribution where there is no hierarchy between $\tau-\tau_{\rm min}$ and $\mhat_t^2$. The collinear modes (which contain the top-quark decay products with four-momentum $q^\mu$) exhibit invariant mass fluctuations scaling as $(q^2-m_t^2)/m_t\sim m_t$ while the soft modes have a much lower virtuality. This SCET factorization formula may be formulated in the context of a $6$-flavor QCD theory. In	 the resonance region defined by $\tau-\tau_{\rm min} \sim \mhat_t^2 \Gamma_t/m_t\ll \mhat_t^2$, one has $(q^2-m_t^2)/m_t\sim \Gamma_t$, enforcing an additional factorization using bHQET. Off-shell (mass-mode) fluctuations of the top quark are integrated out such that the collinear dynamics only contains radiation involving momenta scaling like $k_{uc}^{{\rm rest},\mu}\sim\Gamma_t$ in the top-quark rest frame, denoted as ultra-collinear modes. In the peak region the virtuality of the large-angle soft radiation is also lowered and involves momenta scaling like $k_s^\mu\sim\mhat_t\Gamma_t\gtrsim\Lambda_{\rm QCD}$ in the $e^+e^-$ c.m.\ frame. Here the ultra-collinear and large-angle soft dynamics are described in a 5-flavor scheme (treating all other quarks as massless). The fixed-order perturbative description of this process exhibits large logarithms of ratios of these momentum scales yielding the hierarchy $Q\gg m_t\gg \Gamma_t > \mhat_t\Gamma_t\gtrsim\Lambda_{\rm QCD}$. The dominant (also called singular) tower of these logarithms with respect to ratios of scales is summed in the SCET/bHQET framework, see Tab.~\ref{tab:orders} for the naming convention of the logarithmic resummation orders. 

\begin{table}[t]
	\centering
	\begin{tabular}{ l | l | c c c c c c}
		order & log terms & cusp & non-cusp & matching & $\beta[\alpha_s]$ & $\gamma_R$ & $\delta$\\
		\hline
		LL & $\alpha_s^nL^{n+1}$ & 1 & - & tree & 1 & - & -\\
		NLL+LO & $\alpha_s^nL^{n}$ & 2 & 1 & tree & 2 & 1 & -\\
		N\textsuperscript{2}LL+NLO & $\alpha_s^{n+1}L^{n}$ & 3 & 2 & 1 & 3 & 2 & 1
	\end{tabular}
	\caption{Required loop orders for the resummation of logarithmic terms at N$^k$LL of the type $\alpha_s^nL^{n+1-k}\sim\mathcal{O}(\alpha_s^{k-1})$ (considering $L\sim\mathcal{O}(\alpha_s^{-1})$) and fixed-order $\text{N}^{k-1}\text{LO}$ matrix element and matching corrections~\cite{Abbate:2010xh}. Cusp and non-cusp anomalous dimensions and \mbox{$\beta$-function} coefficients are given in App.~\ref{app:evolfact}. The R-anomalous dimension $\gamma_R$ and the renormalon subtraction series $\delta$ refer to both soft-gap and pole-mass renormalons.}
	\label{tab:orders} 
\end{table}

The factorization formula in the resonance region $\tau-\tau_{\rm min} \sim \mhat_t^2 \Gamma_t/m_t\ll \mhat_t^2$ has the form
\begin{align}
	\label{eq:bhqetdistribution}
		\left. \frac{1}{\sigma_0^C} \dv{\sigma^C_\mathrm{bHQET}}{\tau}\right|_\mathrm{strict} ={}& m_t Q^2 H_Q\nflav{6}(Q,\mu_H) U_{H_Q}\nflav{6}(Q,\mu_H,\mu_m) H_m\nflav{6}(m_t,\varrho,\mu_m) U_{v}\nflav{5}(\varrho,\mu_m, \mu) \nonumber\\
		&\times \!\!\int\! \dd{\ell} \dd{\shat} \dd{\shat'} \delta(\shat_\tau - \shat - \varrho \ell) % \nonumber\\
		%&\times \int \dd{\shat'}
		U_B\nflav{5}(\shat- \shat', \mu, \mu_J) J_{B,\tau}^{(5)}(\shat', \Gamma_t, \delta m_t, \mu_J) \nonumber\\
		&\times\!\! \int\! \dd{\ell'} \dd{k} U_S\nflav{5}(\ell - \ell', \mu, \mu_S) \hat{S}_\tau\nflav{5}(\ell' - k, \bar{\delta}, \mu_S) F(k - 2\hat{\Delta}) \,,
\end{align}
where $\sigma_0^{C}$ stands for the vector ($C=V$) and axial-vector ($C=A$) massless quark Born cross sections, see Eqs.~\eqref{eq:bornXSection}, and the factorization formula shown on the RHS is the same for $V$ and $A$. The superscripts $(6)$ and $(5)$ of the various functions indicate the number of active flavors, and we have defined the off-shellness variable
\begin{equation}
\label{eq:shattauold}
	\shat_\tau \equiv \frac{Q^2(\tau - \taumin)}{m_t}\,.
\end{equation}
The ratio 
	\begin{equation}
\label{eq:rhodef}
	\varrho\equiv \frac{Q}{m_t}\,,
\end{equation}
is the leading term of the on-shell top quark Lorentz factor for the boost that relates the c.m.\ and top/antitop rest frames in the resonance region. It is tied to the definition of the velocity labels~\cite{Bachu:2020nqn} of the heavy quarks in bHQET. These labels are controlled by a reparametrization invariance when subleading power corrections are included. The term $m_t$ appearing in $\varrho$ is therefore not tied to a particular renormalization scheme, and should in practice be set to a kinematic mass compatible with the invariant mass of the top (or antitop) system~\cite{Bachu:2020nqn} such as the pole mass $m_t^{\rm pole}$ or the MSR mass $m_t^{\rm MSR}(R\sim 1-2\,\mbox{GeV})$. Possible variations of $m_t$ in $\varrho$ are of order $\Gamma_t$~\cite{Bachu:2020nqn} and lead to tiny effects which are irrelevant in our analysis, and for this quantity we use the pole mass determined from the $\overline{\rm MS}$ mass $\overline{m}_t(\overline{m}_t)$ at three-loops. The term $\taumin$ is the lower endpoint $\tau$ value for stable top quarks for which we always use the exact  expressions quoted in Sec.~\ref{sec:observables}. This already provides the treatment of the most important $\mhat_t$ power corrections, but is not yet sufficient for the precision of our analysis, as we discuss in Sec.~\ref{sec:absorbconcept}.

The term $H^{(6)}_Q$ is the SCET hard function, which is the modulus squared of the Wilson coefficient obtained by matching the QCD and SCET top-antitop currents at leading order in $\mhat_t$. It contains the short-distance dynamics at the scale $Q=E_{\rm cm}\gg m_t$ that are integrated out in SCET and reads~\cite{Fleming:2007xt}\footnote{In Ref.~\cite{Gracia:2021nut} the hard and jet functions in SCET and bHQET have been computed to all orders in the large-$\beta_0$ approximation, which entails that terms proportional to $\alpha_s^{n+1}n_\ell^n$ are known for any $n\geq0$. In the same reference, it was also found that the SCET-bHQET current matching function $H_m^{(6)}$ in Eq.~(\ref{eq:massmatching}) has a  $\mathcal{O}(\Lambda_{\rm QCD})$ renormalon, which is, however, power-suppressed by the top quark mass.}
\begin{equation}
\label{eq:hardfunction}
H_Q^{(6)}(Q,\mu_H)=1+\CFalphasix{\mu_H}\Bigg[6\ln\biggl(\frac{Q^2}{\mu_H^2}\biggr)-2\ln^2\biggl(\frac{Q^2}{\mu_H^2}\biggr)-16+\frac{7\pi^2}{3}\Biggr].
\end{equation}
The natural scaling for its renormalization scale is $\mu_H\sim Q$, so that no large logarithms arise.

The term $H_m^{(6)}$ is the current matching coefficient between SCET and bHQET. It contains top quark fluctuations that are off-shell in the resonance region and therefore integrated out~\cite{Fleming:2007xt,Hoang:2015vua}. It has the form 
\begin{align}
\label{eq:massmatching}
H_m^{(6)}(m_t,\varrho, \mu_m)={}&1+\CFalphasix{\mu_m}\biggl(2L_m^2-2L_m+8+\frac{\pi^2}{3}\biggl)\\
&+C_F T_F\Biggl[\frac{\alpha_s^{(6)}(\mu_m)}{4\pi}\Biggr]^{\!2}\ln\biggl(\frac{m_t^2}{Q^2}\biggr)\biggl(\frac{8}{3}L^2_m+\frac{80}{9}L_m+\frac{224}{27}\biggr),\nonumber
\end{align}
\begin{equation}
\label{eq:Lmdef}
L_m\equiv\ln\Bigl(\frac{m_t^2}{\mu_m^2}\bigg)\,.   
\end{equation}
The 2-loop term, which is enhanced by a so-called \textit{rapidity logarithm}, is formally counted as $\alpha_s^2\ln{\mhat_t^2}\sim \mathcal{O}(\alpha_s)$ and is therefore included at N$^2$LL. This term appears since there are two types of fluctuations at the mass scale, collinear and soft mass modes, which have the same invariant mass but different rapidities with respect to the top-antitop axis. The N$^2$LL rapidity logarithms can be resummed to all orders~\cite{Hoang:2015vua} (see also Ref.~\cite{Hoang:2019fze}), but the numerical effect is negligible and therefore not included here. The natural scaling for the renormalization scale is the top quark mass, $\mu_m\sim m_t$, also called the mass-mode scale.
For $H_m^{(6)}$ one may also use the 5-flavor scheme for the strong coupling at the order we consider. Numerically, the difference of the two choices is orders of magnitude smaller than our perturbative uncertainties~\cite{Bachu:2020nqn}. Note that the scheme choice for the top mass $m_t$ appearing in $L_m$ is not relevant at this order either. Here we use the pole mass as obtained for $\varrho$. Using a different scheme leads to tiny effects as well. 

The bHQET jet function $J_{B,\tau}^{(5)}$ describes the ultra-collinear dynamics of the decaying top-antitop system. For stable top quarks it has the form
\begin{equation}
\label{eq:jetfunction}
m_t^2 J_{B,\tau}^{(5)}(\shat,\Gamma_t=0,\delta m_t=0,\mu) =\delta(\shat)+\CFalphafive{\mu}\Bigl[\bigl(8-\pi^2\bigr)\delta(\shat)+16\plusFunc{1}{\mu}{\shat}-8\plusFunc{0}{\mu}{\shat}\Bigr],\\
\end{equation}
where $\mathcal{L}_i$ are the standard plus distributions defined in Eq.~(\ref{eq:plusdist}). The bHQET jet function accounts for the leading double-resonant contributions in the peak region. The top quark finite-width effects, which we treat in the leading double resonant approximation as well, are described via a convolution of the stable-quark jet function with a Breit-Wigner function
\begin{equation}
\label{eq:breitwigner}
J_{B,\tau}^{(5)}(\shat, \Gamma_t, \delta m_t, \mu_J) = \int \dd{\shat'} G(\shat - \shat',\Gamma_t) J_{B,\tau}^{(5)}(\shat', \Gamma_t = 0, \delta m_t, \mu_J)\,,
\end{equation}
where
\begin{equation}
\label{eq:BWfunction}
G(\shat, \Gamma_t) = \frac{1}{\pi} \frac{2 \Gamma_t}{\shat^2 + (2 \Gamma_t)^2}\,.
\end{equation}
The factors of $2$ in $G(\shat, \Gamma_t)$ arise because $J_{B,\tau}^{(5)}$ accounts for the top and antitop quarks. As was shown in Ref.~\cite{Fleming:2007xt}, this treatment is equivalent to having an explicit imaginary width term in the (anti)top quark HQET propagator, $\sim 1/(v\cdot k+i\Gamma_t/2)$ with $v^\mu$ the top quark velocity label. The natural scaling for the bHQET jet function renormalization scale is $\mu_J\sim \shat_\tau =Q^2(\tau - \taumin)/m_t$, which is linearly increasing with $\tau$ to the right of the peak and of order $\Gamma_t$ on the resonance region and below. The residual mass term $\delta m_t\equiv m_t-m_t^{\rm pole}$ specifies the renormalization scheme that is used for the top mass $m_t$, and enters through the replacement  $\shat \rightarrow \shat - \frac{Q^2}{m_t}\dv{\taumin}{m_t}\delta m_t$. In the pole mass scheme we have $\delta m_t=0$. In general $\delta m_t$ is a series starting at ${\cal O}(\alpha_s)$ and one has to consistently expand to ${\cal O}(\alpha_s)$ to obtain the bHQET jet function in any other top quark mass scheme. The mass schemes used in this work are explained in Sec.~\ref{sec:massSchemes}.

The soft function $\hat{S}_\tau\nflav{5}$ accounts for the effects of large-angle soft radiation with respect to the top-antitop jet axis at parton-level. It has the form
\begin{equation}
\label{eq:softfunction}
	\hat{S}_\tau^{(5)}(\ell,\bar{\delta}=0,\mu_S)=\delta(\ell)+\CFalphafive{\mu_S}\biggl[\frac{\pi^2}{3}\delta(\ell)-16\plusFunc{1}{\mu}{\ell}\biggr],
\end{equation}
with the natural scaling $\mu_S\sim \mu_J m_t/Q$ for its renormalization scale. In the resonance region $\mu_S\sim \Gamma_t m_t/Q=\Gamma_t/\varrho$, but the renormalization scale must be chosen such that $\mu_S$ still remains sufficiently perturbative. This also implies that $\mu_J$ is always set larger than the top quark width.

The large-angle soft radiation also has a non-perturbative component featuring scales of order $\Lambda_{\rm QCD}$, which arise from hadronization effects related to the soft exchange between the two hemispheres. In the resonance region they are implemented through the convolution of $\hat{S}_\tau^{(5)}$ with a non-perturbative model function $F(k)$~\cite{Hoang:2007vb}, referred to as the shape function,
\begin{equation}
\label{eq:softshapeconvolution}
S(\ell,\mu_S)=
\int \dd{k}\hat{S}_\tau^{(5)}(\ell-k,\bar{\delta},\mu_S)F(k-2\hat\Delta)\,,
\end{equation}
where the shift parameter $\hat\Delta$ accounts for the average minimum hadronic energy deposit in each hemisphere originating from hadron masses and is also referred to as the ``gap''~\cite{Hoang:2007vb}. More details on the gap and the concrete treatment of the dependence on $\bar{\delta}$ are given in Sec.~\ref{sec:gapsubtractions}. The form of Eq.~(\ref{eq:softshapeconvolution}) with the convolution of the partonic soft and shape functions provides a {\it first-principle QCD description} of the hadronization effects associated to the large-angle soft radiation tied to the hemisphere prescription of the shape variables we consider in our analysis. It has the advantage that the partonic component of the cross section, which is obtained setting $F(k)=\delta(k)$, is not modified. This entails in particular that all infrared properties of the parton-level cross section such as its renormalon structure remain intact and that the treatment of subleading power corrections is straightforward. In this context the shape function $F(k)$ has a form that peaks at $k \sim \lqcd$ and is normalized to unity. The model character of Eq.~(\ref{eq:softshapeconvolution}) arises from the particular form of the ansatz (including the gap parameter $\hat\Delta$) and the parametrization of the shape function in practical applications. We use the parametrization developed in Ref.~\cite{Ligeti:2008ac}, which has support for $k \ge 0$ and has the following form:
\begin{equation}
\label{eq:shapefunc}
F(k; \lambda,\{c_i\},N) = \frac{1}{\lambda}\Biggl[\,\sum_{n=0}^{N}c_n f_n\biggl(\frac{k}{\lambda}\biggr)\Biggr]^{\!2}\,,
\end{equation}
with
\begin{align}
\label{eq:shapefunc2}
f_n(z) ={}& 8\sqrt{\frac{2z^3(2n+1)}{3}}e^{-2z}P_n(g(z))\,,\\
g(z) ={}& \frac{2}{3}\Bigl[3-e^{-4z}\bigl(3+12z+24z^2+32z^3\bigr)\Bigr]-1\,,\nonumber
\end{align}
where $P_n$ are the Legendre polynomials and the normalization is fixed by $\sum_i c_i^2 = 1$. We truncate the sum over basis functions $f_n$ at $N=3$, which is sufficient to describe corrections to the peak shape due to non-perturbative effects. The function $f_0$ appearing in Eq.~(\ref{eq:shapefunc}) is positive definite and has one peak, while the functions $f_{n\ge 1}$ have $n$ zeros. The latter are less important for the shape of the cross section's peak, because the details of the shape function are smeared by the convolution. The width of the region where the $f_n$ functions have a sizable contribution is determined by the parameter $\lambda$, which is adjusted such that the series in $n$ converges rapidly and truncation in $N$ still allows to describe all relevant non-perturbative features in the resonance region. The most important quantity specifying the impact of the shape function on the peak distribution is the shape function's first moment 
\begin{equation}
\label{eq:O1def}
\Omega_1(\lambda,\hat\Delta,N)=\frac{1}{2}\int_0^\infty {\rm d}k\, k \,F(k-2\hat\Delta; \lambda,\{c_i\},N)\,,
\end{equation}
which reads 
\begin{align}
\label{eq:omega1formula1}
\Omega_1(\lambda,\hat\Delta,3) = \hat\Delta+\lambda\bigl(0.5\,c_0^2+ 0.47360764\,c_0c_1  + 0.10067713\,c_0c_2+ 0.094954074\,c_0c_3\\
	+ 0.54502418\,c_1^2 + 0.50700667\,c_1c_2 + 0.12507929\,c_1c_3  \nonumber\\
	+ 0.55015667\,c_2^2 + 0.50982331\,c_2c_3 + 0.55170015\,c_3^2\bigr)\,,\nonumber
\end{align}
and provides a quantitative measure of where the shape function peaks. We stress that the shape function, $\Omega_1$ and all other moments have a rigorous non-perturbative matrix element definition in QCD and are not model parameters~\cite{Korchemsky:1998ev,Lee:2006nr}. It is only the parametrization of the shape function with the truncation order $N$ that introduces model character in practical applications. As can be seen from the form of the factorization formula~(\ref{eq:bhqetdistribution}), the shape function shifts the peak location of the $\tau$ distribution by an amount $\Delta \tau\sim \Omega_1/Q$. For the top quark mass dependence this corresponds to a shift of \mbox{$\Delta m_t\sim\Omega_1 Q/m_t$}, which increases with $Q$.\footnote{We note that the physical first moment entering the calibration fits contains additional modifications explained in more detail in Sec.~\ref{sec:gapsubtractions}.}  Next to the top quark mass, the first moment of the shape function is therefore the other essential parameter that needs to be accounted for in the calibration fits. This dependence also illustrates the need to include MC samples produced for different $Q$ values in order to lift the degeneracy of the peak location concerning its dependence on $m_t$ and $\Omega_1$. We also note that away from the resonance peak, in the tail region of the $\tau$ distribution where $\ell \gg \lqcd$, it is in principle sufficient to use an operator product expansion (OPE) where the leading non-perturbative correction is related to $\Omega_1$. However, we always describe the non-perturbative effects through the convolution with the shape function, since this is fully compatible with the OPE description. 

The renormalization group (RG) evolution factors $U_{H_Q}\nflav{6}$, $U_B\nflav{5}$ and $U_S\nflav{5}$ appearing in the factorization formula~(\ref{eq:bhqetdistribution}) describe the renormalization-scale dependence of the hard $H_Q^{(6)}$, bHQET jet $J_{B,\tau}^{(5)}$ and soft $\hat{S}_\tau^{(5)}$ functions, respectively. The RG factor $U_{v}\nflav{5}$ describes the (5-flavor) evolution of the top-antitop production current matching in bHQET, which  compensates the combined $\mu$ dependence of the bHQET jet and soft functions. These evolution factors sum up large logarithms of ratios of the different physical scales arising in the resonance region. Due to RG consistency relations~\cite{Fleming:2007xt} not all of them are independent quantities. In Eq.~(\ref{eq:bhqetdistribution}), the (6-flavor) SCET current evolution only proceeds until the mass mode scale $\mu_m$ where the top quark off-shell mass modes are integrated out. The global scale $\mu$ should therefore be formally chosen below $\mu_m$. However, the dependence on $\mu$ cancels exactly and its specific value is irrelevant. The concrete expressions for the evolution factors are for convenience collected in App.~\ref{eq:evolutions}. Overall, we determine all evolution factors at N$^2$LL order using the inputs indicated in Tab.~\ref{tab:orders}. Here we use 4-loop running and 3-loop matching of $\alpha_s$ for the evolution of the strong coupling provided by the REvolver library~\cite{Hoang:2021fhn}.

\subsection{Renormalon Subtractions}
\label{sec:renormalonsubtractions}

\subsubsection{MSR Mass Scheme}
\label{sec:massSchemes}

For the top quark mass in our calibration analysis we employ the pole $m_t^{\rm pole}$ and MSR $m_t^{\rm MSR}(R)$ renormalization schemes. For the $m_t^{\rm pole}$ calibration all instances of $m_t$ are the pole mass without further modification. At the precision level of our N$^2$LL$\,+\,$NLO calibration analysis, which can reach $200$\,MeV, the size of the pole mass ${\cal O}(\Lambda_{\rm QCD})$ renormalon ambiguity already matters. Thus using $m_t^{\rm MSR}(R)$, which is a short-distance mass free of the ${\cal O}(\Lambda_{\rm QCD})$ renormalon, leads to a higher level of stability and smaller theoretical uncertainties~\cite{Butenschoen:2016lpz}. The MSR mass~\cite{Hoang:2008yj,Hoang:2017suc} is defined from the perturbative series for the difference between $m_t^{\rm pole}$ and the renormalon-free $\msbar$ mass at the $\msbar$ mass scale \mbox{$\overline{m}_t\equiv \overline{m}^{(6)}_t(\overline{m}^{(6)}_t)$} which reads
\mbox{$\mpole - \overline{m}_t=\overline{m}_t\sum_{n=1}a_n^{\overline{\mathrm{MS}}}(n_\ell=5,n_h=1)[\alpha_s^{(6)}(\overline{m}_t)/(4\pi)]^n$,} where the coefficients $a^{\overline{\mathrm{MS}}}_n(n_\ell,n_h)$ are known up to 4-loops~\cite{Tarrach:1980up,Gray:1990yh,Melnikov:2000qh,Chetyrkin:1999ys,Chetyrkin:1999qi,Marquard:2007uj}. The scale-dependent top $\msbar$ mass $\overline{m}^{(6)}_t(\mu)$ is a 6-flavor quantity. Here $n_\ell$ stands for number of massless flavors appearing in closed fermion loops and $n_h$ for those with mass $m_t$. The MSR mass (which is called `natural' MSR mass in Ref.~\cite{Hoang:2017suc}) is a 5-flavor quantity defined by integrating out all virtual top mass loops,
\begin{equation}
\delta m_t^{\rm MSR}(R)=\mpole - m_t^{\rm MSR}(R)=R\sum_{n=1}a_n^{\overline{\mathrm{MS}}}(n_\ell=5,0)\biggl[\frac{\alpha_s^{(5)}(R)}{4\pi}\biggr]^n\,. \label{eq:pMSR}
\end{equation}
The appearance of the scale $R$, which yields a linear RG $R$-evolution in contrast to the logarithmic $\mu$ evolution of the $\msb$ mass, is essential at low virtualities in the resonance bHQET region, where all radiation effects are governed by momentum scales much smaller than $m_t$. In the bHQET jet function $J_{B,\tau}^{(5)}$ this $R$ scaling is crucial since the absence of large logarithms implies the natural scale choice $R\sim\shat_\tau$ that cannot be realized for the $\msb$ mass. Since $\shat_\tau\sim\Gamma_t$ is small in the resonance region, the MSR mass $m_t^{\rm MSR}(R)$ with some scale $R\sim\Gamma_t$ is numerically close to the pole mass and therefore constitutes a kinematic mass like $m_t^{\rm pole}$.\footnote{Kinematic top quark mass schemes are sometimes also referred to as ``schemes consistent with the top quark's Breit Wigner line shape''~\cite{Schwienhorst:2022yqu}.} Note that for a complete cancellation of the ${\cal O}(\Lambda_{\rm QCD})$ renormalon it is mandatory to expand $\delta m_t^{\rm MSR}(R)$ in powers of $\alpha_s^{(5)}(\mu_J)$, where $\mu_J$ is the renormalization scale of the bHQET jet function.

At N$^2$LL$\,+\,$NLO we need the residual mass term at ${\cal O}(\alpha_s)$ which reads $\delta m_t^{\rm MSR}(R)=4\alpha_s^{(5)}(R)/(3\pi)$, and we employ $3$-loop $R$-evolution and $2$-loop matching to the $\overline{\rm MS}$ mass for $m_t^{\rm MSR}(R)$. For the MSR top mass calibration we employ $m_t^{\rm MSR}(1\,\mbox{GeV})$ as the input reference mass, following the convention used in the original calibration~\cite{Butenschoen:2016lpz}. Note that the MSR mass renormalization scale $R$ used in the theoretical description is tied to the jet function scale $\mu_J$, see Eq.~(\ref{sec:MSRRsetting}), which is typically in the range of $10$ to $20$\,GeV. Therefore, the choice of reference scale does not have any particular physical meaning and results at a different reference scale can be obtained using $R$-evolution at $3$-loops. The form of the $R$-evolution equation can be found up to $4$-loops in Ref.~\cite{Hoang:2017suc}, see also the App.~F of Ref.~\cite{Bachu:2020nqn} as well as Tab.~\ref{tab:massConversionTable}. We use the REvolver library~\cite{Hoang:2021fhn} for all RG evolution and the conversion between different mass schemes. REvolver also provides routines to convert to all other common top quark mass short-distance renormalization schemes used in the literature.\footnote{Note that in the original calibration analysis~\cite{Butenschoen:2016lpz} the so-called `practical' MSR mass definition was employed where top quark loop corrections are not fully integrated out. The difference to the `natural' MSR mass is at the level to $10$\,MeV~\cite{Hoang:2017suc} which is insignificant at the level of precision of our calibration framework.}

We note that the $\msb$ mass $\overline{m}^{(6)}_t(\mu)$ is also close to the pole mass for scales around $\mu=80$\,GeV (see e.g.\ Fig.~5 in Ref.~\cite{Hoang:2017suc}). This may erroneously be interpreted as a fact supporting the use of the $\msb$ mass as a low-scale short distance mass in the bHQET jet function. However, the unphysical logarithmic $\mu$-dependence of $\overline{m}^{(6)}_t(\mu)$ for these low scales is much stronger than the linear $m_t^{\rm MSR}(R)$ evolution for $R\sim\Gamma_t$, which at the practical level makes it hard to achieve high precision when scale variations are accounted for. At the conceptual level, the fact $\overline{m}^{(6)}_t(80\,\mbox{GeV})\approx m_t^{\rm pole}$ (in the absence of electroweak corrections) should be viewed as purely accidental as it involves the summation of large logarithmic corrections in $m_t^{\rm pole}-\overline{m}^{(6)}_t(\mu=80\,\mbox{GeV})$ to all orders. In fact, the ${\cal O}(\alpha_s)$ residual mass term for the $\msb$ mass $\delta \overline{m}(\mu)=4\,\overline{m}^{(6)}_t(\mu)\alpha_s^{(6)}(\mu)/(3\pi)$ cannot be consistently used in the bHQET jet function as its size by far exceeds that of dynamical QCD corrections in the peak region, no matter which choice of $\mu$ is adopted. This is related to the fact that the logarithms that are summed in $\overline{m}^{(6)}_t(\mu)$ for $\mu<m_t$ are not compatible with the low-scale bHQET dynamics in the heavy top quark rest frame.

\subsubsection{Soft Gap Subtraction Schemes}
\label{sec:gapsubtractions}

The parton level soft function $\hat{S}_\tau^{(5)}(\ell,\bar{\delta}=0,\mu_S)$ in Eq.~(\ref{eq:softfunction}) has a leading ${\cal O}(\Lambda_{\rm QCD})$ renormalon similar to the bHQET jet function in the pole mass scheme which also leads to instabilities of the partonic threshold. While the pole mass ${\cal O}(\Lambda_{\rm QCD})$ renormalon can be removed by a quark mass scheme change (and is therefore an artificial theoretical issue) the renormalon in the partonic soft function is physical and related to a non-perturbative effect. If we do not deal with this renormalon, eventually, at high orders, we would find instabilities in our calibration fits for the shape function's first moment $\Omega_1$ in Eq.~(\ref{eq:omega1formula1}). Due to the linear dependence of $\Omega_1$ on the non-perturbative gap parameter $\hat \Delta$ we can associate its renormalon instability to $\hat\Delta$. Thus, given a perturbative series $\bar{\delta}(R_s,\mu_S)$ in powers of $\alpha_s^{(5)}(\mu_S)$ that precisely reproduces the soft function ${\cal O}(\Lambda_{\rm QCD})$ renormalon asymptotic behavior, called the \textit{gap subtraction} series, we can remove this renormalon. This is achieved using the gap formalism~\cite{Hoang:2007vb} which starts from the combined perturbative and non-perturbative soft function
\begin{align}
	S(\ell,\mu_S) =\,& 
	\int \dd{k}\hat{S}_\tau^{(5)}(\ell-k,\bar{\delta}=0,\mu_S)F(k-2\Delta)\,,
\end{align}
where both the partonic soft function $\hat{S}_\tau^{(5)}(\ell,\bar{\delta}=0,\mu_S)$ and the shape function \mbox{$F(k-2\Delta)$}, through its dependence on $\Delta$, still contain the ${\cal O}(\Lambda_{\rm QCD})$ ambiguity. We now write $\Delta = \overline{\Delta}(R_s,\mu_S) +\bar{\delta}(R_s, \mu_S)$, where $\Delta$ is strictly scale-independent (in analogy to the pole mass). Since $\Delta$ has the dimension of energy and the soft function renormalon in $\Delta$ scales with \mbox{$\ell\sim \shat_\tau m_t/Q=\shat_\tau/\varrho$}, the gap subtraction series has the dimension of energy as well through an overall factor $R_s$ with the natural scale choice $R_s\sim \shat_\tau/\varrho$~\cite{Hoang:2007vb}:
\begin{equation}
\label{eq:gapdef}
	\bar{\delta}(R_s, \mu_S) =\Delta-\overline{\Delta}(R_s,\mu_S) =R_s\! \sum_{i=1} d_i( R_s,\mu_S)\! \alphaflavpi{5}{\mu_S}^{\!i}\,.
\end{equation}
The scale $R_s$ and the renormalon-free gap parameter $\overline{\Delta}(R_s,\mu_S)$ play roles in close analogy to the scale $R$ and 
the MSR mass $m_t^{\rm MSR}(R)$, where $\overline{\Delta}(R_s,\mu_S)$ also satisfies a linear RG equation in $R_s$. We keep the argument $\mu_S$ in $\overline{\Delta}(R_s,\mu_S)$ since it, depending on the gap choice, may not be RG invariant with respect to $\mu_S$. The gap subtraction series can now be shifted into the partonic soft function in the convolution of Eq.~(\ref{eq:softshapeconvolution}) yielding~\cite{Hoang:2007vb}
\begin{align}
\label{eq:Sgapderivation}
S(\ell,\mu_S) =\,& 
\int \dd{k}\hat{S}_\tau^{(5)}(\ell-k,\bar{\delta}=0,\mu_S)F(k-2\Delta)  \\
=\, &
\int \dd{k}\hat{S}_\tau^{(5)}(\ell-2\bar{\delta}(R_s,\mu_S)-k,0,\mu_S)F(k-2\overline{\Delta}(R_s,\mu_S))\nonumber \\
=\, &
\int \dd{k}\hat{S}_\tau^{(5)}(\ell-k,\bar{\delta}(R_s,\mu_S),\mu_S)F(k-2\overline{\Delta}(R_s,\mu_S))
\,,\nonumber
\end{align}
where the last equality, together with Eqs.~(\ref{eq:Deltahat}) and (\ref{eq:O1formula}) given below, define the form for the soft function shown in Eq.~(\ref{eq:bhqetdistribution}). Note that for the gap subtraction different schemes can be adopted (which are discussed in the following). This scheme dependence is suppressed in this notation. As for the residual mass term, $\overline{\delta}(R_s,\mu_S)$ needs to be consistently expanded out together with the soft function in powers of $\alpha_s^{(5)}(\mu_S)$ such that the corresponding renormalon is removed order-by-order. The renormalon-free gap parameter $\overline{\Delta}(R_s,\mu_S)$, which depends on the scheme choice for $\bar{\delta}(R_s,\mu_S)$ and obeys a RG evolution equation in $R_s$ (and potentially also in $\mu_S$) remains in the shape function. Since $R_s$ and $\mu_S$ are in general \mbox{$\tau$-dependent} in order to properly sum all logarithms, see Sec.~\ref{sec:profiles}, we adopt $\Delta_0\equiv \overline{\Delta}(R_\Delta,R_\Delta)$ at the reference scale $R_\Delta=2$\,GeV as the specified input and determine $\overline{\Delta}(R_s,\mu_S)$ through its $R_s$ (and potentially $\mu_S$) evolution equation(s). 

A general parametrization for suitable subtraction schemes, collectively referred to as R-gap schemes, has been introduced in Ref.~\cite{Bachu:2020nqn} by imposing a general condition on the soft function at a point in position space,
\begin{equation}
	\label{eq:softFourier}
	\tilde{S}_\tau^{(5)}(y, \mu) = \int\dd{\ell} e^{-i y \ell} \hat{S}_\tau^{(5)}(\ell, \mu) =  \exp\Biggl\{\sum_{i=1} \alphaflavpi{5}{\mu}^{\!i} \sum_{j=0}^{i+1} s_{ij} \ln^j(i e^{\gamma_E} y \mu) \Biggr\}.
\end{equation}
The solution is given by
\begin{equation}
	\gap(R_s,\mu_S;\mathrm{A},n,\xi)\equiv\begin{cases}
		\frac{R_s}{2\xi}\dv[n]{\ln(iy)}\ln\Bigl[\tilde{S}_\tau(y,\mu_S)\Bigr]_{iy=\frac{\xi}{R_s}}& \text{if A\,$=$\,on} \\
		\frac{R_s}{2\xi}\dv[n]{\ln(iy)}\ln\Bigl[\tilde{S}_\tau(y,R_s)\Bigr]_{iy=\frac{\xi}{R_s}}& \text{if A\,$=$\,off}
	\end{cases}\,.
\end{equation}
A relation to obtain the coefficients $s_{ij}$ in terms of $s_{i0}$, the coefficients of the cusp and non-cusp partonic soft function anomalous dimensions, and the QCD $\beta$-function is provided in App.~C.2 of Ref.~\cite{Bachu:2020nqn}. The switch A turns the non-trivial anomalous dimension in $\mu_S$ on or off. When A$\,=\,$on the scale of the strong coupling in the subtraction series is $\mu_S$ by construction, such that $\overline{\Delta}(R_s, \mu_S)$ and the gap series $\gap(R_s, \mu_S)$ satisfy RG equations in $R_s$ and $\mu_S$. For A$\,=\,$off a gap subtraction series is defined such that it only depends on $R_s$ so that $\overline{\Delta}(R_s, \mu_S)$ and $\gap(R_s, \mu_S)$ satisfy an RG equation in $R_s$, but are $\mu_S$-invariant. 
%We remind the reader that the strong coupling of the subtraction series for the latter case has to be expanded in terms of $\alpha_s(\mu_S)$, so that the renormalon can be properly canceled. 
In this work we employ three different gap subtraction schemes to test the gap scheme dependence of the calibration results:
\begin{align}
\label{eq:gap1defshort}
		\gap^{(1)}(R_s,\mu_S)\equiv{}&\gap(R_s,\mu_S;\text{on},1,e^{-\gamma_E})\,,\\
		\gap^{(2)}(R_s,\mu_S)\equiv{}&\gap(R_s,\mu_S;\text{off},0,e^{5\gamma_E})\,,\nonumber\\
		\gap^{(3)}(R_s,\mu_S)\equiv{}&\gap(R_s,\mu_S;\text{off},0,1)\,\nonumber.
\end{align}
Scheme 1 was the first realization of a gap subtraction and originally devised in Ref.~\cite{Hoang:2008fs}. It was then applied for strong coupling determinations from $e^+e^-$ event-shape data in Refs.~\cite{Abbate:2010xh,Abbate:2012jh,Hoang:2014wka,Hoang:2015hka}. It was also used in the original \pythia top mass calibration of Ref.~\cite{Butenschoen:2016lpz}. The gap subtraction series reads
\begin{equation}
\label{eq:gap1def}
%		\gap_1(R_s,\mu_S)={}&\frac{R_s e^{\gamma_E}}{2}\sum_{i=1}\alphaflavpi{5}{\mu_S}^i\sum_{j=0}^{i}(j+1)s_{i,j+1}\ln^j{\frac{\mu_S}{R_s}}
		\gap^{(1)}(R_s,\mu_S)=\frac{R_s e^{\gamma_E}}{2}\sum_{i=1}\alphaflavpi{5}{\mu_S}^{\!i}\sum_{j=0}^{i}(j+1)s_{i,j+1}L_R^j\,,
\end{equation}
where $L_R\equiv \ln(\mu_S/R_s)$. Explicit results for $\gap^{(1)}$ and the $R_s$ as well as $\mu_S$ evolution equations can e.g.\ be found in Section~2.F of Ref.~\cite{Abbate:2010xh}. The choice $n=1$ concerning the number of $y$-derivatives in Eq.~\eqref{eq:gap1defshort} sets the non-logarithmic coefficient to zero since ${s_{11}}=0$, so that $d_1^{(1)}(R_s,\mu_S)=-18.9981 L_R$. This implies that at ${\cal O}(\alpha_s)$ the gap subtraction in scheme 1 is zero for the choice $R_s=\mu_S$. A subtraction with the proper sign is only achieved if $R_s<\mu_S$. Therefore, in this scheme $R_s$ has to be strictly set below the soft renormalization scale $\mu_S$ to achieve a useful subtraction term with the proper sign at ${\cal O}(\alpha_s)$ in the peak region.

Gap scheme 3 was devised in Ref.~\cite{Bachu:2020nqn} in a phenomenological analysis of the bHQET factorization formula~(\ref{eq:bhqetdistribution}) at N$^3$LL to allow for the setting $R_s=\mu_S$, since using $R_s<\mu_S$ in the peak region can lead to an unstable behavior of the N$^3$LL corrections due to larger values of $\alpha_s(\mu_S)$. This is achieved by using the position-space partonic soft function in Eq.~(\ref{eq:gap1defshort}) without any $y$-derivative (i.e.\ $n=0$). The subtraction series has the form
\begin{equation}
\label{eq:gap3subtractionseries}
	\gap^{(3)}(R_s,\mu_S)={}\frac{R_s}{2}\sum_{i=1}\alphaflavpi{5}{R_s}^{\!i}\sum_{k=0}^{i+1}s_{ik}\gamma_E^k\,.
\end{equation}
The gap subtraction series of $\gap^{(3)}$ has a sizable ${\cal O}(\alpha_s)$ term $d_1^{(3)}(R_s,\mu_S)=-8.35669$, see Eq.~(\ref{eq:gapdef}). Gap scheme 3 is $\mu_S$-invariant, but retains a residual dependence on the soft scale $\mu_S$ at any finite order once the strong coupling is expanded in powers of $\alpha_s^{(5)}(\mu_S)$ as required by renormalon cancellation. We have noticed in our numerical studies that gap scheme~3 can yield some unphysical behavior of the $\tau$ distribution in the transition from the resonance peak to the tail region when paired together with the pole mass scheme and using profile functions with fast changing scales. This is caused by the sizable constant ${\cal O}(\alpha_s)$ term $d_1^{(3)}(R_s,\mu_S)$ in $\gap^{(3)}$ which in turn leads to a quite rapid evolution of $\overline{\Delta}_3(R_s,R_s)$ in $R_s$. For a strongly increasing profile for $R_s(\tau)=\mu_S(\tau)$ to the right of the peak region this can give rise to a severe cancellation of the $\tau$-dependence in $\shat_\tau$ and $\overline{\Delta}_3(R_s(\tau),R_s(\tau))$ in the factorization formula~(\ref{eq:bhqetdistribution}), so that the distribution does not show any more a falling tail. As we show in Secs.~\ref{sec:shapefctfits}, \ref{sec:pcanalysis} and \ref{sec:finalresults}, this can result in larger calibration uncertainties and instabilities for the top quark pole mass which are, however, an artifact of gap scheme~3. If the MSR mass scheme is adopted, this feature is absent, since the $\tau$ dependence of $m_t^{\rm MSR}(R(\tau))$ through its profile $R(\tau)$ partly cancels the $\tau$ dependence of $\overline{\Delta}_3(R_s(\tau),R_s(\tau))$, see also Sec.~5.B of Ref.~\cite{Bachu:2020nqn}. Even though one may argue that this is yet another argument that disfavors the use of $m_t^{\rm pole}$, we do not adopt this point of view because this feature does not arise in general.

The problematic feature of gap scheme~3 in the pole mass scheme motivates the introduction of gap scheme~2
\begin{equation}
		\gap^{(2)}(R_s,\mu_S)=\frac{R_s}{2e^{5\gamma_E}}\sum_{i=1}\alphaflavpi{5}{R_s}^{\!i}\sum_{k=0}^{i+1}(6\gamma_E)^k s_{ik}\,,
\end{equation}
which differs from gap 3 by setting $\xi$ to $e^{5\gamma_E}$ instead of 1. For this $\xi$ value the non-logarithmic ${\cal O}(\alpha_s)$ term $d_1^{(2)}( R_s,\mu_S)=-3.9363$ is substantially smaller than for gap~3 such that the glitch mentioned above does not arise. One can consider gap scheme~2 to be halfway between gap schemes~1 and 3, which also motivates our numbering. Nevertheless, for $R_s=\mu_S$ gap scheme~2 is very effective in removing the soft function renormalon and will therefore be the gap scheme we use for quoting the final calibration results. Complete formulae for $\gap^{(i)}(R_s,\mu_S)$ for the three gap schemes and the resulting $R_s$-evolution equations for $\overline{\Delta}_{1,2,3}(R_s,\mu_S)$, which we employ at 2-loops, are given in App.~\ref{app:Revo}. The subtraction series $\gap^{(i)}$ are only needed to one-loop at N$^2$LL$\,+\,$NLO order. 

Through the shape function's dependence on $\overline{\Delta}^{(i)}(R_s,\mu_S)$, where $i$ stands for the gap scheme, the gap parameter $\hat{\Delta}$ in the shape function in Eqs.~(\ref{eq:bhqetdistribution}) and (\ref{eq:softshapeconvolution}) gains scheme dependence and evolves with $R_s$ and (potentially) $\mu_S$, which themselves are $\tau$-dependent as well. The concrete expression for $\hat{\Delta}$  reads~\cite{Abbate:2010xh,Abbate:2012jh,Hoang:2014wka,Hoang:2015hka,Bachu:2020nqn}
\begin{equation}
\label{eq:Deltahat}
\hat\Delta  \equiv \hat\Delta^{(i)}(R_s,\mu_S)  =\Delta_0 + \bigl[\,\overline{\Delta}^{(i)}(R_s,\mu_S)-\overline{\Delta}^{(i)}(R_\Delta,R_\Delta)\bigr]\,,
\end{equation}
where $\Delta_0$ is a free parameter that agrees with the reference value $\overline{\Delta}^{(i)}(R_\Delta,R_\Delta)$, and the difference $\bigl[\,\overline{\Delta}^{(i)}(R_s,\mu_S)-\overline{\Delta}^{(i)}(R_\Delta,R_\Delta)\bigr]$ is obtained from solving the evolution equation(s). This also results in a scale-dependent first shape-function moment
\begin{equation}
\label{eq:O1formula}
	\Omega_1^{(i)}(R_s,\mu_S) \equiv \Omega_1(\lambda,\hat{\Delta}^{(i)}(R_s,\mu_S),3)\,,
\end{equation}
where the expression for $\Omega_1(\lambda,\Delta,3)$ is given in Eq.~(\ref{eq:omega1formula1}). 

We note that the term $\Delta_0$ represents an additional parameter of the shape function besides $\lambda$ and the coefficients $c_i$, see Eq.~(\ref{eq:shapefunc}). Both parameters are in principle redundant if the coefficients $c_i$ provide sufficient flexibility in the calibration fits. For a large value of $N$ this would be automatically ensured, but in phenomenological applications $N$ must be chosen sufficiently small to be practical. In Refs.~\cite{Abbate:2010xh,Abbate:2012jh,Hoang:2014wka,Hoang:2015hka,Bachu:2020nqn} and the original $m_t^{\rm MC}$ calibration analysis~\cite{Butenschoen:2016lpz}, where gap scheme~1 was employed, $\Delta_0=0.05$\,GeV and $\lambda=0.5$\,GeV were used (i.e.\ they were not fit parameters), and it was checked that the coefficients $c_i$ with a proper choice of $N$ provide sufficient flexibility for carrying out phenomenologically meaningful fits. For other gap schemes, this flexibility needs to be reinvestigated, which is the topic of Sec.~\ref{sec:shapefctfits}. We also note that $\Omega_1$ without any soft function renormalon subtraction (i.e.\ for $\bar\delta=0$) was referred to as $\overline\Omega_1$ in Refs.~\cite{Abbate:2010xh,Abbate:2012jh,Hoang:2014wka,Hoang:2015hka,Bachu:2020nqn}.

It is the first moment at the reference scale $R_\Delta=2$\,GeV, namely 
\begin{equation}
\label{eq:O1ref}
\Omega_1^{(i)}(R_\Delta) \equiv \Omega_1^{(i)}(R_\Delta,R_\Delta)\,,
\end{equation}
which we quote in the presentation of the results for the $m_t^{\rm MC}$ calibration. To show the outcome of our analyses in the different gap schemes, and to visualize the gap-scheme independence of the calibration, it is useful to convert the results for the $\Omega_1^{(i)}(R_\Delta,R_\Delta)$ to a common reference scheme. Since gap scheme~1 was the first available in the literature, we pick it as our reference. The corresponding conversion formulae are obtained from the relation $\Omega_1^{(i)}(R_\Delta)-\Omega_1^{(1)}(R_\Delta) = \bar{\delta}^{(1)}(R_\Delta,R_\Delta)-\bar{\delta}^{(i)}(R_\Delta,R_\Delta)$ and read
\begin{align}
\label{eq:gapconversion}
\Omega_1^{(3)}(R_\Delta)-&\Omega_1^{(1)}(R_\Delta)
= R_\Delta\,\Biggl\{ 8.3567 \Biggl[\frac{\alpha_s^{(5)}(R_\Delta)}{4\pi}\Biggr]+ 28.49 \Biggl[\frac{\alpha_s^{(5)}(R_\Delta)}{4\pi}\Biggr]^{\!2}+\ldots \Biggr\} ,\\
\Omega_1^{(2)}(R_\Delta)-&\Omega_1^{(1)}(R_\Delta)
 = R_\Delta\,\Biggl\{ 3.9363\Biggl[\frac{\alpha_s^{(5)}(R_\Delta)}{4\pi}\Biggr] + 50.92 \Biggl[\frac{\alpha_s^{(5)}(R_\Delta)}{4\pi}\Biggr]^{\!2}+\ldots \Biggr\} .
\nonumber
\end{align}

\subsection{Profile Functions}
\label{sec:profiles}
The bHQET $\tau$ distribution in the resonance region depends on the natural renormalization scales $\mu_H$, $\mu_m$, $\mu_J$ and $\mu_S$ of the hard, mass-mode, bHQET jet and partonic soft functions, as well as on the soft renormalon subtraction scale $R_s$ and, if applicable, the MSR top mass scale $R$. Formally, at the all-order level, these scale dependences would vanish, but at any finite order a residual dependence remains, which we utilize as a quantification for the theoretical uncertainty of our N$^2$LL$\,+\,$NLO description. While all scales can be considered as $\tau$-independent directly on the peak, where the scale hierarchy is the largest, only $\mu_H$ and $\mu_m$ are also constant away from the peak. The scales $\mu_J$, $\mu_S$, $R_s$ and $R$, on the other hand, are in general $\tau$-dependent as already explained in Sec.~\ref{sec:facttheorem}. While these scales should be varied  to obtain an adequate theory uncertainty estimate, they also need to obey some physical correlations so that the natural scaling  hierarchy is not upset. This is achieved by profile functions for all renormalization scales. For the differential distribution for massive quark production  in the entire $\tau$ spectrum, an efficient parametrization of these profile functions was designed in Ref.~\cite{Dehnadi:2016snl}, which is a generalization of the profile functions used for massless event-shape distributions designed and employed earlier in Refs.~\cite{Abbate:2010xh,Abbate:2012jh,Hoang:2014wka,Hoang:2015hka}. This profile parametrization applies to top and bottom quark production. The formulae for the profile functions of Ref.~\cite{Dehnadi:2016snl} in the resonance region, which we need for the calibration analysis, were also presented in Ref.~\cite{Bachu:2020nqn}. Here, we review some basic aspects of these profile functions in the resonance region and point out some differences concerning the range of variations of the profile function parameters used in this article compared to the original calibration work of Ref.~\cite{Butenschoen:2016lpz} and to the N$^3$LL analysis of Ref.~\cite{Bachu:2020nqn}.

The $\tau$-independent hard function and mass-mode matching scales are $\mu_H=e_H Q$ and $\mu_m=\sqrt{e_H}\,m_t$, where the rescaling parameter $e_H$ is varied in the interval $[0.5,2]$ with a default value $e_H=1$. They are correlated to retain the correct scale hierarchy. The mass parameter $m_t$ used for $\mu_m$ and also in formulae below is fixed to the MSR mass $m_t^{\rm MSR}(5\,\mbox{GeV})$ such that the profile functions can properly adapt to the top quark mass dependence of the peak-region $\tau$ range. Within the profile function variations we adopt, the scale choice of $5$\,GeV is simply a matter of convenience, and any other low scale larger than $1$\,GeV would yield equivalent results.

The general form of the $\tau$-dependent jet and soft profile functions are given by piece-wise functions, which describe the \textit{non-perturbative} ($\tau<t_0$), \textit{resummation} ($t_1<\tau<t_2$) and \textit{fixed-order} ($\tau>t_s$) regions, where  $t_0<t_1<t_2<t_s$. In the non-perturbative region the scales are frozen at a low but still perturbative value. In the resummation region the profiles grow steadily and in the fixed-order region they merge with the hard function scale $\mu_H$. These three regions are connected by transition regions, which allow the piece-wise functions $F(\tau<t_a)$ and $G(\tau>t_b)$ to be smoothly connected by a double quadratic function $\zeta(F(\tau),G(\tau),t_a,t_b,\tau)$ for $t_a<\tau<t_b$, which has been given e.g.\ in Eq.~(74) of Ref.~\cite{Hoang:2014wka}.
Since the calibration only concerns the resonance region, where the bHQET description is sufficient, we only need the profile functions in the non-perturbative and the transition to resummation regions, so that only $t_0$ and $t_1$ are relevant. The boundary $t_0$ is located to the right of the peak position and the condition $\tau>t_0$ roughly indicates the region where the OPE description with the first moment $\Omega_1$ and the effects of the shape function agree to better than $2\%$. The boundary $t_1$ is located in the tail, where the distribution reaches about half of the peak height. They read~\cite{Dehnadi:2016snl}
\begin{align}\label{eq:profBoundaries}
t_0&{}=\frac{2}{Q/(\SI{1}{\GeV})}+\frac{d_0}{[Q/(\SI{1}{\GeV})]^{0.5}}+\taumin\biggl(\frac{m_t^2}{Q^2}\biggr),\\
t_1&{}=\frac{2.25}{[Q/(\SI{1}{\GeV})]^{0.75}}+\frac{d_1}{[Q/(\SI{1}{\GeV})]^{0.5}}+\taumin\biggl(\frac{m_t^2}{Q^2}\biggr),\nonumber
\end{align}
where $\taumin(m_t^2/Q^2)$ refers to the minimal stable quark $\tau$ values for the different shape variables given in Sec.~\ref{sec:observables} as a function of the top mass. This introduces two additional profile parameters $d_{0,1}$ which are varied in the interval $[-0.05,+0.05]$, with zero as their default value.

The canonical scaling for the soft function is $\mu_S\sim Q(\tau - \tau_{\rm min})\sim \mu_H (\tau - \tau_{\rm min})$. To allow for small logarithms in the soft function, it is sufficient to consider the soft scale proportional to its canonical form, being the overall scaling factor $(1+e_S)r_{\rm slope}\sim \mathcal{O}(1)$ in the massless quark limit.  Numerical investigations~\cite{Dehnadi:2016snl} indicate that the proportionality factor needs to depend on the quark mass such that the same profile can be used for top and bottom quarks. To that end, we add a mass-dependent prefactor which depends on the free parameter $n_s$ parametrizing the somewhat arbitrariness of the choice. At the peak location and for $\tau$ values below, non-perturbative effects set in, which modify the parton-level motivated canonical scaling, resulting in a low constant value which should be chosen larger than $\Lambda_{\rm QCD}$. The two regions are matched with a double quadratic $\zeta$ function. All in all, the soft function scale profile is given by
\begin{equation}
\mu_S(\tau)=
\begin{cases}
\mu_0 & \tau< t_0\\
\zeta\left[\mu_S(\tau< t_0),\mu_S(\tau> t_1),t_0,t_1,\tau\right] & t_0\le\tau\le t_1\\
\Bigg[ 1+\frac{n_s e_S}{n_s+\frac{m_t}{Q}-\taumin\bigl(\frac{m_t^2}{Q^2}\bigr)}\Bigg]r_\mathrm{slope}\,\mu_H\!\Bigl[\tau-\taumin\Bigl(\frac{m_t^2}{Q^2}\Bigr)\!\Bigr] & t_1<\tau 
\end{cases}.
\end{equation}
This incorporates four more profile parameters $\mu_0$, $n_s$, $e_s$ and $r_{\rm slope}$, where $n_s\in[0.375,0.425]$ with default value $n_s=0.4$ and $e_S\in[1/1.13-1,1.13-1]$ with default value $e_S=0$. The parameter $r_{\rm slope}$ determines the soft-scale slope in the SCET region and, guided by the studies carried out in Ref.~\cite{Hoang:2014wka}, it is fixed to $r_{\rm slope}=2$. The form of $\mu_S(\tau>t_1)$ affects the calibration results only indirectly through the transition function $\zeta$, since our fit ranges only involve values $\tau<t_1$. The parameter $\mu_0$ has the largest influence and sets the soft scale in the non-perturbative region. It is varied in the interval $\mu_0\in[2.5,3.5]$\,GeV with the default value $\mu_0=3$\,GeV. These choices of the default values (including those discussed below), which are compatible with the canonical scaling, mostly affect the NLL theory predictions, but have only very little impact on the $\tau$ distribution at N$^2$LL$\,+\,$NLO due the reduced renormalization scale sensitivity at that order.

The bHQET jet-function scale profile reads
\begin{equation}
\mu_J(\tau)=
\begin{cases}
\bigl[ 1+\tilde{e}_J(t_s-t_0)^2\bigr]\tilde{\mu}_J(t_0) & \tau< t_0\\
\zeta\left[\mu_J(\tau< t_0),\mu_J(\tau> t_1),t_0,t_1,\tau\right] & t_0\le\tau\le t_1\\
\bigl[ 1+\tilde{e}_J(t_s-\tau)^2\bigr]\tilde{\mu}_J(\tau) & t_1<\tau
\end{cases},
\end{equation}
where $\tilde{\mu}_J(\tau)=\sqrt{e_H}\mu_S(\tau) Q/m_t$, $t_s=n_s+m_t/Q$ and
\begin{equation}
\tilde{e}_J=e_J\!\left\{\frac{n_s-\Bigl[t_0-\taumin\!\Bigl(\frac{m_t^2}{Q^2}\Bigr)\Bigr]}{t_s-t_0}\right\}^{\!\!2}.
\end{equation}
It is built on the generic bHQET jet scale function $\tilde{\mu}_J(\tau)$, which encodes the natural relation of the hard, jet and soft scales, with modulations controlled by the parameter \mbox{$e_J\in[-3,0]$} with default value \mbox{$e_J=-1.5$}, that is constructed to have no effect in the fixed-order region far above the resonance. We refer to Ref.~\cite{Dehnadi:2016snl} for more details. The additional fixed-order region parameter $n_s\in[0.375,0.425]$ has very little impact, and its default value is $n_s=0.4$. 

The soft function renormalon subtraction scale $R_s$ has to be close to the soft scale $\mu_S$, but we need two different prescriptions, one for gap scheme~1, where one should use $R_s<\mu_S$, and another one for gap schemes~2 and 3, where we use $R_s=\mu_S$. For gap scheme~1 we employ
\begin{equation}
\label{eq:Rs_scale1}
R^{(1)}_S(\tau)=
\begin{cases}
0.75\mu_0 & \tau< t_0\\
\zeta\Bigl[R^{(1)}_S(\tau< t_0),R^{(1)}_S(\tau> t_1),t_0,t_1,\tau\Bigr] & t_0\le\tau\le t_1\\
\mu_S(\tau) & t_1<\tau  
\end{cases},
\end{equation}
while for gap scheme 2 and 3 we use
\begin{equation}
\label{eq:Rs_scale23}
R^{(2,3)}_S(\tau)= \mu_S(\tau) \,.
\end{equation}
The renormalization scale of the MSR mass $m_t^\mathrm{MSR}(R)$ is always set to the jet scale:
\begin{equation}
\label{sec:MSRRsetting}
R(\tau)=\mu_J(\tau)\,.
\end{equation}
The renormalization scale for the remaining fixed-order QCD corrections at NLO that are not accounted for in the bHQET and SCET factorization formula, see Sec.~\ref{sec:non-singular}, is denoted by $\mu_\mathrm{ns}$. It is set to a weighted average of the hard and jet scales,
\begin{equation}
\label{eq:munsdef}
\mu_\mathrm{ns}(\tau)=\mu_H+\frac{n_\mathrm{ns}}{2}\bigl[\,\mu_J(\tau)-\mu_H\bigr]\,,
\end{equation}
where for the label $n_\mathrm{ns}$ we adopt the values $\{-1,0,1\}$ with the default value zero. The ranges of variations of all profile function parameters are collected in Tab.~\ref{tab:scalevariation}. For the calibration we use $501$ sets of profile function parameters randomly chosen in the ranges shown in Tab.~\ref{tab:scalevariation} (with flat probability distributions), where for the first profile all parameters are set to their default values.

\begin{table}[t]\label{tab:profile_params}
	\centering
	\begin{tabular}{ c | c | c }
		\hline
		\hline
		parameters & default value & range of values\\ 
		\hline
		$n_s$ & $0.4$ & $0.375$ to $0.425$\\
		$d_0$ & $0$ & $-0.05$ to $0.05$\\
		$d_1$ & $0$ & $-0.05$ to $0.05$\\
		$\mu_0$ & $3\mathrm{\,GeV}$ & $2.5\mathrm{\,GeV}$ to $3.5\mathrm{\,GeV}$\\
		$r_\mathrm{slope}$ & $2$ & --\\
		$e_H$ & $1$ & $0.5$ to $2$\\
		$e_S$ & $0$ & $1/1.13-1$ to $1.13-1$\\
		$e_J$ & $-1.5$ & $-3$ to $0$\\
		$n_\mathrm{ns}$ & $0$  & $-1, 0, 1$\\
		\hline
		\hline
	\end{tabular}
	\caption{Parameter ranges for the profile function scale variation. }
	\label{tab:scalevariation} 
\end{table}

The profile function formulae and parameters employed here are identical to the ones used for the original calibration~\cite{Butenschoen:2016lpz} and in the analyses of Ref.~\cite{Dehnadi:2016snl}, except for the gap~2 and 3 renormalon subtraction scales $R^{(2,3)}_S$ since there only gap~1 was considered. In Ref.~\cite{Dehnadi:2016snl} the parameter ranges have been tested extensively at N$^2$LL$\,+\,$NLO, where also the SCET and QCD non-singular corrections were accounted for. In the analysis of Ref.~\cite{Bachu:2020nqn} the different variations $\mu_0\in [3/\sqrt{2},3\sqrt{2}]$, $e_J\in [-1.5,1.5]$ and $n_s\in [0.475,0.525]$  were adopted. For gap~1, which was not analyzed in Ref.~\cite{Bachu:2020nqn}, the larger $\mu_0$ variation is not suitable since $R^{(1)}_S$ in Eq.~(\ref{eq:Rs_scale1}) can become too low. Furthermore, in the  analysis of Ref.~\cite{Bachu:2020nqn} the singular bHQET factorization formula of Eq.~(\ref{eq:bhqetdistribution}) was determined and analyzed at N$^3$LL order, but did not account for the non-singular SCET or QCD corrections. The different variation ranges for $e_J$ and $n_s$ used there yielded better convergence for these singular contributions. The difference is associated to the non-singular corrections, which, as we show in the subsequent section, are not small.

\subsection{Non-singular Corrections}
\label{sec:non-singular}

The bHQET factorization formula for the resummed singular $\tau$ distribution valid in the resonance region discussed in Sec.~\ref{sec:facttheorem} and shown in Eq.~(\ref{eq:bhqetdistribution}) contains the leading distributional and non-perturbative corrections in an expansion in $\mhat_t=m_t/Q$, $\lqcd/Q$ and $\Gamma_t/m_t$~\cite{Fleming:2007qr,Fleming:2007xt}. For reliable phenomenological applications, however, formally subleading power corrections need to be accounted for since they are not negligible. These can be included by recovering contributions that have been integrated out in the two-step matching from QCD to SCET at the scale $Q$ and then from SCET to bHQET at the scale $m_t$. The procedure to recover and include these subleading power corrections, which are called non-singular or matching corrections, is in general not unique since one may absorb some of them already in the singular bHQET factorization formula. At this point we remind the reader that using the term ``non-singular'' is somewhat misleading for the case of massive quark production, since the distributional terms contained in the leading singular bHQET cross section do not encode the entire singular distributional terms (i.e.\ delta-functions and plus-distributions) which have coefficients containing $\mhat_t^2$ power corrections. Since the difference to an approach where the singular cross section is treated in a strict power counting approach, where no subleading power contributions are absorbed, is associated to the resummation of formally power-suppressed logarithms of certain types of massive power corrections, any absorption prescription should be based on physical arguments. An essential guiding principle is that fixed-order final matched formulae reproduce the fixed-order full QCD result.

In the factorization formula~(\ref{sec:facttheorem}) one such absorption prescription has been applied by using the exact kinematic stable-top quark expression for the minimal $\tau$ value $\taumin$. This prescription resums kinematic $\mhat_t$ power corrections beyond a strict power counting approach to all orders and is crucial for the phenomenological reliability of the factorization theorem, as we already mentioned in Sec.~\ref{sec:observables}. It is physically sensible since the higher power $\mhat_t^2$ terms contained in $\taumin$ represent a global shift with respect to which the singular dynamical QCD effects unfold in a universal and observable-independent way. It is therefore physically unreasonable to treat the higher-power $\mhat_t$ terms in $\taumin$ in an expansion. Beyond the absorption concerning $\taumin$, however, the factorization formula~(\ref{eq:bhqetdistribution}) applies strict power counting. We therefore label it with the subscript `strict'. In the original 2-jettiness calibration analysis of Ref.~\cite{Butenschoen:2016lpz} the same strict approach was applied and the non-singular corrections were included in two steps by first matching back to SCET and then to full QCD. In Sec.~\ref{sec:non-singularstrict} we review the `strict' approach of Ref.~\cite{Butenschoen:2016lpz}. Since this approach does not yield consistent calibration results for the three observables 2-jettiness $\tau_2$, sJM $\tau_s$ and mJM $\tau_m$, as we shall show in Sec.~\ref{sec:pcanalysis}, we discuss an improved procedure in Sec.~\ref{sec:absorbconcept}. Note that the presentations in this subsections still use the generic shape variable $\tau$ which can stand for $\tau_2$, $\tau_s$ or $\tau_m$. The shape-variable dependent NLO fixed-order results, which are used to determine the QCD non-singular contributions are given in App.~\ref{app:FONLOQCD}. We also note that much more details on the matching procedure to achieve a reliable description for all values of $\tau$ can be found in Ref.~\cite{Dehnadi:2016snl}.

\subsubsection{QCD and SCET non-singular Distributions: Strict Power Counting}
\label{sec:non-singularstrict}

The full parton-level, stable-top, pole-mass and non-renormalon subtracted SCET and QCD matched resonance region cross section in the strict approach has the form
\begin{equation}
	\label{eq:sumallstrict}
	\dv{\hat\sigma^{C}_{\rm full, strict}(\tau)}{\tau} =\left.\dv{\sigma_\mathrm{bHQET}^C}{\tau}\right|_{\mathrm{strict},\delta_m=\gap=\Gamma_t=0,F(k-2\hat\Delta)=\delta(k)} +
	\left.\dv{\sigma_\mathrm{nsb}^{C}}{\tau}\right|_{\mathrm{strict}} + \left.\dv{\sigma_\mathrm{ns}^{C}(\mu_\mathrm{ns})}{\tau}\right|_\mathrm{strict}\,,
\end{equation}
where the strict bHQET cross section from Eq.~(\ref{sec:facttheorem}) is used for $\delta_m=\gap=\Gamma_t=0$ and $F(k-2 \hat \Delta)=\delta(k)$.
The SCET non-singular cross section $\dd{\sigma}^C_\mathrm{nsb}/\dd{\tau}|_\mathrm{strict}$ is defined from the fact that the bHQET factorization theorem emerges from the SCET factorization theorem valid for $(q^2-m_t^2)/m_t\sim m_t$ when the off-shellness $(q^2-m_t^2)/m_t$ reaches values below $m_t$~\cite{Fleming:2007qr,Fleming:2007xt}. As already explained at the beginning of Sec.~\ref{sec:facttheorem}, apart from the resulting modified RG evolution factors in the 5-flavor scheme, this only affects the collinear sector, where the SCET jet function $J_\mathrm{SCET}(s,\mu)$ splits in the mass-mode matching function $H_m$ times the bHQET jet function $J_{B,\tau}^{(5)}$ plus a contribution that is power suppressed, non-singular and also integrable in $(q^2-m_t^2)/m_t$,
\begin{equation}
\label{eq:SCETjetmatchingstrict}
	J_\mathrm{SCET}(s,m_t,\mu)=m_t H_m^{(6)}(m_t,\varrho,\mu) J_{B,\tau}^{(5)}(s/m_t,0,0,\mu) + J^{(5)}_\mathrm{nsb}(s,m_t,\mu)\,,
\end{equation}
where $q^2=s+m_t^2$ is the inclusive invariant mass of the collinear radiation described by the SCET jet function. 
The NLO non-singular jet function $J^{(5)}_\mathrm{nsb}$ reads 
\begin{equation}
	\label{eq:nsbdefstrict}
		J^{(5)}_\mathrm{nsb}(s,m_t,\mu)=\CFalphafive{\mu}\biggl[\frac{2s}{(s+m_t^2)^2}-\frac{8}{s}\ln\biggl(1+\frac{s}{m_t^2}\biggr)\biggr]\theta(s)\,.
\end{equation}
At NLO, the specification of the flavor-number scheme for the strong coupling in Eq.~(\ref{eq:SCETjetmatchingstrict}) is not yet relevant, but we indicate the choice implemented in our numerical code. Note that the SCET massive primary quark jet function has recently been computed at 2 loops in Ref.~\cite{Hoang:2019fze}. The SCET non-singular cross section in the resonance region is given by
\begin{equation}
	\label{eq:signsbstrict}
	\left.\dv{\sigma_\mathrm{nsb}^{C}}{\tau}\right|_\mathrm{strict}  ={}\left.\dv{\sigma_\mathrm{bHQET}^C}{\tau}\right|_{\mathrm{strict};\{H_m^{(6)}(m_t,\varrho,\mu_J)J_{B,\tau}^{(5)}(\shat, \Gamma_t, \delta m_t, \mu_J)\rightarrow J^{(5)}_\mathrm{nsb}(m_t\shat,m_t,\mu_J)/m_t\}},
\end{equation}
which means that the bHQET jet function is simply replaced by the non-singular SCET function with the analogue scale setting. 
This implies that the contributions in the non-singular SCET jet function are treated as low-scale dynamical fluctuations. In the original calibration analysis~\cite{Butenschoen:2016lpz} the scale setting $J^{(5)}_\mathrm{nsb}(m_t \shat ,m_t,\mu_m)$ was used, such that the non-singular SCET jet function was treated as an off-shell contribution. The difference is numerically insignificant since the overall contribution of the SCET non-singular cross section turns out to be tiny, and the difference concerning the resummed logarithms is irrelevant as well. Since $J^{(5)}_\mathrm{nsb}$ is a pure ${\cal O}(\alpha_s)$ contribution all other fixed-order matrix elements in $\dd{\sigma}^C_\mathrm{nsb}/\dd{\tau}$ are taken at tree-level. Therefore, the NLO expanded SCET non-singular cross section simply reads $\sigma_0 Q^2 J^{(5)}_\mathrm{nsb}(m_t\shat_\tau,m_t,\mu)$. 

The QCD non-singular cross section $\dd{\sigma}^C_\mathrm{ns}/\dd{\tau}|_\mathrm{strict}$ is obtained by subtracting the bHQET and non-singular SCET cross sections expanded at ${\cal O}(\alpha_s)$ from the NLO full QCD fixed-order cross section $\dd{\sigma}^C_\mathrm{QCD}/\dd{\tau}|_\mathrm{strict}$, all evaluated at the non-singular renormalization scale $\mu_\mathrm{ns}$: 
\begin{align}
	\label{eq:signsstrict}
		 \frac{1}{\sigma^C_0}\dv{\sigma_\mathrm{ns}^C(\mu_\mathrm{ns})}{\tau}\biggr|_\mathrm{strict} &\equiv{} \frac{1}{\sigma^C_0}\Biggl(\dv{\sigma_\mathrm{QCD}^C}{\tau} -  \dv{\sigma_\mathrm{bHQET}^C}{\tau}\biggr|_\mathrm{strict, NLO} - \dv{\sigma_\mathrm{nsb}^{C}}{\tau}\biggr|_\mathrm{strict, NLO}\Biggr) \nonumber \\
		 &={} \Big( R_0^{C}(\mhat_t)-1\Big)\delta(\tau-\taumin)\, +\, C_F\frac{\alpha_s^{(6)}(\mu_\mathrm{ns})}{4 \pi} \\
		  \times \, &\Biggl\{ A_\tau^{C,\mathrm{ns}}(\mhat_t) \, \delta(\tau-\taumin) +  B_\mathrm{plus}^{C,\mathrm{ns}}(\mhat_t) \biggl[\frac{1}{\tau-\taumin}\biggr]_+ \! + F_\tau^\mathrm{NS,C,ns}(\tau, \mhat_t)\Biggr\}.\nonumber
\end{align}
Apart from the expression for $\taumin$ appearing in the bHQET singular cross section, only the QCD non-singular cross section is observable dependent. The functions $A_\tau^{C,\mathrm{ns}}(\mhat_t)$, $B_\mathrm{plus}^{C,\mathrm{ns}}(\mhat_t)$ and $F_\tau^\mathrm{NS,C,ns}(\tau, \mhat_t)$ are obtained from the corresponding QCD functions shown in Eq.~(\ref{eq:QCDdistribution}) upon the subtractions from the expanded singular bHQET and non-singular SCET cross sections. The NLO expanded singular bHQET cross section reads
\begin{align}
	\label{eq:sigFObHQETstrict}
		\frac{1}{\sigma_0}&  \dv{\sigma_\mathrm{bHQET}^C}{\tau} \Biggr|_\mathrm{strict,NLO}={} \delta(\tau-\taumin) \\
		&+ \frac{C_F \alpha_s}{4\pi}  \biggl( {A}_\tau^{\mathrm{bHQET}}(\mhat_t)  \delta(\tau-\taumin)  + {B}_\textrm{plus}^{\mathrm{bHQET}}(\mhat_t) \biggl[\frac{1}{\tau-\taumin}\biggr]_+\biggr)
		+ \mathcal{O}(\alpha_s^2)\,,\nonumber
\end{align}
where $A_\tau^{\mathrm{bHQET}}(\mhat_t)$ and $B_\mathrm{plus}^{\mathrm{bHQET}}(\mhat_t)$ are given in Eqs.~(\ref{eq:ABbHQET}) for $L_s=0$. This yields the following results for the QCD non-singular functions
\begin{align}
	\label{eq:nscoeffstrict}
		A_\tau^{C,\mathrm{ns}}(\mhat_t)={}& R_0^{C}(\mhat_t) A_\tau^{C}(\mhat_t)- A_\tau^{\mathrm{bHQET}}(\mhat_t)\,,\\
		B_\mathrm{plus}^{C,\mathrm{ns}}(\mhat_t)={}&  R_0^{C}(\mhat_t) B_\mathrm{plus}^{C}(\mhat_t)- B_\mathrm{plus}^{\mathrm{bHQET}}(\mhat_t)\,,\nonumber
\end{align}
and
\begin{equation}
	\label{eq:Fnsstrict}
		F_\tau^\mathrm{NS,C,ns}(\tau, \mhat_t)=F_\tau^\mathrm{NS,C}(\tau, \mhat_t) - 
		Q^2 \biggl[\frac{2m_t \shat_\tau}{(m_t \shat_\tau+m_t^2)^2}-\frac{8}{m_t \shat_\tau}\ln\biggl(1+\frac{m_t \shat_\tau}{m_t^2}\biggr)\biggr]\theta(\shat_\tau)\,.
\end{equation}
The NLO fixed-order functions $R_0^{C}$, $A_\tau^{C}$, $B_\mathrm{plus}^{C}$ and $F_\tau^\mathrm{NS,C}(\tau, \mhat_t)$ are defined in Eq.~(\ref{eq:QCDdistribution}). 

\subsubsection[Absorption of $\mhat_t^2$ Power Corrections]{Absorption of $\mathbf{\mhat_t^2}$ Power Corrections}
\label{sec:absorbconcept}
As we demonstrate in Sec.~\ref{sec:pcanalysis}, the strict approach to define the bHQET cross section (including the exact expression for $\taumin$) and to construct the non-singular cross sections still yields a sizable residual observable dependence on the top quark mass calibration results, which arise from $\mhat_t^2$ power corrections not contained in $\tau_{\rm min}$. This motivates the absorption of additional $\mhat_t^2$ power corrections in the singular bHQET differential distribution. In this section we discuss three kinds of absorption prescriptions, which remove the observable dependence for the calibration results. We emphasize that the discussions presented in this subsection do not constitute a comprehensive and complete treatment of $\mhat_t^2$ power corrections. However, we believe that we have identified the ones most relevant for phenomenological applications and implemented a reasonable way to estimate the remaining uncertainties due to $\mhat_t^2$ power corrections that are not yet accounted for. We also mention that in the context of our analysis it turns out that the 2-jettiness distribution, which was used in the original calibration analysis~\cite{Butenschoen:2016lpz}, is largely insensitive to the treatment of $\mhat_t^2$ corrections indicating its robustness with respect to power-suppressed effects. 

We start the discussion concerning the $\mhat_t^2$ power corrections with the observation that the non-perturbative shape function has a sizable impact on the location of the resonance peak position. This sensitivity on non-perturbative effects parametrized by the shape function is encoded in the measurement delta function $\delta(\shat_\tau - \shat - \varrho \ell)$ appearing in the factorization formula~\eqref{eq:bhqetdistribution}. This corresponds to a generic modification of the kinematic variable of order $\delta \shat_\tau\sim (Q/m_t) \Omega_1$, which implies that the resonance peak position (with respect to the top mass) is shifted by the shape function effects by an amount $\Delta m_t\sim \delta \shat_\tau/2 \sim  Q \Omega_1/(2m_t)$. For $\Omega_1$ in the range of $0.5$\,GeV to $1$\,GeV, which covers the typical values we obtain for $\Omega_1$ from our calibration analysis, this corresponds to a contribution to the fitted top quark mass of around $1$ to $2$\,GeV for $Q$ in the range of $600$ to $1400$\,GeV. This means that $\mhat_t^2$ power corrections to the measurement delta function of the form $\delta[\shat_\tau - \shat - r_{\tau,s}(\mhat_t) \varrho \ell]$ with $r_{\tau,s}(\mhat_t)=(1+\mbox{const}\times\mhat_t^2)$ can still lead to shifts at the level of $250$ to $300$\,MeV, larger than the uncertainties expected for the top quark mass at N$^2$LL$\,+\,$NLO order~\cite{Butenschoen:2016lpz}. It is therefore reasonable to include the rescaling factor $r_{\tau,s}(\mhat_t)$ for the shape variables we consider.  

To that end, let us consider generic soft momenta $k_s$ and $k_{\bar s}$ arising from large-angle soft radiation in the top ($n$) and antitop ($\bar n$) hemispheres, respectively. In the absence of any ultra-collinear radiation one has for the four-momenta flowing in each hemisphere the following expressions: 
\begin{align}
\label{eq:pnpnbar}
	p_n^{\mu}= m_t v_+^{\mu} + k_s^{\mu}\,, \qquad
	p_{\bar{n}}^{\mu}= m_t v_-^{\mu} + k_{\bar{s}}^{\mu}\,,
\end{align}
where $v^\mu_{\pm}=1/\sqrt{1-v^2}(1,0,0,\pm v)$ with $v=\sqrt{1-4\mhat_t^2}$  are the (stable) top and antitop velocities without large-angle soft radiation, which we assume to be in the $z$-direction.
For the 2-jettiness variable $\tau_2$ defined in Eq.~(\ref{eq:tau2def}) it is easy to see that soft momenta may modify the thrust axis which is along the $z$-direction in the absence of soft radiation, but this modification is of order $k_s\sim k_{\bar s}$ leading to effects quadratic in $k_{s,\bar s}$. Let us now define $n^\mu=(1,0,0,1)$ and $\bar{n}^\mu=(1,0,0,-1)$, with the thrust axis pointing in the $z$-direction and use the usual light-cone decomposition of momenta $p^\mu =p^+\frac{\bar{n}^\mu}{2}+p^-\frac{n^\mu}{2} + p^\mu_\perp$. As a result we obtain
\begin{equation}
\label{eq:tau2power}
\tau_2 = 1-v+\frac{k_s^++k_{\bar s}^-}{Q}=\tau_{2,{\rm min}}+\frac{k_s^++k_{\bar s}^-}{Q}\,,
\end{equation}
so that $\shat_{\tau_2}=\varrho (k_s^++k_{\bar s}^-)$. We see that there are no ${\cal O}(\mhat_t^2)$ power corrections to the soft rescaling factor, and we therefore have 
\begin{equation}
\label{eq:rstau2}	
r_{\tau_2,s}(\mhat_t)=1\,. 
\end{equation}

For the sum of jet masses variable (sJM) $\tau_s$ defined in Eq.~(\ref{eq:tausdef}) the situation is more complicated since invariant masses exhibit a non-linear dependence on the soft momenta $k_{s,\bar s}$. We apply the following heuristic consideration, neglecting again any soft modification of the thrust axis along with contributions quadratic in $k_{s,\bar s}$. We obtain that 
\begin{equation}
\tau_s= 
\frac{1}{Q^2}(p_n^+ p_n^- + p_{\bar{n}}^- p_{\bar{n}}^+)\,.
\end{equation}
We can now write the $p_n^-$($p_{\bar{n}}^+$) momentum components in terms of $p_n^+$($p_\nbar^-$) using the relations
\begin{align}
\label{eq:conserveenergy}
		p_n^-+p_n^+={}&Q+\Delta E\,,\\
		p_\nbar^-+p_\nbar^+={}&Q-\Delta E\,,\nonumber
\end{align}
which arise from energy conservation, and where $\Delta E = k_s^++k_s^-=-k_{\bar s}^+-k_{\bar s}^-$ represents the soft energy imbalance between the two hemispheres. 
Together with Eq.~(\ref{eq:pnpnbar}) this yields
\begin{eqnarray}
\label{eq:taussoft}
\tau_s &=&\frac{1}{Q^2}\Big[(p_n^+ (Q-p_n^++\Delta E) + p_{\bar{n}}^- (Q-p_{\bar{n}}^--\Delta E)\Big] \\ \nonumber 
&=& 2\mhat_t^2 + \frac{v}{Q} \,(k_s^++k_{\bar{s}}^-)+{\cal O}(k_{s,\bar s})^2 = 
\tau_{s,\mathrm{min}}+\frac{\sqrt{1-4\mhat_t^2}}{Q}\,(k_s^++k_{\bar{s}}^-)+{\cal O}(k_{s,\bar s})^2\,,
\end{eqnarray}
where the linear soft contribution $\propto \Delta E$ cancels between the two hemispheres and we have neglected all contributions quadratic in soft momenta or energies.  As a result we have $\shat_{\tau_s}=r_{\tau_s,s}(\mhat_t)\varrho (k_s^++k_{\bar s}^-)$ with 
\begin{equation}
\label{eq:rssJM}
r_{\tau_s,s}(\mhat_t)=\sqrt{1-4\mhat_t^2}=1-2\mhat_t^2+{\cal O}(\mhat_t^4)\,.
\end{equation}
Note that the large-angle soft momenta $k_{s,\bar s}$ appearing in Eqs.~(\ref{eq:conserveenergy}) and (\ref{eq:taussoft}) are not exclusively related to on-shell gluons, but also account for the recoil effects on the top and antitop quarks, so that $\Delta E$ can have any sign. The result for $r_{\tau_s,s}(\mhat_t)$ thus accounts for the effects that radiation in one hemisphere has on the entire event. We furthermore emphasize that the renormalization scheme for the top mass $m_t$ appearing in the rescaling factor $r_{\tau_s,s}(\mhat_t)$ cannot be fixed since, as already stated above, our considerations do not represent a complete treatment of power corrections. Since the power corrections encoded in the rescaling factor are of purely kinematical origin, it is reasonable to adopt the MSR mass at a low scale. The exact choice of scale has, however, no impact for our phenomenological analyses since variations of a few GeV in the value of $m_t$ only lead to tiny variations in the value of the rescaling factor. As a matter of convenience we adopt the MSR mass $m_t^{\rm MSR}(5\,\mathrm{GeV})$, which is also the choice we adopted for the $m_t$ dependence in the profile functions discussed in Sec.~\ref{sec:profiles}.

The modified jet mass variable (mJM) $\tau_m= \tau_s + \tau_s^2/2$ is designed such that the soft rescaling factor does not have a quadratic $\mhat_t^2$ term. Using the result on the second line of Eq.~\eqref{eq:taussoft} we obtain
\begin{equation}
\label{eq:taumpower}
\tau_m = 2\mhat_t^2 + 2\mhat_t^4 + (1+2\mhat_t^2)\frac{v}{Q}(k_s^++k_{\bar{s}}^-)+{\cal O}(k_{s,\bar s})^2\,,
\end{equation}
such that we arrive at $\shat_{\tau_m}=r_{\tau_m,s}(\mhat_t)\varrho (k_s^++k_{\bar s}^-)$ with
\begin{equation}
\label{eq:rsmJM}	
r_{\tau_m,s}(\mhat_t)=(1+2\mhat_t^2)\sqrt{1-4\mhat_t^2}=1+{\cal O}(\mhat_t^4)\,.
\end{equation}
We use mJM as a diagnostic shape variable to cross check that the sizable $\mhat_t^2$ power corrections associated to $r_{\tau_s,s}(\mhat_t)$, which are present in the sJM variable are indeed absent in mJM.  

The second absorption prescription is related to the observation that, as was observed in Ref.~\cite{Lepenik:2019jjk}, the NLO fixed-order results given in App.~\ref{app:FONLOQCD} exhibit a universal observable-independent coefficient $B_\mathrm{plus}(\mhat_t)$ multiplying the plus-distribution term $[1/(\tau-\taumin)]_+$ once the tree-level cross section term $R_0^C(\mhat_t)$ is factored out, see Eq.~(\ref{eq:QCDdistribution}). The plus distribution coefficient $B_\mathrm{plus}(\mhat_t)$ is also universal concerning vector (V) or axial-vector induced top-antitop production. This universality does not only concern the three shape variables considered here, but applies to any global and infrared-safe event-shape variable~\cite{Lepenik:2019jjk}. It is therefore reasonable to assume that including the tree-level cross section term $R_0^C(\mhat_t)$ as a global factor multiplying the singular bHQET factorization formula resums another set of important power corrections. Together with the soft rescaling factor this motivates the following modified form of the  parton-level, stable-top, pole-mass and non-renormalon-subtracted bHQET factorization formula 
\begin{align}
\label{eq:bhqetdistributionres}
 \frac{1}{\sigma_0^C} \dv{\sigma^C_\mathrm{bHQET}}{\tau}\Biggr|_\mathrm{pow\,1} \!\!\!=\,& R_0^{C}(\mhat_t) m_t Q^2 H_Q\nflav{6}(Q,\mu_H) U_{H_Q}\nflav{6}(Q,\mu_H,\mu_m) H_m\nflav{6}(m_t,\varrho,\mu_m) U_{v}\nflav{5}(\varrho,\mu_m, \mu) \nonumber\\
&\times\!\! \int \dd{\ell} \dd{\shat} \dd{\ell'} \delta[\shat_\tau - \shat - r_{\tau,s}(\mhat_t) \varrho \ell]\,% \nonumber\\
U_S\nflav{5}(\ell - \ell', \mu, \mu_S) \hat{S}_\tau\nflav{5}(\ell', \bar{\delta}=0, \mu_S)\nonumber\\
&\times\!\! \int \dd{\shat'} U_B\nflav{5}(\shat- \shat', \mu, \mu_J) J_{B,\tau}^{(5)}(\shat', \Gamma_t=0, \delta m_t=0, \mu_J)\,, 
\end{align}
which differs from the strict formula of Eq.~(\ref{eq:bhqetdistribution}) concerning the overall factor $R_0^{C}(\mhat_t)$ and the additional factor $r_{\tau,s}(\mhat_t)$ in the measurement delta-function. Expanded to ${\cal O}(\alpha_s)$, which we need to determine the non-singular cross section this yields
\begin{align}
\frac{1}{\sigma_0^C}&  \dv{\sigma_\mathrm{bHQET}^{C}}{\tau} \Biggr|_\mathrm{pow\,1, NLO}={} R_0^C(\mhat_t)\biggl\{
\delta(\tau-\taumin) \\
&  + \CFalpha{\mu}  \biggl( A_\tau^{\mathrm{bHQET}}(\mhat_t)\,  \delta(\tau-\taumin)  + B_\textrm{plus}^{\mathrm{bHQET}}(\mhat_t) \biggl[\frac{1}{\tau-\taumin}\biggr]_+\biggr) + \mathcal{O}\bigl(\alpha_s^2\bigr)\biggr\}\,, \nonumber
\end{align}
where
\begin{align}
\label{eq:ABbHQET}
A_\tau^{\mathrm{bHQET}}(\mhat_t)={}& 2 \pi^2 + 4 L_{\hat m} + 16 L_{\hat m} ^2
- 8 ( L_s^2 + 2 L_s  L_\mu)\,,  \\
B_\textrm{plus}^{\mathrm{bHQET}}(\mhat_t)={}&  - 8  ( 1 + 2 L_{\hat m} )
+ 16  L_s\,,\nonumber
\end{align}
with 
\begin{equation}
L_s \equiv \ln(r_{\tau,s}(\mhat_t))\,,\qquad L_{\hat m}  \equiv \ln(\mhat_t)\,,\qquad L_\mu \equiv \ln\Bigl(\frac{\mu}{Q}\Bigr)\,.
\end{equation}
Note that $L_s$ is not a large logarithm, but ${\cal O}(\mhat_t^2)$ power-suppressed, and that $L_{\hat m}$ should not be confused with $L_m=\ln(m_t^2/\mu_m^2)$ defined in Eq.~(\ref{eq:Lmdef}). Since the modification of the measurement delta-function also applies in the context of SCET, the SCET non-singular cross section adopts the form 
\begin{align}
	\label{eq:nsbdistribution}
	\frac{1}{\sigma_0^C} \dv{\tilde{\sigma}^C_\mathrm{nsb}}{\tau} ={}& Q^2 H_Q\nflav{6}(Q,\mu_H) U_{H_Q}\nflav{6}(Q,\mu_H,\mu_m)  U_{v}\nflav{5}(\varrho,\mu_m, \mu)\\
	&\times \int \dd{\ell} \dd{\shat} \dd{\shat'} \delta[\shat_\tau - \shat - r_{\tau,s}(\mhat_t) \varrho \ell]\,
	U_B\nflav{5}(\shat- \shat', \mu, \mu_J) J^{(5)}_\mathrm{nsb}(m_t\shat,m_t,\mu_J) \nonumber\\
	&\times \int \dd{\ell'} \dd{k} U_S\nflav{5}(\ell - \ell', \mu, \mu_S) \hat{S}_\tau\nflav{5}(\ell' - k, \bar{\delta}, \mu_S) F(k - 2\hat{\Delta})\,. \nonumber
\end{align}
Note that for the SCET non-singular cross section we do not factor out the tree-level factor $R_0^C(\mhat_t)$ since it leaves the structure of Eq.~(\ref{eq:Fnsstrict}) intact, given our parametrization of the non-singular contribution $F_\tau^\mathrm{NS,C}(\tau, \mhat_t)$ in the NLO fixed-order full QCD distribution shown in Eq.~(\ref{eq:QCDdistribution}). The numerical impact is, however, tiny anyway, as we have already mentioned above in Sec.~\ref{sec:non-singularstrict}.

If we had stopped here, the coefficient of the delta-function, $h_\tau^C(\mhat_t)$, and the coefficient of the plus-distribution $b(\mhat_t)$ in the QCD non-singular cross section (with the phase space function $R_0^C(\mhat_t)$ factored out) would have the form
\begin{align}
\label{eq:bhcorr}
b(\mhat_t)={}& B_\mathrm{plus}(\mhat_t)- B_\mathrm{plus}^{\mathrm{bHQET}}(\mhat_t)\,,\\
h_\tau^C(\mhat_t)={}&A_\tau^C(\mhat_t)-A_\tau^{\mathrm{bHQET}}(\mhat_t) +\ln(\mhat_t^2) b(\mhat_t)\,.\nonumber
\end{align}
We now adopt a third prescription where these contributions are absorbed into the singular bHQET cross section as well. However, since we do not have any compelling physical argument supporting this prescription, we implement it with scaling factors which we vary in our calibration fits to estimate the uncertainty concerning our treatment of the $\mhat_t^2$ power corrections. The final form of the bHQET cross section with all three absorption prescription implemented reads
\begin{equation}
\label{eq:sigbHQETabsorbdef}
\dv{\tilde{\sigma}_\mathrm{bHQET}^C}{\tau} ={}\left.\dv{\sigma_\mathrm{bHQET}^C}{\tau}\right|_{\mathrm{pow\,1},  \{H_Q\rightarrow\tilde{H}_Q,H_m\rightarrow\tilde{H}_m, J_{B,\tau}^{(5)}\rightarrow\tilde J_{B,\tau}^{(5)}\}},
\end{equation}
where
\begin{align}
\label{eq:xiimplementation}
\tilde{H}_Q(\mu_H)={}&H_Q(\mu_H)+C_F\frac{\alpha_s(\mu_H)}{4\pi}(1-\xi_J-\xi_B)\xi_{A1}\,h^C_\tau(\mhat_t)\,,\\
\tilde{H}_m(\mu_m)={}&H_m(\mu_m)+\CFalpha{\mu_m}\xi_J \xi_{A1}\,h^C_\tau(\mhat_t)\,,\nonumber\\
m_t^2\tilde J_{B,\tau}^{(5)}(\shat,\mu_J)={}& m_t^2 J_{B,\tau}^{(5)}(\shat,\mu_J)+\CFalpha{\mu_J}\biggl\{\xi_B \xi_{A1} h^C_\tau(\mhat_t)\delta(\shat)+\xi_{B1}b(\mhat_t)\frac{1}{m_t}\biggl[\frac{1}{
	\shat/m_t}\biggr]_+\!\biggr\}.\nonumber
\end{align}
The scaling parameters $\xi_{A1}$ and $\xi_{B1}$ determine the fractions of the coefficients $b(\mhat_t)$ and $h_\tau^C(\mhat_t)$ being absorbed into the bHQET cross section, where $\xi_{A1}=\xi_{B1}=1$ refers to full absorption and $\xi_{A1}=\xi_{B1}=0$ refers to the treatment where  $b(\mhat_t)$ and $h_\tau^C(\mhat_t)$ are fully contained in the QCD non-singular cross section. In our calibration fits we vary $\xi_{A1}$ and $\xi_{B1}$ independently in the interval $[0,2]$. The scaling parameters $\xi_J$ and $\xi_B$ reflect how the delta-function coefficient $h_\tau^C(\mhat_t)$ is redistributed into the constant non-logarithmic contributions of the hard, mass-mode and jet bHQET functions. In our calibration fits they are varied independently in the interval $[0,1]$ with the constraint $\xi_J+\xi_B\le 1$. For the calibration fits, in order to quantify the uncertainty of our treatment of the $\hat m_t^2$ power corrections, the values of $\xi_{J}$, $\xi_{B}$, $\xi_{A1}$ and $\xi_{B1}$ are chosen randomly in the ranges given above. Specifically, we pick the points $\{\sqrt{\xi_J}, \sqrt{\xi_B},\sqrt{1-\xi_J-\xi_B}\}$ to be uniformly distributed on the unit-sphere in the first octant. This ensures a symmetrical distribution among the three coefficients. For $\xi_{A1}$ and $\xi_{B1}$ we independently use the Beta distribution $N(x/2)^{-0.5}(1-x/2)^{-0.5}$ in the interval $x\in [0,2]$, which conservatively enhances the population of the boundary regions close to $0$ and $2$. When the absorption prescription for the treatment of $\mhat_t^2$ power corrections is used, the random variations of the $\xi$ parameters is implemented in parallel to the 501 random profile function parameter variations. Thus, the variation of both types of parameters combined constitutes our estimate of the perturbative uncertainties.

Overall, the NLO expanded expression for the modified bHQET factorization formula with the three absorption prescriptions reads
\begin{align}
\label{eq:sigFObHQETabsorb}
\frac{1}{\sigma_0}  \dv{\tilde{\sigma}_\mathrm{bHQET}^C}{\tau} \Biggr|_\mathrm{FO}={} R_0^C(\mhat_t)\biggl\{& \delta(\tau-\taumin)  + \CFalpha{\mu}  \biggl(\! \tilde{A}_\tau^{C,\mathrm{bHQET}}(\mhat_t,\xi_{A1},\xi_{B1})  \delta(\tau-\taumin) \nonumber \\
&  + \tilde{B}_\textrm{plus}^{\mathrm{bHQET}}(\mhat_t,\xi_{B1}) \biggl[\frac{1}{\tau-\taumin}\biggr]_+\biggr)+ \mathcal{O}(\alpha_s^2)\! \biggr\}\,,
\end{align}
where
\begin{align}
\label{eq:AtildeBtilde}
\tilde{A}_\tau^{C,\mathrm{bHQET}}(\mhat_t,\xi_{A1},\xi_{B1})={}&A_\tau^{\mathrm{bHQET}}(\mhat_t) +\xi_{A1}h^C_\tau(\mhat_t)-\xi_{B1}\ln(\mhat_t^2) b(\mhat_t)\\
={}&\xi_{A1}A_\tau^C(\mhat_t)+(1-\xi_{A1})A_\tau^{\mathrm{bHQET}}(\mhat_t)\nonumber\\
&+(\xi_{A1}-\xi_{B1})\ln(\mhat_t^2)b(\mhat_t)\,,\nonumber\\[2mm]
\tilde{B}_\textrm{plus}^{\mathrm{bHQET}}(\mhat_t,\xi_{B1})={}& B_\mathrm{plus}^{\mathrm{bHQET}}(\mhat_t) + \xi_{B1}b(\mhat_t)\nonumber\\
={}&\xi_{B1} B_\mathrm{plus}(\mhat_t)+(1-\xi_{B1})B_\mathrm{plus}^{\mathrm{bHQET}}(\mhat_t)\,.\nonumber
\end{align}
Note that $\tilde{A}_\tau^{C,\mathrm{bHQET}}(\mhat_t,\xi_{A1},\xi_{B1})$ and $\tilde{B}_\textrm{plus}^{\mathrm{bHQET}}(\mhat_t,\xi_{B1})$ do not depend on 
the scaling parameters $\xi_J$ and $\xi_B$ since these only specify how $h_\tau^C(\mhat_t)$ is distributed among the hard, mass-mode and bHQET jet functions. The QCD non-singular cross section then adopts the form
\begin{align}
\label{eq:QCDnsdef}
\frac{1}{\sigma^C_0}\dv{\tilde{\sigma}_\mathrm{ns}^C(\mu)}{\tau} \equiv{}& \frac{1}{\sigma^C_0}\Biggl(\dv{\sigma_\mathrm{QCD}^C}{\tau} -  \dv{\tilde{\sigma}_\mathrm{bHQET}^C}{\tau}\biggr|_\mathrm{FO} -  \dv{\tilde{\sigma}_\mathrm{nsb}^{C}}{e}\biggr|_\mathrm{FO}\Biggr) \\
={}& C_F \frac{\alpha^{(6)}_s(\mu)}{4 \pi} \biggl\{R_0^{C}(\mhat_t)\biggl(  \tilde{A}_\tau^{C,\mathrm{ns}}(\mhat_t,\xi_{A1},\xi_{B1}) \delta(\tau-\taumin) \nonumber\\
&\ +  \tilde{B}_\mathrm{plus}^{C,\mathrm{ns}}(\mhat_t,\xi_{B1}) \biggr[\frac{1}{\tau-\taumin}\biggr]_+ \biggr) + F_\tau^\mathrm{NS,C,ns}(\tau, \mhat_t)\biggr\}\,, \nonumber
\end{align}
where
\begin{align}\label{eq:nscoeffs}
\tilde{A}_\tau^{C,\mathrm{ns}}(\mhat_t,\xi_{A1},\xi_{B1})={}& A_\tau^{C}(\mhat_t)- \tilde{A}_\tau^{C,\mathrm{bHQET}}(\mhat_t,\xi_{A1},\xi_{B1})\,,\\
\tilde{B}_\mathrm{plus}^{C,\mathrm{ns}}(\mhat_t,\xi_{B1})={}&  B_\mathrm{plus}(\mhat_t)- \tilde{B}_\mathrm{plus}^{\mathrm{bHQET}}(\mhat_t,\xi_{B1})\,,
\nonumber
\end{align}
and $F_\tau^\mathrm{NS,C,ns}(\tau, \mhat_t)$ is already given in Eq.~(\ref{eq:Fnsstrict}).

The final expression for the full parton-level, stable-top, pole-mass and non-renormalon-subtracted SCET- and QCD-matched resonance region cross-section in the absorption approach has the form
\begin{equation}
\label{eq:sumallabs}
\dv{\hat\sigma^{C}_{\rm full, absorb}(\tau)}{\tau} =\dv{\tilde{\sigma}_\mathrm{bHQET}^C}{\tau} + \dv{\tilde{\sigma}_\mathrm{nsb}^{C}}{\tau} + \dv{\tilde{\sigma}_\mathrm{ns}^{C}(\mu_\mathrm{ns})}{\tau}.
\end{equation}
We remind the reader that $\mathrm{d}\tilde{\sigma}_\mathrm{bHQET}^C/\mathrm{d}\tau$ and $\mathrm{d}\tilde{\sigma}_\mathrm{nsb}^{C}/\mathrm{d}\tau$ depend on the $\tau$-dependent profiles for the renormalizations scales $\mu_H$, $\mu_m$, $\mu_J$ and $\mu_S$. Furthermore, $\mathrm{d}\tilde{\sigma}_\mathrm{bHQET}^C/\mathrm{d}\tau$ depends on the scaling parameters $\xi_{A1}$, $\xi_{B1}$, $\xi_J$ and $\xi_B$,
and $\mathrm{d}\tilde{\sigma}_\mathrm{ns}^{C}/\mathrm{d}\tau$ depends on the scaling parameters $\xi_{A1}$ and $\xi_{B1}$. This dependence is suppressed in the arguments to avoid cluttering.

\subsection{Combining Ingredients}
\label{sec:combination}

In Sec.~\ref{sec:non-singular} we have derived the full parton-level resonance $\tau$ distributions including the singular bHQET and non-singular cross sections in the limit of a stable top quark and without any renormalon subtractions. So the formulae for $\mathrm{d}\sigma^{C}_{\rm full, strict}(\tau)/\mathrm{d}\tau$ in Eq.~(\ref{eq:sumallstrict}) and for $\mathrm{d}\sigma^{C}_{\rm full, absorb}(\tau)/\mathrm{d}\tau$ in Eq.~(\ref{eq:sumallabs}) are in the pole mass scheme and without any soft gap subtraction. For the event-shape distributions used in the calibration fits, the non-perturbative effects parametrized in the shape function $F(k)$, the top quark width effects and the renormalon subtractions still need to be implemented. This is achieved by the following additional convolutions involving the shape function $F(k-2\hat\Delta)$ of Eq.~(\ref{eq:shapefunc}) and the Breit-Wigner function $G(\hat{s}, \Gamma_t)$ of Eq.~(\ref{eq:BWfunction}):
\begin{align}
\label{eq:fulleventshape}
\dv{\sigma^C_{\rm full, strict/absorb}(\tau)}{\tau} ={}&
 \int \dd{\hat{s}} \dd{k} \dv{\hat{\sigma}^C_{\rm full, strict/absorb}}{\tau}\biggl( \tau -\Big(\dv{\taumin}{m_t}\Big)\delta_m- \frac{m_t \hat{s}}{Q^2} - \frac{k+2\bar{\delta}}{Q}\biggr) \nonumber\\
& \times G(\hat{s}, \Gamma_t) F(k-2\hat{\Delta})\,,
\end{align}
where the residual mass $\delta_m$ and the gap subtraction $\gap$ terms (for the three gap schemes we use) are discussed in Sec.~\ref{sec:renormalonsubtractions}. The mass $m_t$ appearing in the argument on the RHS refers to $m_t^\mathrm{MSR}(R)$ in the MSR mass scheme and to $m_t^\mathrm{pole}$ in the pole mass scheme. The same is true also for the top mass appearing in the denominator of $\shat_{\tau}$ in Eq.~(\ref{eq:shattauold}). Note that for the top mass appearing in the soft rescaling factor $r_{\tau,s}(\mhat_t)$ we always adopt the MSR mass $m_t^{\rm MSR}(5\,\mathrm{GeV})$, as explained in the comment after Eq.~(\ref{eq:rssJM}). The MSR-mass and gap subtractions are expanded strictly in $\alpha_s$ at the same respective renormalization scales together with the bHQET jet and soft functions to guarantee a correct order-by-order cancellation of the renormalons. 

We stress that the finite top width and non-perturbative corrections as well as the renormalon subtractions also affect the non-singular cross sections through the global convolution in Eq.~(\ref{eq:fulleventshape}). This implementation is important, since the final cross section can otherwise show severe instabilities when the singular delta-function or plus-distribution terms are not fully absorbed into the bHQET cross section. We finally mention that for the final expressions entering the calibration analysis the vector- (V) as well as axial-vector- (A) induced cross sections are added up:
\begin{equation}
\label{eq:fulleventshapeVA}
\dv{\sigma_{\rm full, strict/absorb}(\tau)}{\tau} 
={}  \dv{\sigma^V_{\rm full, strict/absorb}(\tau)}{\tau}  \, + \, \dv{\sigma^A_{\rm full, strict/absorb}(\tau)}{\tau} \,.
\end{equation}

\section{Fitting and Data Processing}
\label{sec:fitImprovements}

In this section we provide details on the fit procedure and the data handling, as well as the theory grid we use in order to carry out the fits in a timely manner. They have been carried out as described in the original calibration analysis~\cite{Butenschoen:2016lpz} and realized in the same way in this update. All routines, however, have been coded anew to replace the custom-written in-house calibration software framework of~\cite{Butenschoen:2016lpz} by a workflow that supports current state-of-the-art libraries and data formats.

\subsection{Basic Fit Procedure}
\label{sec:fitprocedure}

We use a standard $\chi^2$ fit procedure for the top quark mass $m_t$ (in either pole or MSR mass schemes) and the non-perturbative model parameters $\{c_0,c_1,c_2,c_3\}$ (and in principle also $\Delta_0$ and $\lambda$), which we outline in the following. The shape function coefficients $c_i$, see Eq.~(\ref{eq:shapefunc}), are restricted by $\sum_{i=0}^3 c_i^2 = 1$, so the actual fit parameters are three euclidean angles $\{a\} = (a_0,a_1,a_2)$. 

The reference data are binned distributions of either 2-jettiness, sJM or mJM, which we simply refer to as $\tau$, obtained from the MCs for the process $e^+e^-\rightarrow t\bar{t}$, where the top quarks decay through all allowed leptonic or hadronic channels, and {\it each histogram contains $10^7$ events}. For this number of events, statistical uncertainties become irrelevant and the MC shape distribution curves can be considered as smooth functions, as can be seen in Fig.~\ref{fig:MCdistributions}.  We use three different fit ranges around the peak of the distribution. These are denoted by $(x,y)$, with the minimum and maximum value $\tau^\mathrm{fit}_\mathrm{min}$ and $\tau^\mathrm{fit}_\mathrm{max}$ defined as the position where the distribution drops to a fraction $x$ and $y$, respectively, of the maximal peak height:
\begin{equation}
\label{eq:binrange}
	\dv{\sigma(\tau^\mathrm{fit}_\mathrm{min})}{\tau}=x\dv{\sigma(\tau_\mathrm{peak})}{\tau},\qquad \dv{\sigma(\tau^\mathrm{fit}_\mathrm{max})}{\tau}=y\dv{\sigma(\tau_\mathrm{peak})}{\tau}\,.
\end{equation}
The three ranges used are $(0.6,0.8)$, $(0.7,0.8)$ and $(0.8,0.8)$.
To break the degeneracy of the peak position with respect to the top quark mass and the shape function (mostly due to the top mass independent value of $\Omega_1$ for the latter) it is necessary to simultaneously include distributions at multiple c.m.\ energies $Q$. We used five different sets of $Q$ values. In GeV units they read: $(700, 1000, 1400)$, $(800, 1000, 1400)$, $(700 - 1400)$, $(600, 1000, 1400)$ and $(600 - 1400)$, where the ranges are in steps of $100$\,GeV. This gives $3$ (ranges around peak) $\times \,5$ ($Q$ sets) $= 15$ different ``fit settings'' (labeled with the subscript $s$ below) of bins included in the $\chi^2$ analyses. For a perfect theoretical description (and assuming that the MC data is equally perfect) these settings should have no influence on the outcome of the fits. The spread of the fit results for the various settings is therefore a quantification for the ``incompatibility'' between theory and MC. Since the theoretical (perturbative and power correction) uncertainties are already estimated through the variations of the profile-function and power-correction \mbox{$\xi$-parameters} ---\,see below Eqs.~(\ref{eq:munsdef}) and (\ref{eq:xiimplementation}), and Tab.~\ref{tab:scalevariation}\,--- the variation of the fit results with the choice of fit setting quantifies the uncertainty of the MC event generator. We therefore include the fit setting dependence as a separate source of uncertainty in addition to the perturbative one.

We use the following procedure to obtain a central value and uncertainties for the top mass $m_t$ (and analogously for the shape function's first moment $\Omega_1$):
\begin{enumerate}
	\item For one fit setting, labeled by $s$, remove $1.5\%$ of the upper and $1.5\%$ of the lower $m_t$ values of the $501$ best-fit values from the variation over the profiles (and $\xi$ parameters when the absorption prescription for $\mhat_t^2$ power corrections is used) to remove potential outliers. Let us call this cleaned up set of masses $\{m_t\}_s$.  
	\item Then take the middle value $m^{\mathrm{set}}_{t,s}=[\max(\{m_t\}_s)+\min(\{m_t\}_s)]/2$ as the central result for this fit setting and half the range as the scale uncertainty $\Delta m^{\mathrm{set}}_{t,s}=[\max(\{m_t\}_s)-\min(\{m_t\}_s)]/2$.
	\item Take the central value of the results for the $15$ fit settings as the final result for $m_t$: $m_{t}^{\mathrm{fit}}=[\max_s(m^{\mathrm{set}}_{t,s})+\min_s(m^{\mathrm{set}}_{t,s})]/2$.
	\item Take the mean of the scale uncertainties for the 15 fit settings as the final \textit{perturbative uncertainty}: $\Delta m_{t,\mathrm{pert}}=[\sum_s\Delta m^{\mathrm{set}}_{t,s}]/15$.
	\item Take half the range of the individual results for the 15 fit settings as \textit{incompatibility uncertainty}: $\Delta m_{t,\mathrm{incomp}}=[\max_s(m^{\mathrm{set}}_{t,s})-\min_s(m^{\mathrm{set}}_{t,s})]/2$.  
\end{enumerate}
When quoting final combined uncertainties we quadratically add the perturbative and incompatibility uncertainties. Note that for $\Omega_1$ the removal of outliers described in bullet point~1 is carried out {\it independently}. 

The best-fit value for a single profile and one fit setting is obtained by minimizing the $\chi^2$ function with respect to the fit parameters using the program \textsc{Minuit}~\cite{James:310399}, with
\begin{equation}
\label{eq:chi2func}
	\chi^2(m_t;\{a\},\Delta_0,\lambda) = \sum_{Q}\sum_{\tau_\mathrm{min}\le\tau_i<\tau_\mathrm{max}}\frac{\left[f_{Q,i}^\mathrm{theo} (m_t;\{a\},\Delta_0,\lambda)-f_{Q,i}^\mathrm{MC}\right]^2}{\sigma_{Q,i}^2}\,.
\end{equation}
The theory bin $f_{Q,i}^\mathrm{theo} (m_t;\{a\},\Delta_0,\lambda)$ at observable value $\tau_i$ is defined as the differential cross section integrated between $\tau_i$ and $\tau_{i+1}$, which we call $\hat{f}_{Q,i}^\mathrm{theo} (m_t;\{a\},\Delta_0,\lambda)$, divided by the norm $\sum_i\hat{f}_{Q,i}^\mathrm{theo} (m_t;\{a\},\Delta_0,\lambda)$, where the sum is over the $\tau$ range of the fit setting:
\begin{equation}
\label{eq:fQitheory}
f_{Q,i}^\mathrm{theo} (m_t;\{a\},\Delta_0,\lambda) = \frac{\hat{f}_{Q,i}^\mathrm{theo} (m_t;\{a\},\Delta_0,\lambda)}{\sum_i\hat{f}_{Q,i}^\mathrm{theo} (m_t;\{a\},\Delta_0,\lambda)} = 
\frac{\int_{\tau_i}^{\tau_{i+1}} {\rm d}\tau\, \dv{\sigma_{\rm full}(\tau)}{\tau}}{\int_{\tau_{\rm min}}^{\tau_{\rm max}} {\rm d}\tau\, \dv{\sigma_{\rm full}(\tau)}{\tau}}\,.
\end{equation}
Likewise, the MC generator bin $f_{Q,i}^\mathrm{MC}$ is defined as the sum of events with $\tau_i<\tau<\tau_{i+1}$, which we call $\hat{f}_{Q,i}^\mathrm{MC}$, divided by the norm $\mathcal{N}_Q^\mathrm{MC}=\sum_i\hat{f}_{Q,i}^\mathrm{MC}$. So theory and MC histograms are normalized to 1 across the fit range $(\tau_\mathrm{min},\tau_\mathrm{max})$. The uncertainty $\sigma_{Q,i}$ is the {\it statistical error} of the event generator bin $f_{Q,i}^\mathrm{MC}$ obtained by naively dividing the bin errors $\Delta\hat{f}_{Q,i}^\mathrm{MC}$ of the unnormalized bins $\hat{f}_{Q,i}^\mathrm{MC}$ by the norm $\mathcal{N}_Q^\mathrm{MC}$. This ``naive'' bin error $\sigma_{Q,i}$ ignores correlations between bins that are introduced by using histograms normalized to the fit range. We also tested the strict statistical treatment of performing the fits with the $\chi^2$ values obtained by using the full covariance matrix for the normalized bins. The differences to the naive treatment of Eq.~(\ref{eq:chi2func}) for the fitted mass are at the sub-MeV level for individual profile fits. In light of the negligible differences, we adopt the naive treatment. We note that the size of the resulting numerical values of $\chi^2$ do by themselves not have any physical meaning since MC modeling uncertainties are not included in the $\chi^2$-function. However, the relative size of the resulting minimal fit values, $\chi^2_\mathrm{min}$, quantifies the quality of the respective fit. In our results we therefore quote the mean and the standard deviation of $\chi^2/\mathrm{dof}$ over all 501 profiles and the $15$ fit settings.

In principle, one may consider that also the strong coupling $\alpha_s$ can be fitted as a theoretical parameter (with the same fundamental meaning as the top quark mass) in the calibration fits. However, as was already pointed out in the original calibration analysis of Ref.~\cite{Butenschoen:2016lpz}, the $\chi^2$ function has a very flat dependence on $\alpha_s$ so that the strong coupling cannot be constrained in the calibration. This is because the primary top mass dependence of the shape variables is of kinematic origin and already captured at tree-level. The QCD effects, on the other hand, only yield corrections of several GeV to the $M_J$ shape distribution shown in Fig.~\ref{fig:MCdistributions} so that variations of $\alpha_s^{(5)}(m_Z)$ within its percent level uncertainty lead to effects that are much smaller than the calibration uncertainties we obtain at N${}^2$LL+NLO. In other words, the parametric uncertainty of strong coupling is negligible within the precision of our top mass calibration approach. In fact, using variations in the value of the input strong coupling value in the range $\alpha_s^{(5)}(m_Z)=0.1181 \pm 0.0013$, which is substantially more conservative than the current world average~\cite{ParticleDataGroup:2022pth}, leads to changes in the top mass results from the calibration at the level of $20$\,MeV, which are negligible in comparison to the uncertainties obtained from the calibration at N$^2$LL$\,+\,$NLO order. In this respect, the top-quark peak region event-shape distributions we consider differ considerably from the massless quark event shape distributions in the tail region used for high precision $\alpha_s$ measurements. Thus the value of $\alpha_s^{(5)}(m_Z)$ has to be taken as an input. For the calibration fits we adopt the value $\alpha_s^{(5)}(m_Z = \SI{91.188}{\GeV})=0.118$.

The top quark width was fixed for theory and event generator to $\Gamma_t=\SI{1.4}{\GeV}$. The generators use a tree-level $e^+e^-\rightarrow t\bar{t}$ matrix element, which goes through their respective internal standard decayer, parton shower and hadronization model. Initial state radiation has been turned off. For \pythia 8.305~\cite{Bierlich:2022pfr} we use the default setting and the standard Monash $e^+e^-$ tune (7). For \herwig 7.2~\cite{Bellm:2015jjp} and \sherpa 2.2.11~\cite{Sherpa:2019gpd}  we use the default settings and tunes.

\subsection{Details on Data Processing and Theory Evaluations}
\label{sec:dataprocessing}

The MC pseudo data are generated with the standard setting of \pythia~8.305~\cite{Bierlich:2022pfr}, \sherpa~2.2.11~\cite{Sherpa:2019gpd} and \herwig 7.2~\cite{Bellm:2015jjp} using the input files given in  App.~\ref{sec:MCsettings}. We use the program \textsc{Rivet}~\cite{Bierlich:2019rhm} paired with a \textsc{python}~\cite{Rossum:1995} custom-written analysis tool to convert per event kinematic information into histograms in the format \textsc{yoda} for our observables. This workflow works with all state-of-the-art MCs that support \textsc{Rivet} directly or the event record format \textsc{HepMC}~\cite{Buckley:2019xhk}. The MC produces events across the full shape-variable range and the choice of the bin specification has no impact on the MC runtime. It is therefore safer to keep a large range and use narrow bins, since wider bins can always be produced by merging smaller ones without loosing information. For the histograms corresponding to a given $Q$ value we use $10000$ evenly spaced bins between 0.0 and 0.5 for each of our observables. This is also the width of the bins we use for the $\chi^2$ function in Eq.~(\ref{eq:chi2func}). The results for the three shape distributions $\tau_2$, $\tau_s$ and $\tau_m$ and the three MCs in the peak region are shown in Fig.~\ref{fig:MCdistributions} exemplarily for $m_t^{\rm MC}=173$\,GeV and for $Q=700$, $1000$ and $1400$\,GeV as a function of the jet mass variable $M_J=Q \sqrt{\tau_{2,s,m}/2}$.

The theory cross section is based on an in-house \textsc{fortran-2008}~\cite{gfortran} object-oriented program called \textsc{Caliper}, written by some of the authors. For the concrete numerical evaluation at the partonic level we compute the bHQET factorization formula in Fourier space since all convolutions turn into easily manageable multiplications. We multiply out all matrix elements appearing in the factorization theorem, along with the gap and MSR mass renormalon subtraction series, and strictly truncate at $\mathcal{O}(\alpha_s)$. On the contrary, resummation factors are fully  multiplied to each of the terms resulting from the expansion  and not expanded in any way with the matrix elements. The final result is then transformed back into momentum space using analytic formulae. All necessary expressions have already been given in Ref.~\cite{Bachu:2020nqn} (see Sec.~V~A and the appendices) and shall not be repeated here. The integration over the Breit-Wigner function is also carried out analytically, while the convolution with the shape function is done numerically in the peak region using the \textsc{quadpack} package~\cite{quadpack}. The RG-evolution of the SCET non-singular contribution involves the evaluation of $_3F_2$ and $_2F_1$ hypergeometric functions, that in the resonance region can be efficiently computed as a Taylor series around the origin, keeping as many terms as necessary to achieve machine precision. The convolution of the QCD and SCET non-singular partonic distributions with the shape function is carried out numerically with \textsc{quadpack}. 

Since the theory cross section cannot be evaluated from scratch during the fits due to performance and speed constraints, extensive grids need to be implemented. We keep track of the dependence on the shape function coefficients $c_i$ exactly: the hadron-level cross section is written as a double sum
\begin{equation}
\label{eq:fkltaudef}
\dv{\sigma_{\rm full}(\tau)}{\tau}  = \sum_{k\ell}c_k c_\ell f_{k\ell}(\tau,m_t,Q,\ldots)\,,
\end{equation}
over distribution functions $f_{k\ell}(\tau,m_t,Q,\ldots)$ since one can factor out the quadratic double sum dependence on the $c_i$ of the shape function over the basis functions given in Eq.~(\ref{eq:shapefunc}). We can therefore treat the dependence on the $c_i$ analytically and only generate grids for the distribution functions $f_{k\ell}(\tau,m_t,Q,\ldots)$ which satisfy $f_{k\ell}= f_{\ell k}$. The ellipses stand for the dependence on the other parameters and will be suppressed from now on. Due to the normalization $\sum_{i=0}^3 c_i^2 = 1$ we express the $c_i$ in terms of euclidean angles $a_{1,2,3}$, such that the cross section will depend on sines and cosines of those. We note that one has to sample multiple starting values to reliably find the true minimum in the $\chi^2$ minimization procedure. For each choice of the $501$ sets of profile functions (including the random values for the power correction $\xi$-parameters) and for fixed values of $\Delta_0$, $Q$ and $\lambda$, we generate grids for all the $f_{k\ell}$ functions in $m_t$ and $\tau$.

The $\tau$ nodes of the grid lie in a range between 0 and $t_1(m_t=\SI{177}{\GeV},Q, d_1 = 0.25)$ which is defined in Eq.~(\ref{eq:profBoundaries}). The value of $t_1$ with the given arguments is larger than any of the upper boundary $\tau_\mathrm{max}$ of our fit ranges, defined by the smallest $y$ parameter given below Eq.~(\ref{eq:binrange}). The range of our MC histogram is also chosen such that $t_1$ always lies within. The $\tau$ values of the theory grid do not have to coincide with the MC histogram bin boundaries, since we compute the integrated bins from the interpolated distribution functions. To determine appropriate $\tau$ values for our grid we first find the peak $t_\mathrm{peak}$ of the $f_{00}$ distribution with the \textsc{fortran} routine \textsc{compass\_search}~\cite{doi:10.1137/S003614450242889} using the tree-level and stable-top threshold $\taumin$ as starting value. We then generate $15$ evenly spaced points in the range $[0, t_\mathrm{peak}-0.4(t_1-t_\mathrm{peak})]$. The next interval $[t_\mathrm{peak}-0.4(t_1-t_\mathrm{peak}), t_\mathrm{peak}+0.4(t_1-t_\mathrm{peak})]$ is filled with $75$ evenly spaced points, and the third interval $[t_\mathrm{peak}+0.4(t_1-t_\mathrm{peak}),t_1]$ has 10 evenly spaced points. We checked, by testing finer $\tau$ grids, that this setting provides an adequate interpolation quality in the actual peak region (with all $f_{k\ell}$ functions included) for the fits. The other dimension of the grids is the top quark mass value (for the pole mass $m_t^\mathrm{pole}$ and the MSR mass $m_t^\mathrm{MSR}(1\,\mbox{GeV})$) in steps of \SI{0.25}{\GeV} between $m_t^\mathrm{MC}-\SI{3}{\GeV}$ and $m_t^\mathrm{MC}+\SI{2}{\GeV}$.

The $\chi^2$ minimizations are performed using a \textsc{python} script. At the beginning of the procedure one determines from the MC histograms the range of bins within the interval $[\tau^\mathrm{fit}_\mathrm{min},\tau^\mathrm{fit}_\mathrm{max}]$ according to a given fit setting, see Eq.~(\ref{eq:binrange}) and the text below. The grids for the $f_{k\ell}(\tau,m_t,Q,\ldots)$ are then turned into a grid $f_{k\ell i}(m_t,Q,\ldots)$, with the $i$ index labeling the MC bins, using the integrals 
\begin{equation}
\label{eq:fklidef}
f_{k\ell i}(m_t,Q)  = \int_{\tau_i}^{\tau_{i+1}} {\rm d}\tau\,f_{k\ell}(\tau,m_t,Q,\ldots)\, ,
\end{equation}
over the spline interpolated $f_{k\ell}(\tau,m_t,Q,\ldots)$ distribution functions. Finally, the $f_{k\ell i}(m_t,Q)$ are spline-interpolated over $m_t$. The $\chi^2$-function in Eq.~(\ref{eq:chi2func}) with
\begin{equation}\label{eq:bintheory}
\hat f_{Q,i}^\mathrm{theo} (m_t,\{a\}) = \sum_{k\ell}c_k(\{a\})c_\ell(\{a\})f_{k\ell i}(m_t,Q)\,,
\end{equation}
is then sampled by \textsc{Minuit}. 

We remind the reader that the procedure just described in this subsection applies for fixed values of $\Delta_0$ and $\lambda$.

\section{Calibration Consistency Test with Previous Results for \pythia and Graphical Representation}
\label{sec:consistencyresults}

\begin{table}[t]
	\renewcommand*{\arraystretch}{1.4}
	\resizebox{\textwidth}{!}{
		\centering
		\begin{tabular}{| l | c  c c c c c c c c c c c c c | c |}
			\hline
			$R \,[\si{\GeV}]$ & 2 & 3 & 4 & 5 & 6 & 7 & 8 & 9 & 10 & 11 & 12 & 13 & 14 & 15 & $\alpha_s^{(5)}(m_Z)$\\
			\hline
			\multirow{3}{*}{$\Delta m_{t}^\mathrm{MSR}(R)\,[\si{\MeV}]$} &
			174 & 308 & 425 & 532 & 633 & 728 & 819 & 908 & 994 & 1077 & 1159 & 1239 & 1317 & 1395 & 0.117\\
			\cline{2-16}
			& 180 & 317 & 437 & 546 & 648 & 745 & 838 & 928 & 1015 & 1100 & 1183 & 1264 & 1343 & 1422 & 0.118 \\
			\cline{2-16}
			& 186 & 327 & 449 & 560 & 664 & 763 & 857 & 948& 1037 & 1123 & 1207 & 1289 & 1370 & 1449 & 0.119\\
			\hline
		\end{tabular}
	}
	\caption{Results for $\Delta m_{t}^\mathrm{MSR}(R)=m_t^{\rm MSR}(1\,\mbox{GeV}) -m_t^{\rm MSR}(R)$ for several $R$ values using the 3-loop \mbox{$R$-evolution} and  $\alpha_s^{(5)}(m_Z = \SI{91.188}{\GeV})=0.117$, $0.118$ and $0.119$ using the REvolver library~\cite{Hoang:2021fhn}. The values for $\Delta m_{t}^\mathrm{MSR}(R)$ depend to a very good approximation linearly on $\alpha_s^{(5)}(m_Z)$.
	}
	\label{tab:massConversionTable} 
\end{table}

The top mass calibration implementation and the results presented in this article constitute an update and generalization of the study carried out in Ref.~\cite{Butenschoen:2016lpz} for \pythia~8.205. Thus, before we enter the discussion of the new analyses, a comparison with the results of Ref.~\cite{Butenschoen:2016lpz} is in order. This also gives us the opportunity to introduce and explain the graphical scheme we employ to represent the results of the different calibration analyses in the following sections. In Ref.~\cite{Butenschoen:2016lpz} the observable 2-jettiness $\tau_2$ was used for the calibration and the gap scheme~1 defined in Eq.~(\ref{eq:gap1def}) was employed for the renormalon subtraction concerning large-angle soft radiation. As already explained in Sec.~\ref{sec:gapsubtractions}, see paragraph below Eq.~(\ref{eq:O1formula}), in Ref.~\cite{Butenschoen:2016lpz} $\Delta_0=0.05$\,GeV and $\lambda=0.5$\,GeV were adopted for the parametrization of the shape function and it was checked that these values provide sufficient flexibility for the shape function fits through the coefficients $c_i$. The analysis was carried out in the pole and MSR mass schemes, adopting $m_t^{\rm MSR}(R=1\,\mbox{GeV})$ as the quoted reference mass for the latter. We note that in the resonance region of the cross section the MSR mass is evaluated at much higher $R$ scales described by the profile function for $R(\tau)$ given in Eq.~(\ref{sec:MSRRsetting}), which are in the range of $10$ to $20$\,GeV. The values quoted for $m_t^{\rm MSR}(1\,\mbox{GeV})$ can be simply converted to other $R$ values using the $R$-evolution equation of the MSR mass, see Sec.~\ref{sec:massSchemes}. We remind the reader that the $R$-evolution of the MSR mass is mass-independent so that the difference 
\begin{equation}
\label{eq:DeltaMSR}
\Delta m_{t}^\mathrm{MSR}(R)\equiv m_t^{\rm MSR}(1\,\mbox{GeV}) -m_t^{\rm MSR}(R)\,,
\end{equation}
only depends on $R$. The conversion of $m_t^{\rm MSR}(1\,\mbox{GeV})$ to a number of other $R$ scales for different values of the strong coupling is given in Tab.~\ref{tab:massConversionTable}.

\begin{table}[t]
	\centering
	\begin{tabular}{|l l c c c || c c c |}
		\hline
		& order & central & perturb. & incomp. & central & perturb. & incomp. \\
		\hline
		\hline
		$m_{t,\SI{1}{\GeV}}^\mathrm{MSR}$ & \nnll & 172.82 & 0.19 & 0.11 & 172.82 & 0.17 & 0.10\\
		$m_{t,\SI{1}{\GeV}}^\mathrm{MSR}$ & NLL & 172.80 & 0.26 & 0.14 & 172.83 & 0.29 & 0.12\\
		$m_{t}^\mathrm{pole}$ & \nnll & 172.43 & 0.18 & 0.22 & 172.40 & 0.18 & 0.20 \\
		$m_{t}^\mathrm{pole}$ & NLL & 172.10 & 0.34 & 0.16 & 172.06 & 0.34 & 0.16 \\
		\hline
		\hline
		$\Omega_1^\mathrm{MSR}$ & \nnll & 0.42 & 0.07 & 0.03 & 0.43 & 0.06 & 0.03\\
		$\Omega_1^\mathrm{MSR}$ & NLL & 0.41 & 0.07 & 0.02 & 0.42 & 0.07 & 0.03\\
		$\Omega_1^\mathrm{pole}$ & \nnll &  & &  & 0.38 & 0.02 & 0.03\\
		$\Omega_1^\mathrm{pole}$ & NLL &  &  &  & 0.31 & 0.05 & 0.04\\
		\hline
	\end{tabular}
	\caption{Calibration results from Ref.~\cite{Butenschoen:2016lpz} for the MSR mass $m_t^{\rm MSR}$($R=\SI{1}{\GeV}$), the pole mass $m_t^{\rm pole}$ and $\Omega_1$($R=\SI{2}{\GeV})$ for \pythia 8.205 with \mbox{$m_t^\mathrm{MC}=\SI{173}{\GeV}$} (left) and our new fits for \pythia 8.305 (right). The results are based on the 2-jettiness distribution $\tau_2$, gap subtraction scheme~1 and the strict treatment of $\mhat_t^2=(m_t/Q)^2$ power corrections. Central values, perturbative and incompatibility uncertainties, are all shown in GeV. The shape function parameters $\Delta_0=0.05$\,GeV and $\lambda=0.5$\,GeV are used.}
	\label{tab:oldresults} 
\end{table}

The calibration results for $m_t^{\rm MC}=173$\,GeV obtained in Ref.~\cite{Butenschoen:2016lpz} (appearing in Tab.~1 of that reference) are shown in the left half of Tab.~\ref{tab:oldresults}. Note that in Ref.~\cite{Butenschoen:2016lpz} values for $\Omega_1$($R=\SI{2}{\GeV})$ were only quoted for the MSR mass analysis. The central values, the perturbative uncertainty (coming from profile function scale variations) and the incompatibility uncertainty as described in Sec.~\ref{sec:fitprocedure} are displayed. The results demonstrate that $m_{t}^{\mathrm{\pythia}}$ is indeed close to the MSR mass at $R=\SI{1}{\GeV}$ at N$^2$LL$+$NLO and NLL$+$LO. On the other hand, there is a significant discrepancy in the pole mass analysis at both orders. The perturbative uncertainties decrease substantially at N$^2$LL$+$NLO in comparison to NLL$+$LO, but the incompatibility uncertainties, which quantify the disagreement of the MC event generator remain comparable. The pole mass calibration results exhibit a large correction between orders, which is associated to the fact that the NLO corrections are larger in the pole mass scheme. The more stable MSR mass results illustrate that in this scheme, and with the proper choice of the MSR mass scale $R$, a sizable fraction of the higher-order QCD corrections related to the top mass sensitivity of the $\tau_2$ distribution in the peak region are absorbed in the mass. Due to the absence of the pole-mass infrared renormalon in this short-distance scheme, the MSR results are expected to be more stable also at higher orders than those of the pole mass.

The right half of Tab.~\ref{tab:oldresults} shows the results of the calibration fits with our new setup and for \pythia~8.305. Up to small differences they are equivalent to the results quoted in Ref.~\cite{Butenschoen:2016lpz}. We have also carried out a calibration with our new setup for \pythia~8.205 which yields numbers that are within $10$\,MeV equivalent to the ones shown in the right half of the table. The agreement between the new results and those from Ref.~\cite{Butenschoen:2016lpz} means that the differences between the old and new fit setup only have a marginal effect. The features in the new setup which have been changed compared to the one of Ref.~\cite{Butenschoen:2016lpz} are:
\begin{enumerate}
	\item The renormalization scale of SCET non-singular term in Eq.~(\ref{eq:signsbstrict}) is $\mu_J$, the renormalization scale of the distributional terms in the bHQET jet function. In the old setup that scale was frozen at the mass mode matching scale $\mu_m$. The effect of this change is tiny because the contribution of this non-singular is small.
	
	\item In the new setup, the interpolation over $m_t$ is at the bin level and a simultaneous fit of all parameters is carried out. The approach of the old setup was to first minimize with respect to the shape function parameters at fixed $m_t$ giving  $\chi^2(m_t,\{a_{\mathrm{min}}(m_t)\})$, then interpolating this marginalized $\chi^2$ over $m_t$ and finding the minimum with respect to $m_t$. Both methods are in principle equivalent if the grid in $m_t$ is fine enough, but the new fit procedure	 is in general more robust.
	
	\item The old fit setup included two additional $Q$ sets: $(600-900)$\,GeV and $(700-1000)$\,GeV. With the new setup, which allows for more freedom in the parametrization of the shape function, it turns out that these two $Q$ sets, which are quite restricted in the range of $Q$ values, are not able to break the degeneracy between $m_t$ and $\Omega_1$. 
	They have therefore been dropped in the new setup for efficiency reasons. We have checked that the removal of these two $Q$ sets only has small effects on the final results quoted in Ref.~\cite{Butenschoen:2016lpz}.
\end{enumerate}
We note that all results in Tab.~\ref{tab:oldresults}, like those quoted in Ref.~\cite{Butenschoen:2016lpz}, are based on the strict approach for the treatment of $\mhat_t^2$ power corrections of Sec.~\ref{sec:non-singularstrict}.

\begin{figure}[t]
	\makebox[\textwidth]{\includegraphics[width=1.05\textwidth]{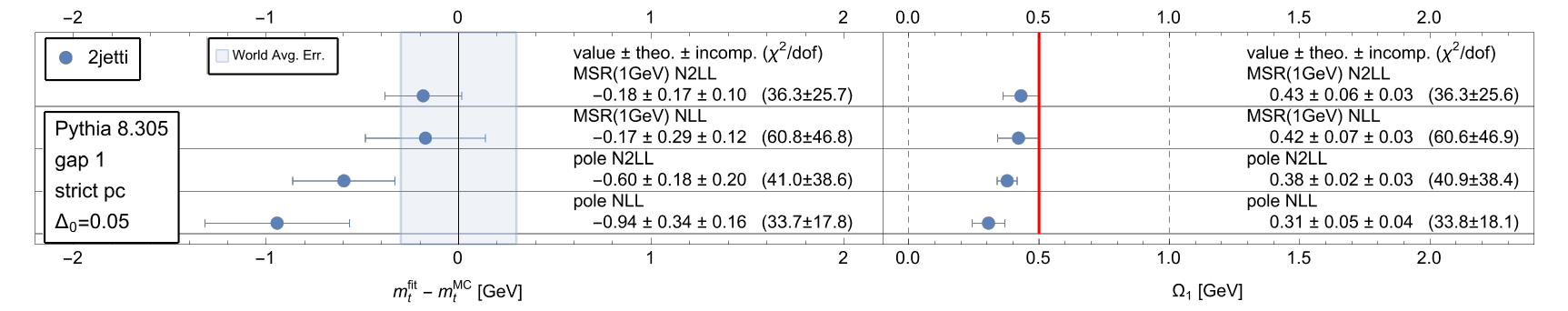}}
	\caption{Calibration results for the MSR top mass $m_t^{\rm MSR}(R=\SI{1}{\GeV})$, the pole mass $m_t^{\rm pole}$ and $\Omega_1$($R=\SI{2}{\GeV})$ in both mass schemes for \pythia 8.305 based on the 2-jettiness distribution $\tau_2$ with $m_t^\mathrm{MC}=\SI{173}{\GeV}$ using either the NLL$+$LO or N\textsuperscript{2}LL$+$NLO theoretical computations in gap scheme~1 with strict treatment of $\mhat_t^2=(m_t/Q)^2$ power corrections. The numbers for the minimal $\chi^2/\mathrm{dof}$ values for the top mass and $\Omega_1$, which are shown in parentheses for each calibration result, differ since the removal of outliers as described in Sec.~\ref{sec:fitprocedure} is carried out independently for both parameters. The blue shaded region visualizes the current world average uncertainty of $300$\,MeV quoted by the review of particle physics for direct top mass measurements~\cite{ParticleDataGroup:2022pth}. The shape function parameters $\Delta_0=0.05$\,GeV and $\lambda=0.5$\,GeV are used. Here and in subsequent figures below all error bars are symmetric.}
	\label{fig:referencepythia8305}
\end{figure}

A graphical representation of the \pythia~8.305 results in the right half of Tab.~\ref{tab:oldresults} is shown in Fig.~\ref{fig:referencepythia8305}. The gap scheme and the approach for the treatment of $\mhat_t^2$ power correction is  indicated in a label box on the left. On the left side the results for $m_t^{\rm MSR}(1\,\mbox{GeV})-m_t^{\rm MC}$ and $m_t^{\rm pole}-m_t^{\rm MC}$ are shown in GeV units. On the right side the results for $\Omega_1^{(1)}(2\,\mbox{GeV})$ in gap scheme~1 for both mass schemes are shown in units of GeV as well. The MSR and pole mass scheme results at N$^2$LL$+$NLO and NLL$+$LO are arranged vertically with the NLL$+$LO pole-mass results at the bottom and the N$^2$LL$+$NLO MSR-mass results at the top. The individual results are graphically displayed with error bars obtained from quadratically adding the perturbative and incompatibility uncertainties. The numerical values for the central values and the two uncertainties are in addition displayed individually to the right of the graphical representation, where the perturbative and incompatibility uncertainties appear in first and second place, respectively. In parentheses we also show the average minimal $\chi^2/\mbox{dof}$ value and standard deviation of all the fits (from the different profile functions and $Q$ sets after removing outliers as described in Sec.~\ref{sec:fitprocedure}). We remind the reader that the $\chi^2$ function defined in Eq.~(\ref{eq:chi2func}) only accounts for statistical MC uncertainties arising from the $10^7$ events we use to generate each MC shape distribution and that at this level of statistics the MC shape distributions are completely smooth functions so that the resulting statistical errors from the MC runs are negligible. This implies that the overall normalization of the minimal $\chi^2/\mbox{dof}$ values depends on the statistics we used for the MC samples. However, since the $10^7$ statistics is globally the same for all MC samples, the relative size of the quoted $(\chi^2/\mbox{dof})_{\rm min}$ values from the different fits still gives us important information concerning their quality. The numbers for $(\chi^2/\mbox{dof})_{\rm min}$ indicate a reasonably good fit.

To ease the interpretation of the results for $\Omega_1(2\,\mbox{GeV})$ obtained in the different gap schemes, the values are always converted to gap~1, $\Omega_1^{(1)}(2\,\mbox{GeV})$, using the formula (see Eq.~(\ref{eq:gapconversion}))
\begin{equation}
\label{eq:omegacorr}
	\Omega_{1,\mathrm{plot}}^{(1)}(2\,\mbox{GeV})=\Omega_{1}^{(i)}(2\,\mbox{GeV}) + [\bar{\delta}^{(i)}(2\,\mbox{GeV},2\,\mbox{GeV})-\bar{\delta}^{(1)}(2\,\mbox{GeV},2\,\mbox{GeV})]_{{\cal O}(\alpha_s)}\,,
\end{equation}
where the gap subtraction series on the RHS are evaluated at the scale $2$\,GeV and truncated at ${\cal O}(\alpha_s)$. We note that the difference between the values for $\Omega_{1}^{(i)}(2\,\mbox{GeV})$ among the gap schemes is quite large. To visualize it we also display the values for $\Omega_{1}^{(i)}(2\,\mbox{GeV})$ with thick red vertical lines. These values are obtained from the gap~1 reference value \mbox{$\Omega_{1,\mathrm{plot}}^{(1)}(2\,\mbox{GeV})=0.5$\,GeV} using the inverse of Eq.~(\ref{eq:omegacorr}):
\begin{equation}
\label{eq:redline}
\Omega_{1,\mathrm{red \, line}}^{(i)}(2\,\mbox{GeV})  = \SI{0.5}{\GeV} - [
\bar{\delta}^{(i)}(2\,\mbox{GeV},2\,\mbox{GeV})-\bar{\delta}^{(1)}(2\,\mbox{GeV},2\,\mbox{GeV})
]_{{\cal O}(\alpha_s)}\,.
\end{equation}
The difference of $\Omega_{1}^{(i)}(R_\Delta)$ values for different schemes $i$ only depends on the scale $R_\Delta$. Note that
in the discussions of the following analyses for brevity we frequently refer to $\Omega_{1,\mathrm{plot}}^{(1)}(2\,\mbox{GeV})$ simply as $\Omega_{1}$.

\section{Refinement for Shape Function Fits}
\label{sec:shapefctfits}

As we have already mentioned in Sec.~\ref{sec:gapsubtractions}, see paragraph below Eq.~(\ref{eq:O1formula}), employing a fit for the model function coefficients $c_0$ to $c_3$ while fixing the shape-function parameters $\Delta_0=0.05$\,GeV and $\lambda=0.5$\,GeV, is adequate only for gap scheme~1. In this section we investigate the modifications needed to carry out reliable shape-function fits for gap schemes~2 and 3, and we explain the fast shape-function fit procedure we adopt for our final calibration analysis.

\subsection{Gap Dependent Fits}
\label{sec:gapdependentfits}

\begin{figure}[t]
	\makebox[\textwidth]{\includegraphics[width=\textwidth]{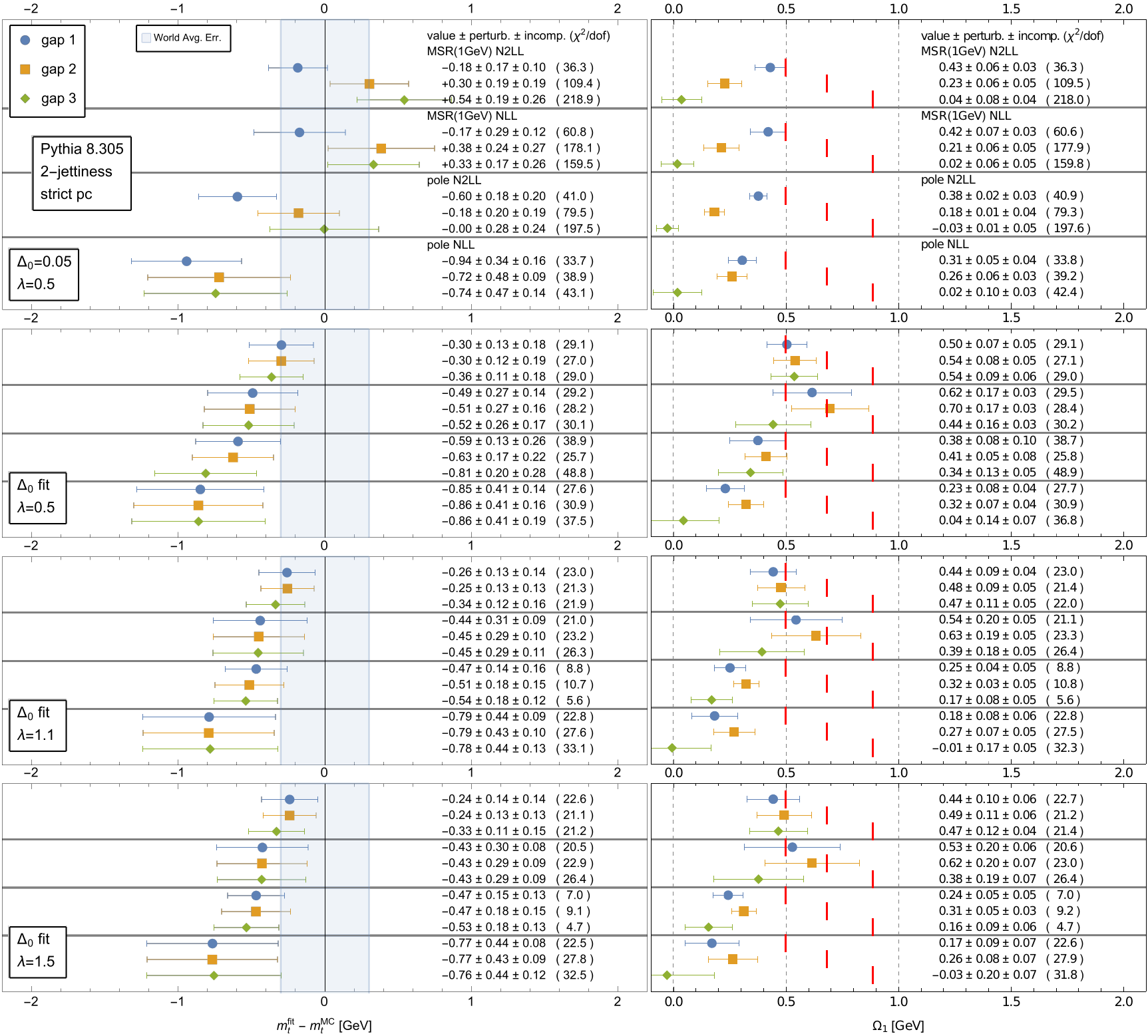}}
	\caption{Gap scheme dependence of the fitted top mass $m_t^\mathrm{fit}$ and the first moment $\Omega_1$ converted to the gap 1 scheme at ${\cal O}(\alpha_s)$, see Eq.~(\ref{eq:gapconversion}). Except for the treatment of $\Delta_0$ and $\lambda$, the calibration results refer to exactly the same setup used in Fig.~\ref{fig:referencepythia8305}. Upper part: Fixed $\Delta_0$ fits do not allow unbiased measurements of $m_\mathrm{fit}$ and $\Omega_1$. Lower parts: Floating $\Delta_0$ fits for three different $\lambda$ values employing $\Delta_0$-independent profiles. The extracted fit values for both parameters are substantially more stable across different gap schemes and for all values of $\lambda$, particularly at N$^2$LL+NLO order and in the MSR top mass scheme. All fits are based on 2-jettiness $\tau_2$.}
	\label{fig:gap23fits1}
\end{figure}

In the upper part of Fig.~\ref{fig:gap23fits1} we display the calibration results for the setup discussed in Sec.~\ref{sec:consistencyresults} for all three gap schemes, based on the 2-jettiness distribution $\tau_2$ and the strict treatment of $\mhat_t^2$ power corrections. The blue bars are the results already displayed in Fig.~\ref{fig:referencepythia8305}, while the orange and green bars refer to gap schemes~2 and 3, respectively. The results for these two schemes differ strongly from one another, but also from gap scheme~1. However, we also observe, that the values for $\chi^2/$dof are significantly larger for gap~2 and even more for gap~3, indicating a much worse fit for these two schemes. The differences in the fit results for $\Omega_1$ for the three gap schemes (even after conversion to gap scheme~1) are furthermore similar to the scheme differences themselves (prior to the conversion to scheme~1) as can be seen from the vertical red lines. This shows that the parametrization of the shape function we used for gap scheme~1 with the fixed values $\Delta_0=0.05$\,GeV and $\lambda=0.5$\,GeV, and using $c_0$\,--\,$c_3$ as fit parameters is not adequate for gap schemes~2 and 3.

The shape function parameters that are naturally connected to the first moment of the shape function $\Omega_1$ in Eq.~(\ref{eq:O1def}) and the effects of the gap scheme, are the renormalon free gap parameter $\hat\Delta =\hat\Delta^{(i)}(R_s,\mu_S)$ and more specifically $\Delta_0$ defined in Eq.~\eqref{eq:Deltahat}. Recall that at the reference scale $R_s=\mu_\delta\equiv R_\Delta=2$\,GeV we have $\hat\Delta =\Delta_0$. As can be seen from Eq.~\eqref{eq:Sgapderivation}, changing the gap scheme, let's say from scheme $i$ to scheme $j$, effectively corresponds to a renormalon-free redefinition of the shape-function's momentum $k\to k + 2\overline{\Delta}^{(i)}(R_s,\mu_S)-2\overline{\Delta}^{(j)}(R_s,\mu_S)$. If we would carry out the sum over all basis functions in Eq.~(\ref{eq:shapefunc}) for the shape function $F(k; \lambda,\{c_i\},N)$, i.e.\ in the limit $N\rightarrow\infty$, this shift could be accurately parametrized by corresponding changes in the infinite sequence of coefficients $\{c_i\}$ for any value of $\Delta_0$ and $\lambda$ without leading to tensions in the fit. In other words, $\Delta_0$ and $\lambda$ would simply specify a choice of basis which would not affect the final outcome. The results shown in the upper part of Fig.~\ref{fig:gap23fits1} indicate that for the truncation value $N=3$ we adopt (since using a larger $N$ is not feasible) this is not any more the case for gap schemes~2 and~3. From the mathematical perspective this means that for the values $\Delta_0=0.05$\,GeV and $\lambda=0.5$\,GeV the quadratic polynomial in Eq.~(\ref{eq:omega1formula1}) is bounded too tightly on the hyper-sphere $\sum_{i=0}^3 c_i^2 = 1$ for these gap schemes. A resolution is to treat $\Delta_0$ as an additional fit parameter as we know that the main issue of the tension is associated to shifts in $k$. 

To efficiently perform the fits we can add an additional $\Delta_0$-dimension to the grid and interpolation procedure described in Sec.~\ref{sec:dataprocessing}. This additional $\Delta_0$ dependence can be handled in the same way as the dependence on the top quark mass. We use steps of the size $\delta\Delta_0=\SI{0.05}{\GeV}$ within the interval $[\SI{-1.00}{\GeV},\SI{1.90}{\GeV}]$ which safely covers all gap and mass schemes at NLL$+$LO and N$^2$LL$+$NLO. The generalization of Eq.~(\ref{eq:bintheory}) then reads
\begin{equation}
\label{eq:bintheory2}
\hat f_{Q,i}^\mathrm{theo} (m_t,\Delta_0,\{a\}) = \sum_{kl}c_k(\{a\})c_l(\{a\})f_{kli}(m_t,\Delta_0,Q)\,,
\end{equation}
where \textsc{MINUIT} is now able to smoothly sample in $\{a\}$, $m_t$, and $\Delta_0$. The outcome of the calibration fits within this extended framework is shown in the three lower sections of Fig.~\ref{fig:gap23fits1} for $\lambda=0.5, 1.1$ and $1.5$ yielding good fits with equivalent top mass and $\Omega_1$ best-fit values within their uncertainties and $\chi^2/$dof values for all settings. The larger uncertainties we observe for  $\Omega_1$ for the pole mass fit results in gap scheme~3 are caused by its large subtraction coefficient, as we already anticipated in Sec.~\ref{sec:gapsubtractions} in the text after Eq.~(\ref{eq:gap3subtractionseries}). Similar observations for gap scheme~3 are also made in the subsequent fit results, and we emphasize that this is an artifact of this gap scheme. The independence concerning the width parameter $\lambda$ indicates that there is a strong degeneracy concerning $\Delta_0$ and $\lambda$, and that $\lambda$ can be safely fixed within a broad interval. The results with $\Delta_0$ as a fit parameter also agree with the original fit setup with fixed values $\Delta_0=0.05$\,GeV and $\lambda=0.5$\,GeV for gap scheme~1, reassuring that the original fit setup is perfectly adequate for this gap scheme.\footnote{The reliability of the soft function fits for gap scheme~1 has already been carefully examined in Ref.~\cite{Butenschoen:2016lpz}.}

\subsection[Fast Fit Procedure with $\Delta_0$ Dependent Profiles]{Fast Fit Procedure with $\mathbf{\Delta_0}$ Dependent Profiles}
\label{sec:Delta0profiles}

Using $\Delta_0$ as a general and independent fit parameter comes with the downside that the size of the interpolation grid increases substantially. This makes the general setup for a floating $\Delta_0$ fit as described in the previous section very costly and time intensive. For calibration studies this setup is only practical if the $\Delta_0$ grid dimension is based on much smaller gap-scheme-dependent ranges and if the one-dimensional spline interpolation that was applied to the $m_t$-dimension before is now replaced by a two-dimensional spline interpolation in the top mass and $\Delta_0$ directions, which leads to lower interpolation precision. For detailed and extended calibration studies that approach turns out to be too slow and expensive. For producing the final results we therefore adopt a physically equivalent, but much faster version of the floating $\Delta_0$ fit approach which, however, also requires setting suitable values for $\lambda$. This fast approach is described in the following subsection.

The fast version of the floating $\Delta_0$ fit procedure is based on the observation that the $\Delta_0$ dependence of the theoretical $\tau$ distributions is formally related to a trivial $Q$ dependent shift in $\tau$, see Eq.~(\ref{eq:bhqetdistribution}). This trivial $\Delta_0$ dependence would, however, only arise if the theory distributions were strictly renormalization-scale independent. In practice, the presence of the profile functions $\mu_i(\tau)$ yields a much more complicated dependence on $\Delta_0$. On the other hand, this complication  diminishes at increasing orders due to a smaller dependence of the $\tau$-distribution on the renormalization scales.

\begin{figure}[t]
	\makebox[\textwidth]{\includegraphics[width=\textwidth]{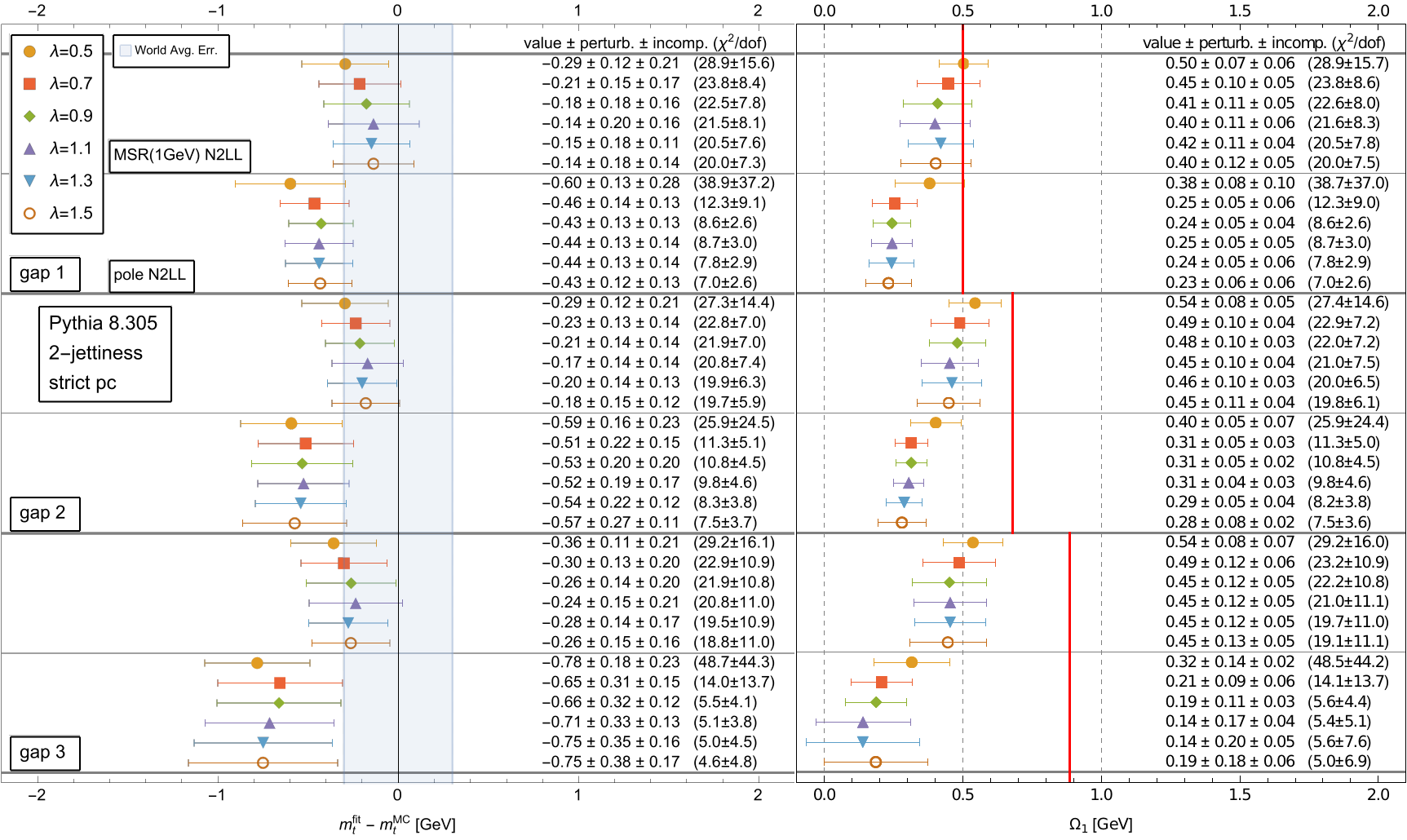}}
	\caption{Dependence of fitted parameters on $\lambda$ values in the range $[0.5-1.5]{\rm GeV}$ using the fast floating $\Delta_0$ fit procedure for \pythia~8.305 with  $m_t^\mathrm{MC}=\SI{173}{\GeV}$. We chose $\lambda=\SI{1.1}{\GeV}$ as default for \pythia.}
	\label{fig:pythialambda}
\end{figure}

For the fast version of the floating $\Delta_0$ calibration fits we make use of this observation and generate our $\tau$-grids with \textit{one} fixed $\Delta_{0,\mathrm{grid}}^{(i)}$ value adequate for each gap scheme $i$ and obtain the distribution for any other $\Delta_0$ by sampling shifted points:
\begin{equation}
\dv{\sigma}{\tau}\bigl(\tau,\Delta_0'\bigr)= \dv{\sigma}{\tau'}\bigl(\tau',\Delta_{0,\mathrm{grid}}\bigr)\biggr|_{\tau'=\tau-2r_s\frac{\Delta_0' - \Delta_{0,\mathrm{grid}}}{Q}}\,.
\end{equation}
This implies that all profile functions are also shifted accordingly:
\begin{equation}
\mu_i(\tau,\Delta_0') \equiv \mu_i\biggl(\tau-2r_s\frac{\Delta_0' - \Delta_{0,\mathrm{grid}}}{Q},\Delta_{0,\mathrm{grid}}\biggr).
\end{equation}
For gap scheme~1 we use $\Delta_{0,\mathrm{grid}}^{(1)}=0.05$\,GeV, while for gap schemes $i=2,3$
\begin{equation}
\Delta_{0,\mathrm{grid}}^{(i)} = 0.05\,\mbox{GeV} - [
\bar{\delta}^{(i)}(2\,\mbox{GeV},2\,\mbox{GeV})-\bar{\delta}^{(1)}(2\,\mbox{GeV},2\,\mbox{GeV})]_{{\cal O}(\alpha_s)}\,,
\end{equation}
is used, which yields $\Delta_{0,\mathrm{grid}}^{(2)}=0.23$\,GeV and $\Delta_{0,\mathrm{grid}}^{(3)}=0.44$\,GeV. To further speed up the code we generate interpolations of the cumulatives of the distribution functions $f_{k\ell}(\tau,m_t,Q)$ in Eq.~(\ref{eq:fkltaudef}) (where the profile functions are evaluated at $\tau'$),
\begin{equation}
\label{eq:Fkltaudef}
F_{k\ell}(\tau,m_t,Q)=\int_{0}^{\tau} {\rm d}\tau'f_{k\ell}(\tau',m_t,Q)\, ,
\end{equation}
for $\Delta_0=\Delta_{0,\mathrm{grid}}^{(i)}$ on the $\tau_i$ grid values\footnote{In Eq.~(\ref{eq:fkelli}) we removed the gap scheme superscript $(i)$ from $\Delta_{0,\mathrm{grid}}^{(i)}$ to avoid confusion with the index $i$ in $\tau_i$ that refers to the bin label.} and determine 2-D spline interpolations of the $F_{k\ell}(\tau,m_t,Q)$ over $m_t$ and $\tau$. The binned distribution functions that enter Eq.~(\ref{eq:bintheory2}) are then determined from the formula 
\begin{equation}
\label{eq:fkelli}
f_{k\ell i}(m_t,\Delta_0,Q)=F_{k\ell}\Bigl(\tau_{i+1}-2r_s\frac{\Delta_0 - \Delta_{0,\mathrm{grid}}}{Q},m_t,Q\Bigr)-F_{k\ell}\Bigl(\tau_i-2r_s\frac{\Delta_0 - \Delta_{0,\mathrm{grid}}}{Q},m_t,Q\Bigr)\,,
\end{equation}
which yields very accurate results due to the small size of our bins. This approach provides a substantial speed gain, since we can use a standard interpolator routine (\python class scipy.interpolate.RectBivariateSpline) that supports vectorization for parallelized evaluation. The creation of the grids as just described is substantially faster and relies on much smaller data files due to the removal of the $\Delta_0$ grid dimension. In addition, this reduces the time required to distribute the grids to each node of the computer cluster needed to carry out the fits.

The fast floating $\Delta_0$ fit approach just described reproduces within errors the results of the general and more flexible but very slow floating $\Delta_0$ fit procedure of Sec.~\ref{sec:gapdependentfits}. But it also reintroduces a dependence on the value of $\lambda$ and the gap scheme in the uncertainties when the calibration is carried out for the pole mass. In Fig.~\ref{fig:pythialambda} the results for the fast floating $\Delta_0$ calibration fits in the MSR and pole mass schemes for the \mbox{2-jettiness} distribution at N$^2$LL$+$NLO order are shown for $\lambda$ between $0.5$\,GeV and $1.5$\,GeV for \pythia~8.305. We see that the results stabilize and yield smaller values for $\chi^2/$dof only for $\lambda\ge 1.1$\,GeV. Compared to the results shown in Fig.~\ref{fig:gap23fits1} the central values are slightly shifted, some of the errors have increased and smaller values for  $\chi^2/$dof can be reached, but the results are fully compatible with those of Fig.~\ref{fig:gap23fits1}. For our final analyses we therefore adopt the fast floating $\Delta_0$ fit procedure with $\lambda=1.1$\,GeV for the calibration fits for \pythia~8.305.  As already anticipated (see also Sec.~\ref{sec:gapsubtractions}), the renormalization scale uncertainties for the pole mass fit results in gap scheme~3 are generally larger than for the other gap schemes. We remind the reader that this is an artifact of gap scheme~3.

\begin{figure}[t]
	\makebox[\textwidth]{\includegraphics[width=\textwidth]{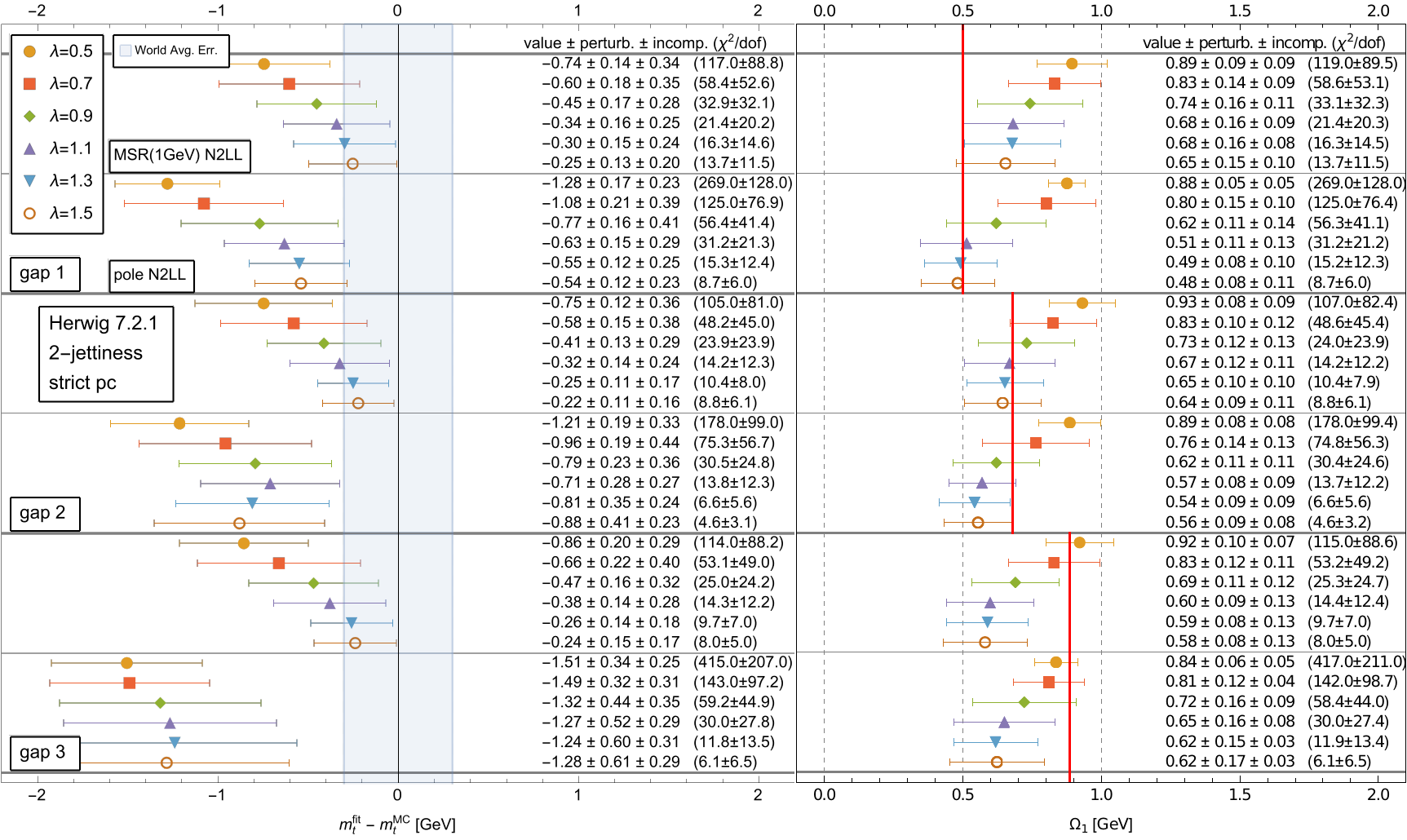}}
	\caption{Dependence of fitted parameters on $\lambda$ values in the range $[0.5-1.5]{\rm GeV}$  using the fast floating $\Delta_0$ fit procedure for \herwig~7.2 with  $m_t^\mathrm{MC}=\SI{173}{\GeV}$. We chose $\lambda=\SI{1.5}{\GeV}$ as default for \herwig.}
	\label{fig:herwiglambda}
\end{figure}

\begin{figure}[t]
	\makebox[\textwidth]{\includegraphics[width=\textwidth]{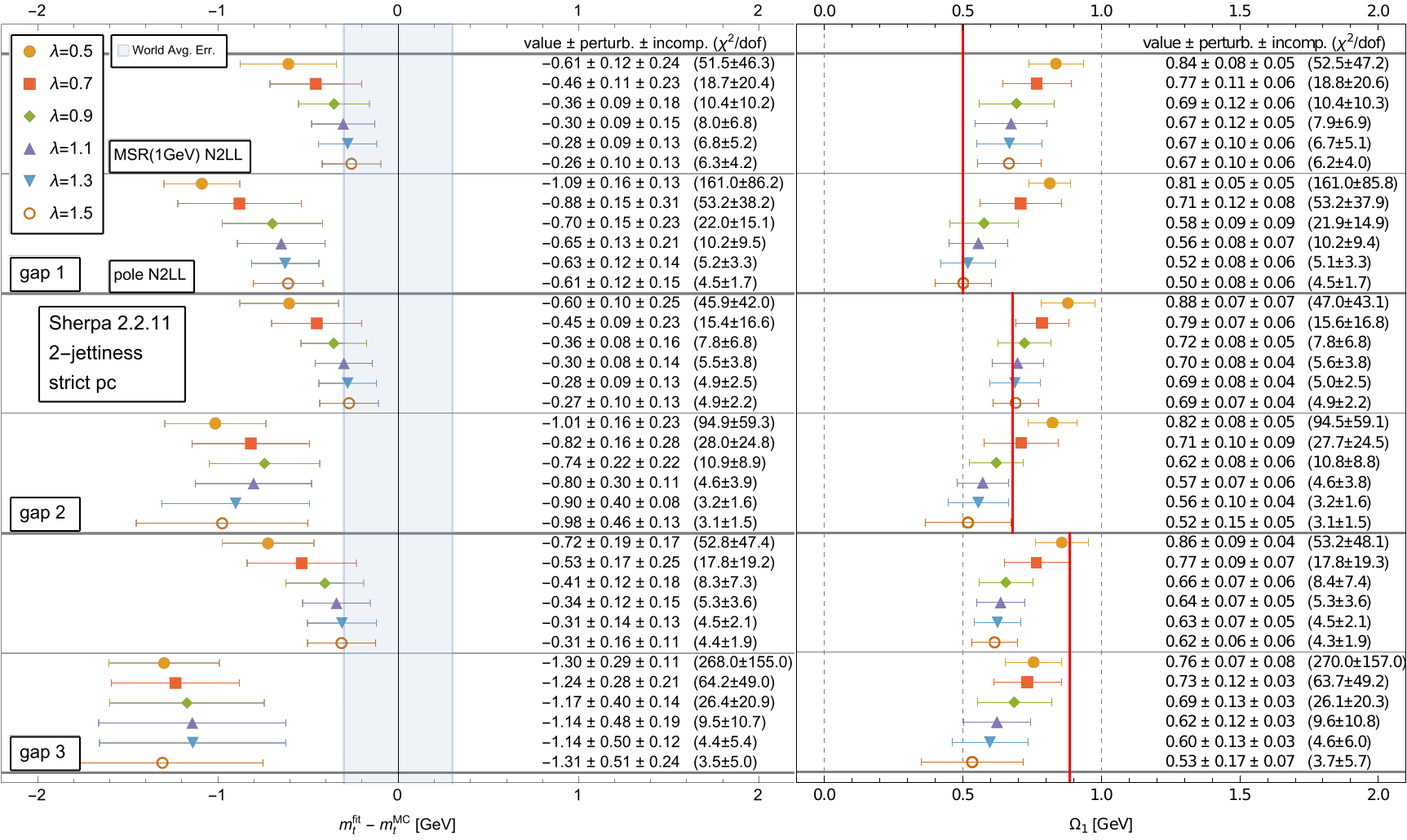}}
	\caption{Dependence of fitted parameters on $\lambda$ values in the range $[0.5-1.5]$\,GeV using the fast floating $\Delta_0$ fit procedure for \sherpa~2.2.11 with  $m_t^\mathrm{MC}=\SI{173}{\GeV}$. We choose $\lambda=\SI{1.3}{\GeV}$ as default for \sherpa.}
	\label{fig:sherpalambda}
\end{figure}

We have carried out analogous comparative analyses for \herwig~7.2 and \sherpa~2.2.11. The results for the fast floating $\Delta_0$ fit approach for \herwig 7.2 and \sherpa 2.2.11 are shown in Figs.~\ref{fig:herwiglambda} and \ref{fig:sherpalambda}, respectively. We observe again a stabilization of the results and improved fits for larger $\lambda$ values, but a much stronger dependence on $\lambda$ than for \pythia. For \herwig and \sherpa using a large value for $\lambda$ is even more important than for \pythia in order to obtain reliable results with the fast floating $\Delta_0$ fit procedure. This can be understood from the fact that the hadron-level distributions generated by \herwig and \sherpa are much broader than those from \pythia, as can be clearly seen in Fig.~\ref{fig:MCdistributions}. As we show in the discussion of our final results in Sec.~\ref{sec:finalresults} this must be attributed to the fact that for the standard tunes we have employed, the hadronization effects (i.e.\ the values for $\Omega_1$) are substantially larger for \herwig and \sherpa than for \pythia. When we apply the fast floating $\Delta_0$ fit procedure for \herwig 7.2 we use $\lambda=1.5$\,GeV while for \sherpa 2.2.11 we adopt $\lambda=1.3$\,GeV. As for the \pythia fits, shown in Fig.~\ref{fig:pythialambda}, we observe particularly sizable uncertainties for the pole mass fits in gap scheme~3, and to a lesser extent also in gap scheme~2. This can be attributed to bigger hadronization effects in \herwig and \sherpa which lower the stability for gap schemes with large gap subtractions.

\section{Observable Universality and Power Corrections}
\label{sec:pcanalysis}

In the preparatory calibration analyses carried out in Secs.~\ref{sec:consistencyresults} and \ref{sec:shapefctfits} based on the 2-jettiness distribution we have used the strict treatment of $\mhat_t^2=(m_t/Q)^2$ power corrections where, apart from incorporating the exact $\mhat_t$-dependent expression for $\tau_{\rm min}$, the leading singular bHQET cross section is defined strictly excluding any formally subleading $\mhat_t^2$ power corrections. This strict treatment of $\mhat_t^2$ power corrections has been explained in Sec.~\ref{sec:non-singularstrict} and was employed in the original calibration analysis of Ref.~\cite{Butenschoen:2016lpz}. In Sec.~\ref{sec:absorbconcept} we have provided conceptual arguments explaining why the strict treatment may not suffice at the precision achieved at N$^2$LL$+$NLO which yields uncertainties of around $200$\,MeV, as it may lead to a discrepancy for shape observables with different sensitivity to $\mhat_t^2$ power corrections. In the following we confirm these arguments by carrying out top mass calibration analyses for all three shape variables, 2-jettiness $\tau_2$, the sum of jet masses (sJM) $\tau_s$ and the modified jet mass (mJM) $\tau_m$. We demonstrate that the strict power correction treatment does not suffice to achieve observable independence and that the absorption prescription laid out in Sec.~\ref{sec:absorbconcept} is mandatory. 

\begin{figure}[t]
	\makebox[\textwidth]{\includegraphics[width=\textwidth]{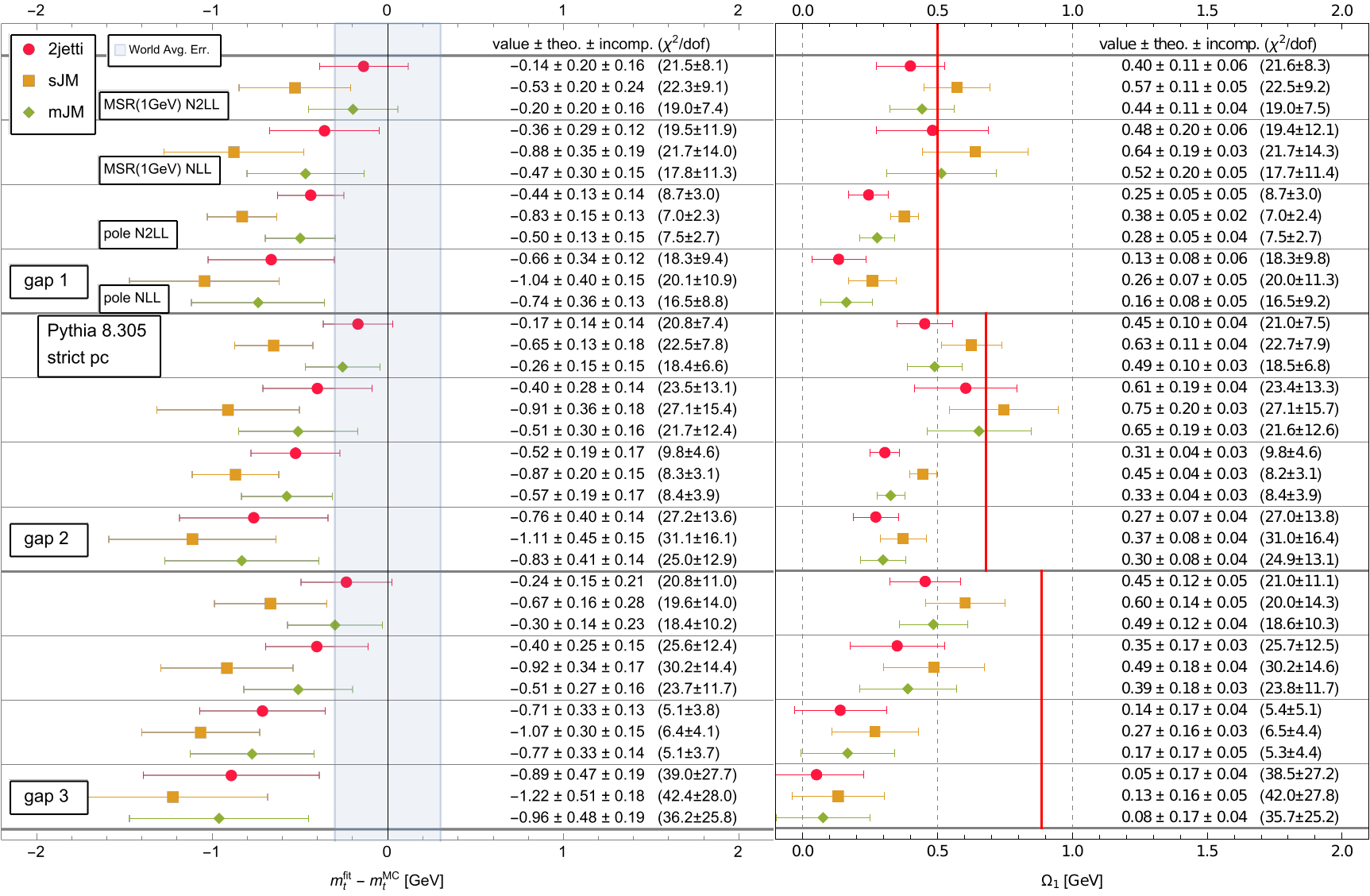}}
	\caption{Dependence of the calibration results on the different observables 2-jettiness, sJM (sum of jet masses) and mJM (modified jet mass) employing the fast floating $\Delta_0$ fit method of Sec.~\ref{sec:Delta0profiles} and strict treatment of $\mhat_t^2$ power corrections for \pythia~8.305 with $m_t^\mathrm{MC}=\SI{173}{\GeV}$. All top mass fit results for sJM are around $200$ to $300$\,MeV lower compared to the shape variables 2-jettiness and mJM.}
	\label{fig:pctreatment1}
\end{figure}

In Fig.~\ref{fig:pctreatment1} the results for the top mass calibration for \pythia~8.305 in the strict power correction treatment is shown for all three shape variables using gap schemes~1, 2 and 3, for the pole as well as the MSR mass and at N$^2$LL$+$NLO and NLL$+$LO. Here and in all subsequent calibration fits we employ the fast floating $\Delta_0$ fit procedure described in Sec.~\ref{sec:Delta0profiles}. It is conspicuous that all top mass results for the sJM variable are systematically lower by around $400$\,MeV compared to the outcome for the 2-jettiness and mJM variables. At the same time, the sJM fit results for $\Omega_1$ are systematically larger by around $200$\,GeV than for 2-jettiness and mJM. On the other hand, the results for 2-jettiness and mJM differ only slightly and are in agreement. The consistency of the results for 2-jettiness and mJM and the discrepancy with the sJM results strongly support the conceptual arguments given in Sec.~\ref{sec:absorbconcept} emphasizing the practical relevance of the $\mhat_t^2$ power corrections and in particular the important role of the soft rescaling factors $r_{\tau,s}(\mhat_t)$ from Eqs.~(\ref{eq:rstau2}), (\ref{eq:rssJM}) and (\ref{eq:rsmJM}) in the measurement $\delta$-function to achieve observable-independent calibration results. 

\begin{figure}[t]
	\makebox[\textwidth]{\includegraphics[width=\textwidth]{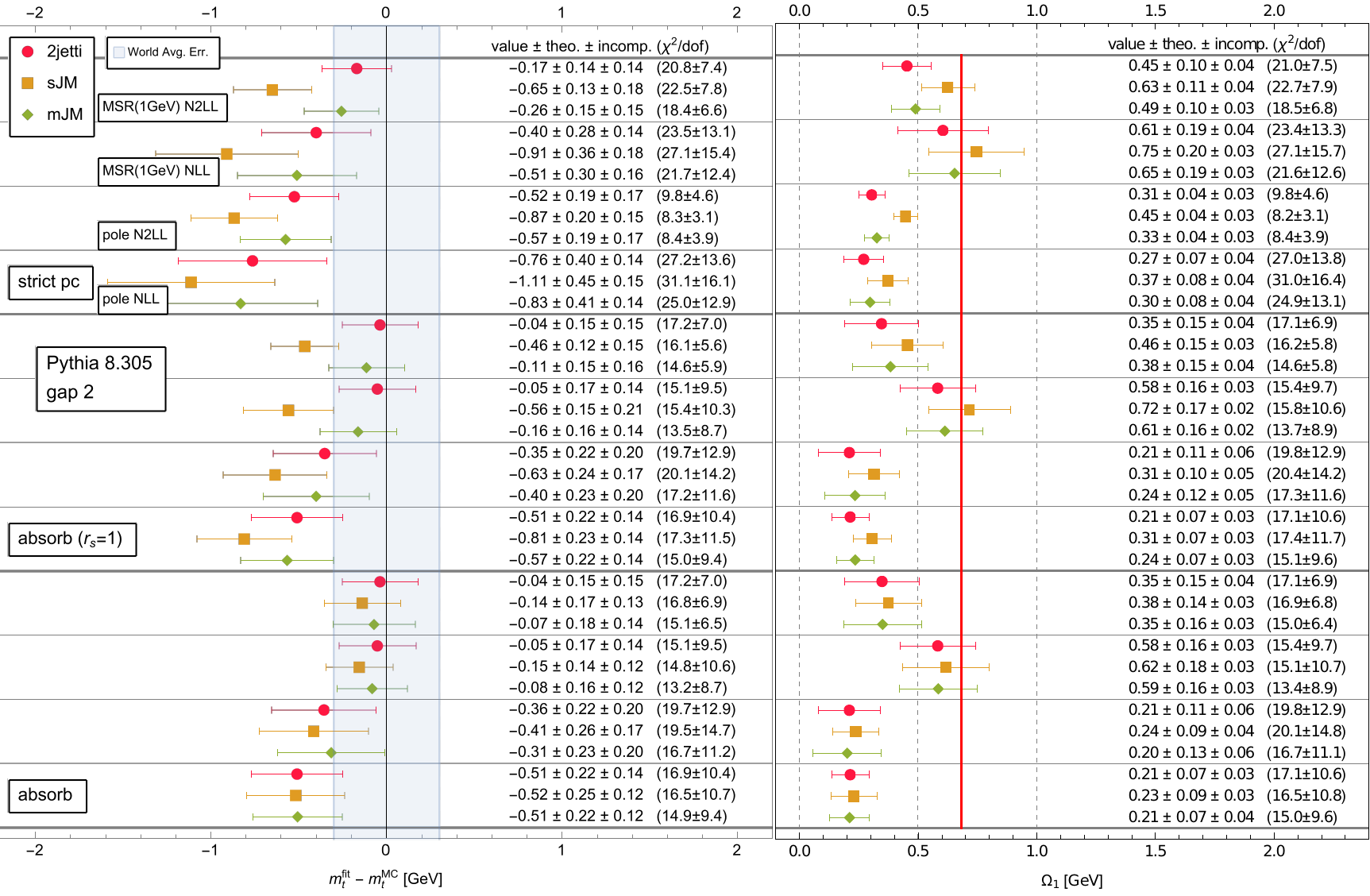}}
	\caption{Fit results for \pythia~8.305 using gap 2 scheme for $m_t^\mathrm{MC}=\SI{173}{\GeV}$. The different sections are: ``strict pc'' uses strict treatment of $(m_t/Q)^2$ power corrections, ``absorb ($r_s=1$)'' absorbs coefficients of distributions from the non-singular contribution into the resummed cross section, and ``absorb'' additionally includes the correct measurement power correction $r_s$.}
	\label{fig:pctreatment2}
\end{figure}

In Fig.~\ref{fig:pctreatment2} we now show the calibration results when the absorption prescriptions for the $\mhat_t^2$ power corrections given in Sec.~\ref{sec:absorbconcept} are employed. Here we only provide the results for gap scheme~2 since the observations for gap schemes~1 and 3 are very similar. In the upper portion of Fig.~\ref{fig:pctreatment2} the results for the strict power correction treatment already given in Fig.~\ref{fig:pctreatment1} are shown as a reference. The middle portion shows the results of our absorption prescription, but setting the soft rescaling factor for all shape variables to unity, $r_{\tau,s}(\mhat_t)=1$. We observe a small increase of about $100$ to $150$\,MeV for the top quark masses (except the NLL$+$LO order pole-mass results) and a comparable decrease for $\Omega_1$. The uncertainties at N$^2$LL$+$NLO are in general a bit larger as a result of the additional $\xi$ parameter variations. However, the discrepancy between the sJM and the 2-jettiness as well as mJM calibrations results remains similar to the strict power correction treatment. In the lower portion of Fig.~\ref{fig:pctreatment2} we use the complete absorption prescription including also the soft rescaling factors as shown in Eqs.~(\ref{eq:rstau2}), (\ref{eq:rssJM}) and (\ref{eq:rsmJM}). Compared to the middle portion, the 2-jettiness results are unchanged since $r_{\tau_2,s}(\mhat_t)=1$. The mJM results only move slightly since $r_{\tau_m,s}(\mhat_t)=1+{\cal O}(\mhat_t^4)$. The sJM results for the top quark masses, on the other hand, increase substantially by around $400$\,MeV and are now fully consistent with the 2-jettiness and mJM calibration results. Likewise, also the $\Omega_1$ results are now in agreement for all three shape variables. Interestingly, we also find that the absorption prescription leads to a general reduction of the perturbative uncertainties at NLL$+$LO order for the MSR and pole mass calibration fits. We have analyzed this behavior in great detail~\cite{OliverJinMasterthesis2022} and found that it is a general feature of  floating $\Delta_0$ fits in combination with the absorption prescription for the $\mhat_t^2$ power corrections visible for gap schemes~1 and 2. We believe that this is related to an accidental interplay between both procedures that leads to an artificial reduction of the profile (and $\xi$) parameter dependence at NLL$+$LO order where the QCD corrections are entirely encoded in renormalization-group evolution factors. These smaller NLL$+$LO perturbative uncertainties should therefore not be considered realistic. At N$^2$LL$+$NLO this effect does not arise. A second feature visible in  Fig.~\ref{fig:pctreatment2} and worth noticing is that the NLL$+$LO values for the pole mass increase by around $350$ to $400$\,MeV when the floating $\Delta_0$ fits are combined with the absorption prescription. Since the pole mass uncertainties at NLL$+$LO order are about $400$\,MeV this is not a point of concern. Still, we have analyzed this behavior as well~\cite{OliverJinMasterthesis2022} and found that half of this shift is caused by using the floating $\Delta_0$ fit and that this only happens for the NLL$+$LO pole-mass fits.

Overall, when using the full absorption prescription for the $\mhat_t^2$ power corrections we find gap scheme and observable independence. We therefore use this prescription for our final calibration analysis which we discuss in the following section.

\section{Final Results}
\label{sec:finalresults}

With all theoretical tools at hand we are now ready to discuss the final results of the NLL$+$LO and N$^2$LL$+$NLO top mass calibration fits for \pythia~8.305, \herwig~7.2 and \sherpa~2.2.11 for the pole and MSR masses based on the three shape variables 2-jettiness $\tau_2$, sJM $\tau_s$ and mJM $\tau_m$, and using the gap subtraction schemes~1, 2 and 3. The fits are based on the updated calibration framework, laid out in detail in the previous sections, which includes an updated shape-function fit procedure and a more sophisticated treatment of $\mhat_t^2$ power corrections. The results for \pythia~8.305 are an update for the results presented in Ref.~\cite{Butenschoen:2016lpz} for \pythia~8.205, where we have checked (see Sec.~\ref{sec:consistencyresults}) that, as far as the shape variables we use in our analysis are concerned, the two \pythia versions are fully equivalent. 

\begin{figure}[t]
	\makebox[\textwidth]{\includegraphics[width=\textwidth]{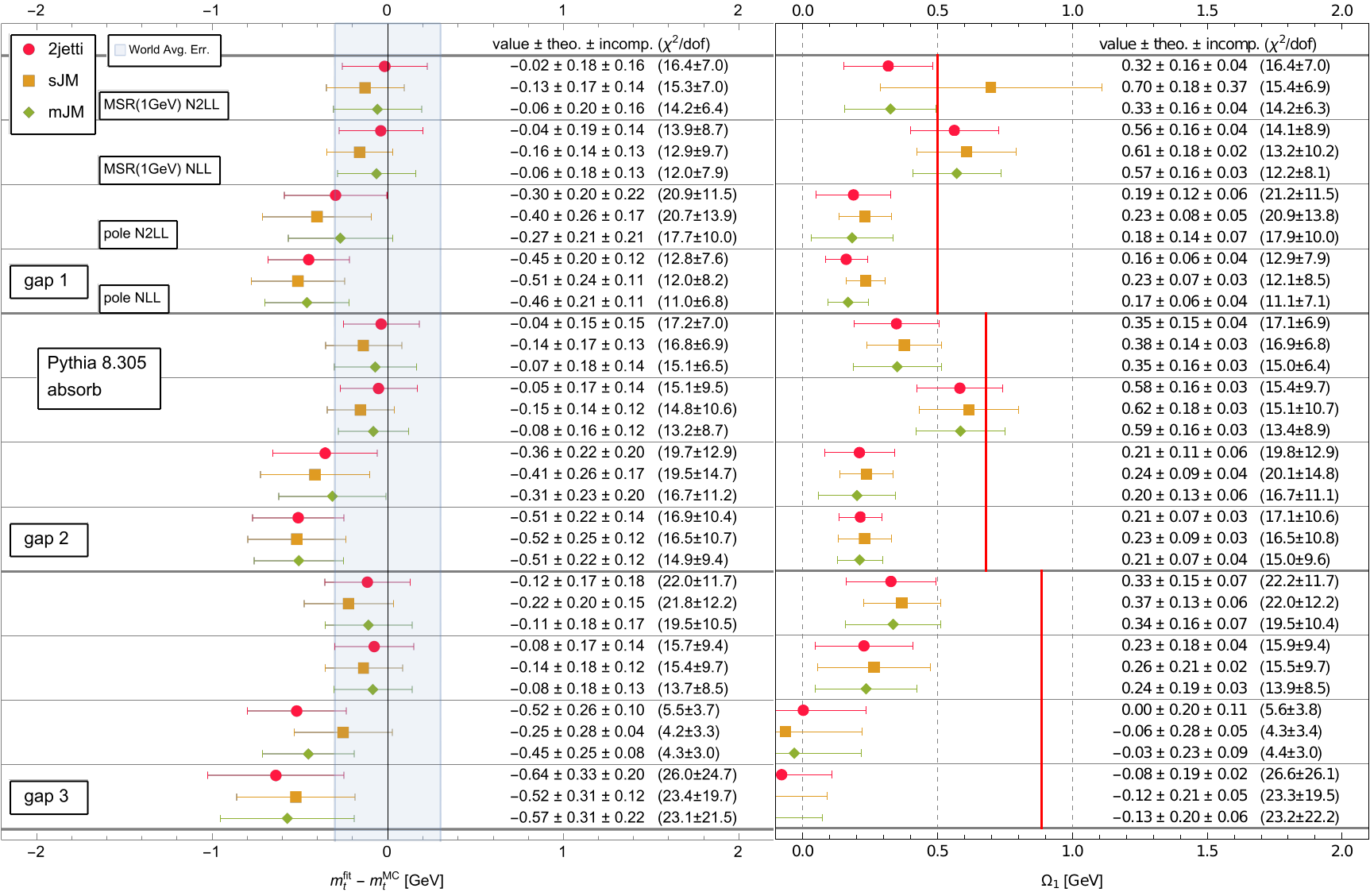}}
	\caption{Summary of final top mass calibration results for \pythia~8.305 with \mbox{$m_t^\mathrm{MC}=\SI{173}{\GeV}$} for three gap subtraction schemes and the shape variables 2-jettiness, sum of jet masses (sJM) and modified jet mass (mJM).}
	\label{fig:pythiafinaloverview}
\end{figure}

The final top mass calibration results for \pythia~8.305 and $m_t^\mathrm{MC}=\SI{173}{\GeV}$ are displayed in Fig.~\ref{fig:pythiafinaloverview}. We observe nicely consistent results for all shape variables in both mass schemes yielding uncertainties of about $200$\,MeV for the MSR mass $m_t^{\rm MSR}(1\,\mbox{GeV})$ and around $300$\,MeV for the pole mass at N$^2$LL$+$NLO order. The smaller uncertainties at NLL$+$LO for the pole mass results are accidental as we have pointed out in Sec.~\ref{sec:pcanalysis} and do not reflect the true uncertainties at this order. The rather large uncertainties (and instabilities for $\Omega_1$) visible for the pole mass calibration results in gap scheme~3 are an artifact of the sizable ${\cal O}(\alpha_s)$ subtraction in this gap scheme, as we have discussed in Sec.~\ref{sec:gapsubtractions}. The results in gap schemes~1 and 2 are very similar, apart from a glitch in the N$^2$LL$+$NLO result for $\Omega_1$ in the MSR scheme fit for sJM, which is caused by some numerical outliers that could not be removed by the procedure described in Sec.~\ref{sec:fitprocedure} (see bullet point~1). This was the only incident in our analysis where our prescription to remove outliers did not suffice. {\it We use the results obtained for the \mbox{2-jettiness} shape variable and in gap scheme~2 at N$^{\,2}\!$LL$+$NLO when quoting the final numbers for the results of our calibration analyses.} 

We have carried out the same analysis for $m_t^\mathrm{MC}$ between $170$ and $175$\,GeV in $1$\,GeV steps and obtained equivalent results for $ m_t^{\rm MSR}(1\,\mbox{GeV})-\mMC$ and $m_t^{\rm pole}-\mMC$. The N$^2$LL$+$NLO MSR mass results are visualized in the left panel of Fig.~\ref{fig:summaryallmtMC} and can be summarized as:
\begin{align}
\label{eq:pythiafinalMSR}
 m_t^\pythia{}&=m^\mathrm{MSR}_t(\SI{1}{\GeV})+\SI{0.03(21)}{\GeV}\,,\\
 \Omega_{1,{\rm MSR}}^\pythia(\SI{2}{\GeV}){}&=\SI{0.35(16)}{\GeV}\,.\nonumber
\end{align}
Note that the fit result for $\Omega_1$, here and in the following, does not depend on the $m_t^\mathrm{MC}$ value within a few MeV.
A comparison between the \pythia 2-jettiness distributions with the N$^2$LL$+$NLO theory cross section using the best MSR mass fit result for $Q=700$, $800$ and $1000$~GeV is shown in the top panels of Fig.~\ref{fig:summarybestfit2jettiness}.
The N$^2$LL$+$NLO pole mass fits, visualized in the left panel of Fig.~\ref{fig:summaryallmtMCpole}, read 
\begin{align}
\label{eq:pythiafinalpole}
m_t^\pythia{}&=m^\mathrm{pole}_t+\SI{0.35(30)}{\GeV}\,,\\
\Omega_{1,{\rm pole}}^\pythia(\SI{2}{\GeV}){}&=\SI{0.21(13)}{\GeV}\,.\nonumber
\end{align}
We remind the reader that the results for $\Omega_{1}(\SI{2}{\GeV})$ we present are always converted to gap scheme~1 via Eq.~(\ref{eq:omegacorr}).

At this point, a comparison to the original calibration analysis of Ref.~\cite{Butenschoen:2016lpz} carried out in the strict power correction approach and displayed (based on our own reanalysis) in Tab.~\ref{tab:oldresults}  and Fig.~\ref{fig:referencepythia8305}, is in order. The N$^2$LL$+$NLO results for the MSR and pole masses obtained in Ref.~\cite{Butenschoen:2016lpz} (based on 2-jettiness and gap scheme~1) were $\mMC = m_t^{\rm MSR}(1\,\mbox{GeV}) + (0.18 \pm 0.23)$\,GeV and $\mMC = m_t^{\rm pole} + (0.57 \pm 0.29)$\,GeV. In our updated analysis the MSR mass result has increased by $150$\,MeV at N$^2$LL$+$NLO (and by almost the same amount at NLL$+$LO). This reduces the difference of $m_t^{\rm MSR}(1\,\mbox{GeV})$ and $m_t^{\rm MC}$ from $180$\,MeV to only $30$\,MeV. The pole mass result at N$^2$LL$+$NLO shows a similar increase. These changes are primarily associated to the treatment of ${\cal O}(\mhat^2_t)$ power corrections in the final results of our new calibration fit. Interestingly, the NLL$+$LO pole-mass result increases by about $400$\,MeV reducing the difference between the calibration results at the two orders from about $350$\,MeV to about $200$\,MeV. The uncertainties for the MSR and pole masses at N$^2$LL$+$NLO are essentially identical to the ones quoted in Ref.~\cite{Butenschoen:2016lpz}. Within uncertainties, all N$^2$LL$+$NLO results are still fully compatible with those of the original calibration analysis, but the updated results presented here should be considered as more reliable.  

\begin{figure}[t]
	\makebox[\textwidth]{\includegraphics[width=\textwidth]{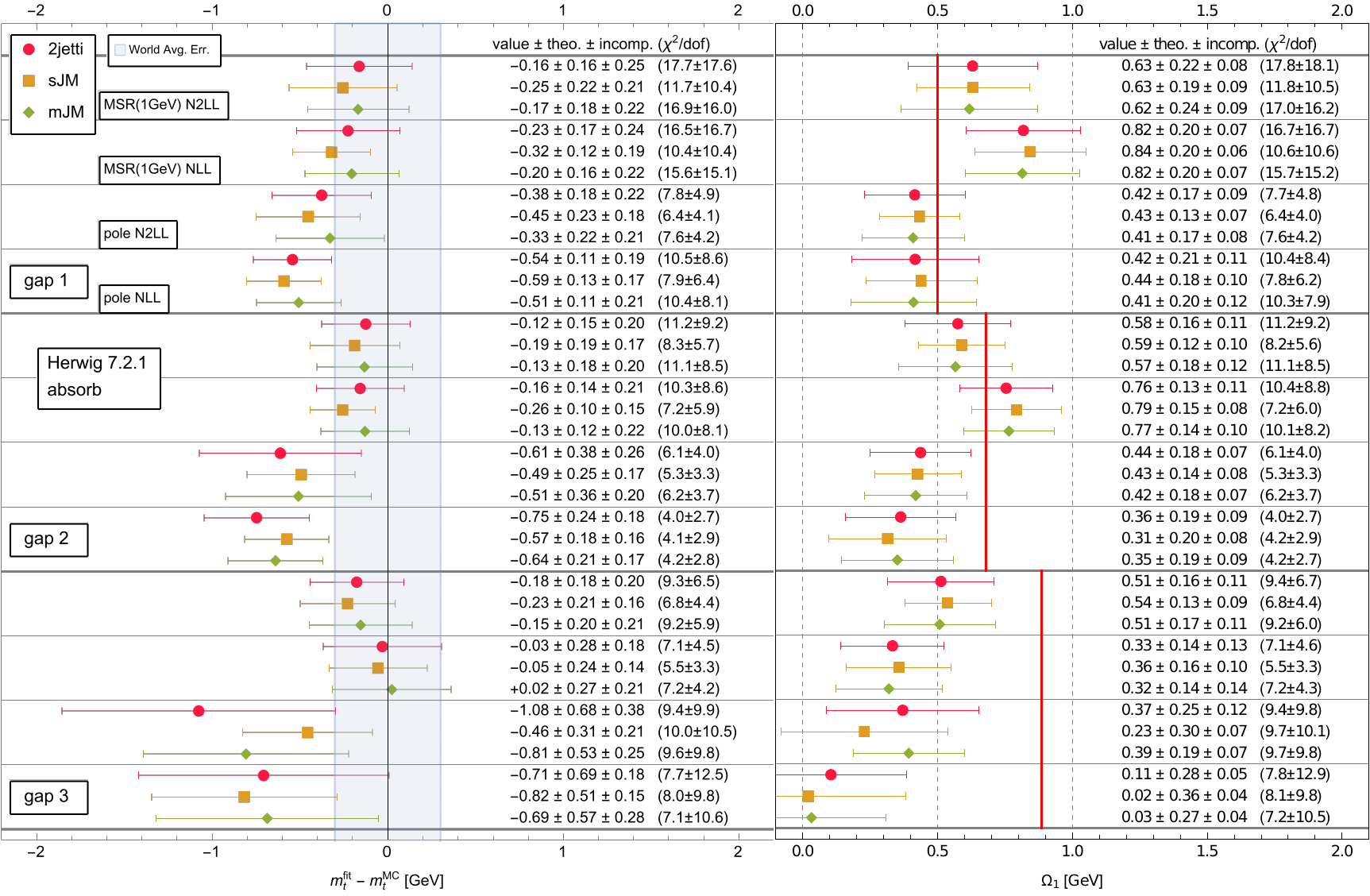}}
	\caption{Summary of final top mass calibration results for \herwig~7.2 with \mbox{$m_t^\mathrm{MC}=\SI{173}{\GeV}$} for three gap subtraction schemes and the shape variables 2-jettiness, sum of jet masses (sJM) and modified jet mass (mJM).}
	\label{fig:herwigfinaloverview}
\end{figure}

The final top mass calibration results for \herwig~7.2 and $m_t^\mathrm{MC}=\SI{173}{\GeV}$ are displayed in Fig.~\ref{fig:herwigfinaloverview}. As for the \pythia analysis we observe nice consistency for the three shape variables 2-jettiness, sJM and mJM and for the three gap schemes albeit with larger uncertainties in gap schemes~2 and 3, particularly in the pole mass scheme. We have again carried out the same analysis for $m_t^\mathrm{MC}$ between $170$ and $175$\,GeV in $1$\,GeV steps and obtained equivalent results for $ m_t^{\rm MSR}(1\,\mbox{GeV})-\mMC$ and $m_t^{\rm pole}-\mMC$. The N$^2$LL$+$NLO MSR mass results are visualized in the central panel of Fig.~\ref{fig:summaryallmtMC} and can be summarized as:
\begin{align}
\label{eq:herwigfinalMSR}
m_t^\herwig{}&=m^\mathrm{MSR}_t(\SI{1}{\GeV})+\SI{0.12(25)}{\GeV}\,,\\
\Omega_{1,{\rm MSR}}^\herwig(\SI{2}{\GeV}){}&=\SI{0.58(19)}{\GeV}\,.\nonumber
\end{align}
A comparison between the \herwig 2-jettiness distributions with the N$^2$LL$+$NLO theory cross section using the best MSR-mass fit result for $Q=700$, $800$ and $1000$\,GeV is shown in the central panels of Fig.~\ref{fig:summarybestfit2jettiness}. The N$^2$LL$+$NLO pole-mass fits, visualized in the middle panel of Fig.~\ref{fig:summaryallmtMCpole}, read 
\begin{align}
\label{eq:herwigfinalpole}
m_t^\herwig{}&=m^\mathrm{pole}_t+\SI{0.61(47)}{\GeV}\,,\\
\Omega_{1,{\rm pole}}^\herwig(\SI{2}{\GeV}){}&=\SI{0.44(18)}{\GeV}\,.\nonumber
\end{align}
The rather large uncertainty of $460$\,MeV for the pole mass calibration is caused by a particularly strong dependence on the $\xi$ parameter variations and is even larger for gap scheme~3, compared to a much smaller uncertainty for gap scheme~1. We believe this is caused by the broadness of the \herwig shape distributions shown in Fig.~\ref{fig:MCdistributions} which makes the fits more unstable for larger gap subtractions at low orders due to the stronger infrared-sensitivity in the cross sections in the pole mass scheme. We make a similar observation for the \sherpa pole-mass results in Eq.~(\ref{eq:sherpafinalpole}) where the effects is, however, less pronounced since the broadness of the \sherpa peak distribution is smaller than for \herwig, see Fig~\ref{fig:MCdistributions}. Since the subtraction of gap scheme~2 lays in the middle between gap schemes~1 and 3, see Sec.~\ref{sec:gapsubtractions}, we consider that the result for gap~2 provides a fair estimate for the pole mass calibration uncertainties for \herwig.

\begin{figure}[t]
	\makebox[\textwidth]{\includegraphics[width=\textwidth]{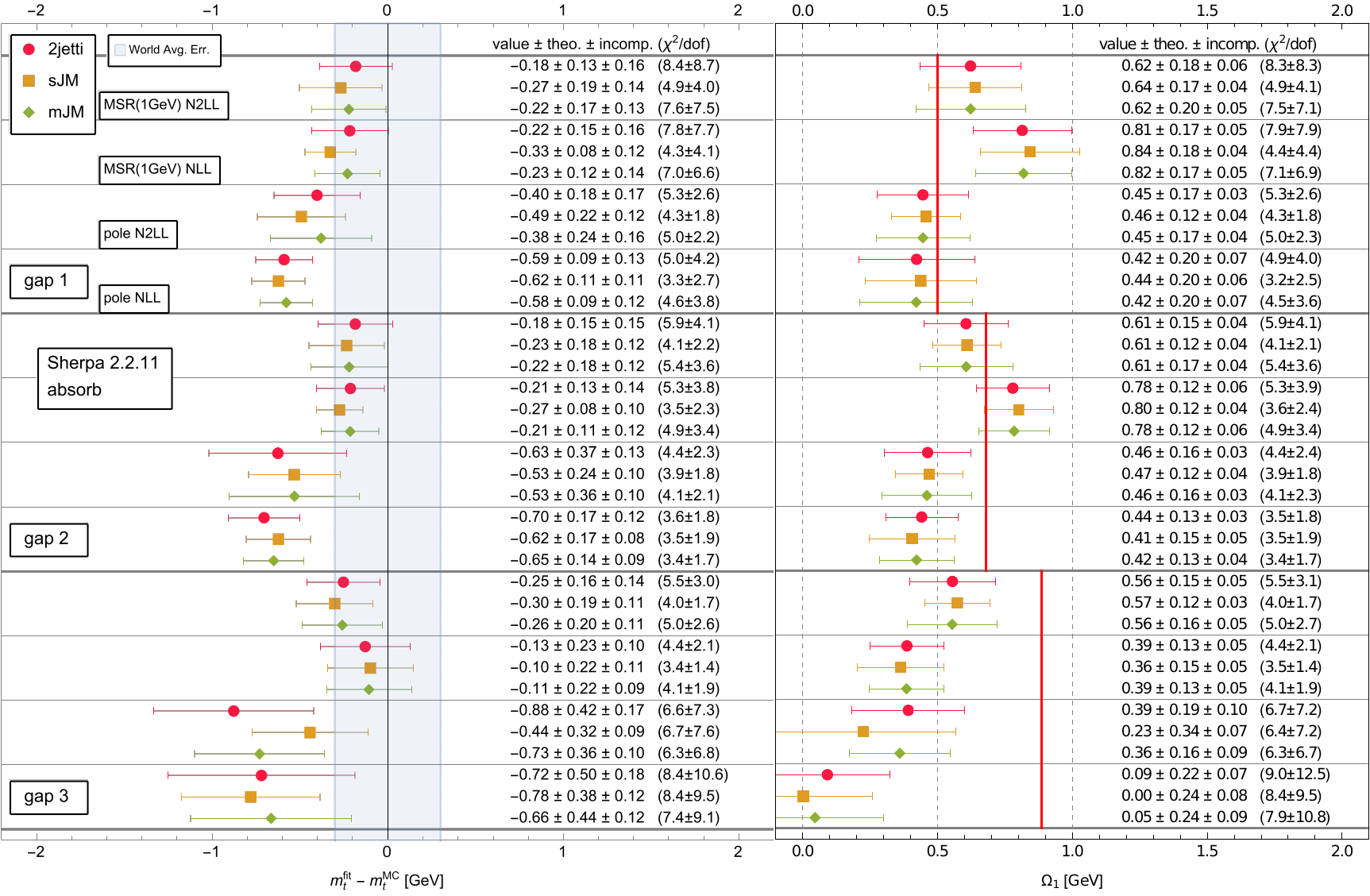}}
	\caption{Summary of final top mass calibration results for \sherpa~2.2.11 with \mbox{$m_t^\mathrm{MC}=\SI{173}{\GeV}$} for three gap subtraction schemes and the shape variables 2-jettiness, sum of jet masses (sJM) and modified jet mass (mJM).}
	\label{fig:sherpafinaloverview}
\end{figure}

The final top mass calibration results for \sherpa~2.2.11 with $m_t^\mathrm{MC}=\SI{173}{\GeV}$ are displayed in Fig.~\ref{fig:sherpafinaloverview}. As for the \pythia and \herwig analyses, we observe nice consistency for the three shape variables 2-jettiness, sJM and mJM, and for the three gap schemes. Compared to the \pythia results, the uncertainties in gap schemes~2 and 3 are again larger, particularly in the pole mass scheme, but they are not as sizable as for \herwig. This is correlated with the fact that the broadness of the \sherpa peak shown in Fig.~\ref{fig:MCdistributions} is in between those of \pythia and \herwig. Once again we have carried out the same analysis for $m_t^\mathrm{MC}$ between $170$ and $175$\,GeV in $1$\,GeV steps and obtained equivalent results for $ m_t^{\rm MSR}(1\,\mbox{GeV})-\mMC$ and $m_t^{\rm pole}-\mMC$. The N$^2$LL$+$NLO MSR mass results are visualized in the right panel of Fig.~\ref{fig:summaryallmtMC} and can be summarized as:
\begin{align}
\label{eq:sherpafinalMSR}
 m_t^\sherpa&{}=m^\mathrm{MSR}_t(\SI{1}{\GeV})+\SI{0.19(21)}{\GeV}\,,\\
	\Omega_{1,{\rm MSR}}^\sherpa(\SI{2}{\GeV})&{}=\SI{0.61(16)}{\GeV}\,.\nonumber
\end{align}
A comparison between the \sherpa 2-jettiness distributions with the N$^2$LL$+$NLO theory cross section using the best MSR mass fit result for $Q=700$, $800$ and $1000$\,GeV is shown in the lower panels of Fig.~\ref{fig:summarybestfit2jettiness}.
The N$^2$LL$+$NLO pole mass fits, visualized in the right panel of Fig.~\ref{fig:summaryallmtMCpole}, read
\begin{align}
\label{eq:sherpafinalpole}
m_t^\sherpa&{}=m^\mathrm{pole}_t+\SI{0.62(39)}{\GeV}\,,\\
\Omega_{1,{\rm pole}}^\sherpa(\SI{2}{\GeV})&{}=\SI{0.47(16)}{\GeV}\,.\nonumber
\end{align}

\begin{figure}[t]
	\makebox[\textwidth]{\includegraphics[width=\textwidth]{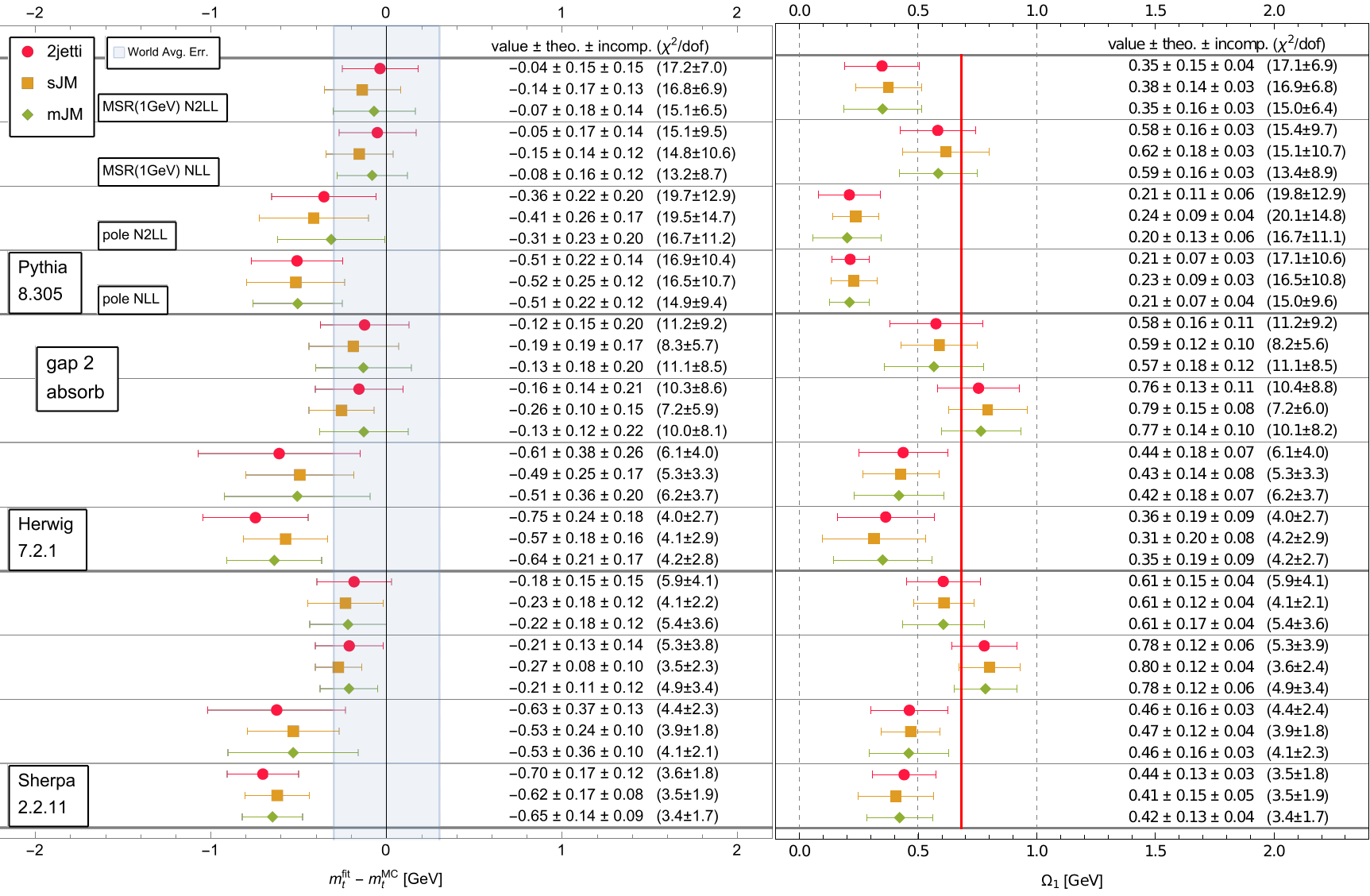}}
	\caption{Comparison between \pythia~8.305, \herwig~7.2 and \sherpa~2.2.11 final results for $m_t^\mathrm{MC}=\SI{173}{\GeV}$ for gap subtraction scheme~2 and the shape variables 2-jettiness, sum of jet masses (sJM) and modified jet mass (mJM).}
	\label{fig:allMCfinalgap2}
\end{figure}

A comparison between the calibration results for \pythia~8.305, \herwig~7.2 and \sherpa~2.211 for all three shape variables for gap scheme~2 is shown in Fig.~\ref{fig:allMCfinalgap2}. The most interesting aspect of the calibration results for the top quark masses is that they are fully compatible among all three MCs. At the same time, the calibration results for $\Omega_1$, which we find to be $m_t^\mathrm{MC}$-independent, are around $250$\,MeV larger for \herwig and \sherpa compared to \pythia. This means the visible discrepancy in the position and the broadness of the peaks for all shape variables shown in Fig.~\ref{fig:MCdistributions} must be attributed to a difference in the modeling of the hadronization effects between \pythia and \herwig, while the conceptual meaning of their top quark mass parameters is within uncertainties (at N$^2$LL$+$NLO) equivalent. While there are general arguments that the exact field-theoretic meaning of $m_t^{\rm MC}$ depends on the parton shower implementation and is therefore different for coherent branching and dipole based parton-shower implementations~\cite{Hoang:2008xm,Hoang:2014oea,Hoang:2018zrp}, this important observation can be interpreted as evidence that these differences may be small numerically at least concerning the meaning of the top quark mass parameter. We emphasize, however, that such statements can be made strict only in the context of observables where all showers are NLL precise and under the assumption that the hadronization models do not interfere in an uncontrolled way. That latter aspect has so far not been investigated yet in the literature and remains an issue that has to be studied carefully.

\begin{figure}[t]
	\makebox[\textwidth]{
		\hspace*{-0.15in}
		\includegraphics[width=0.35\textwidth]{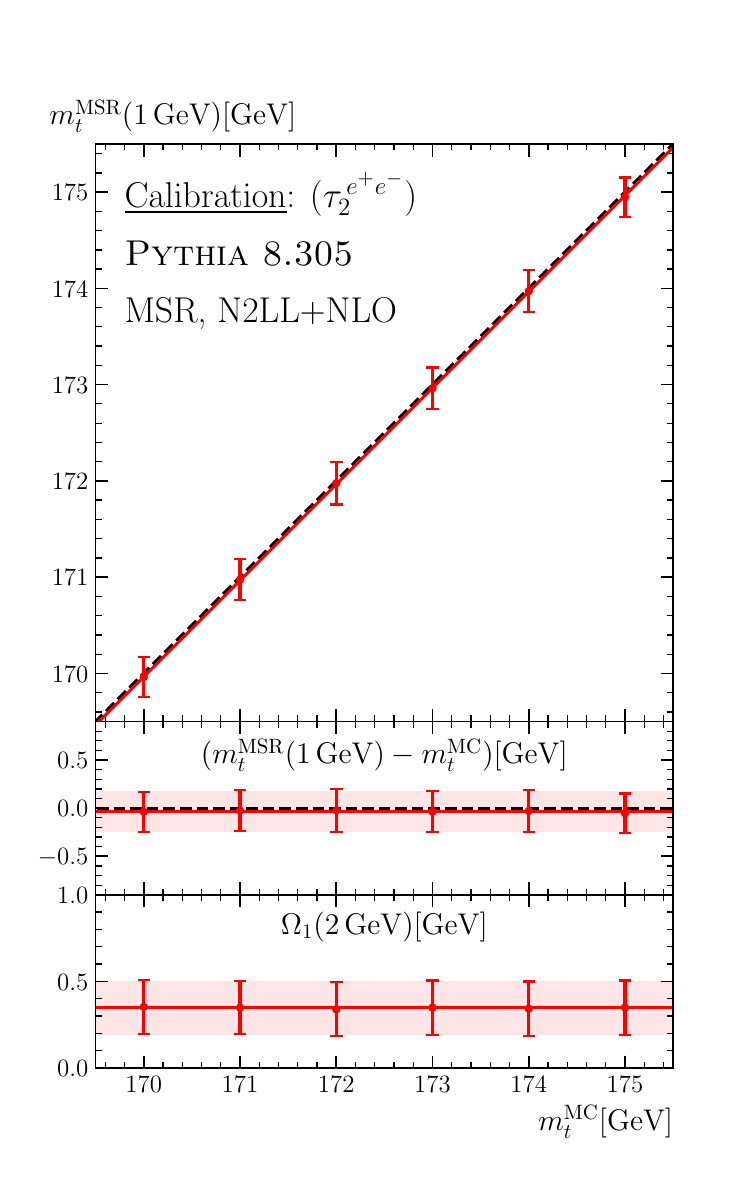}
		\hspace*{-0.3in}
		\includegraphics[width=0.35\textwidth]{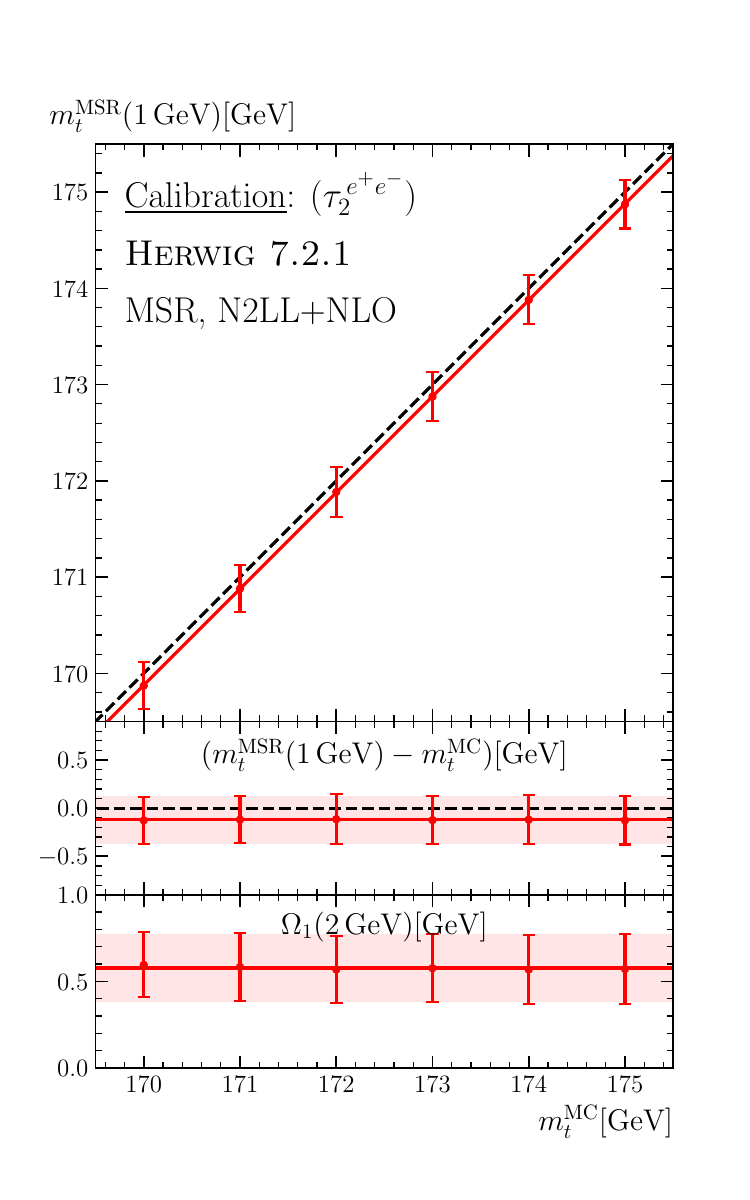}
		\hspace*{-0.3in}
		\includegraphics[width=0.35\textwidth]{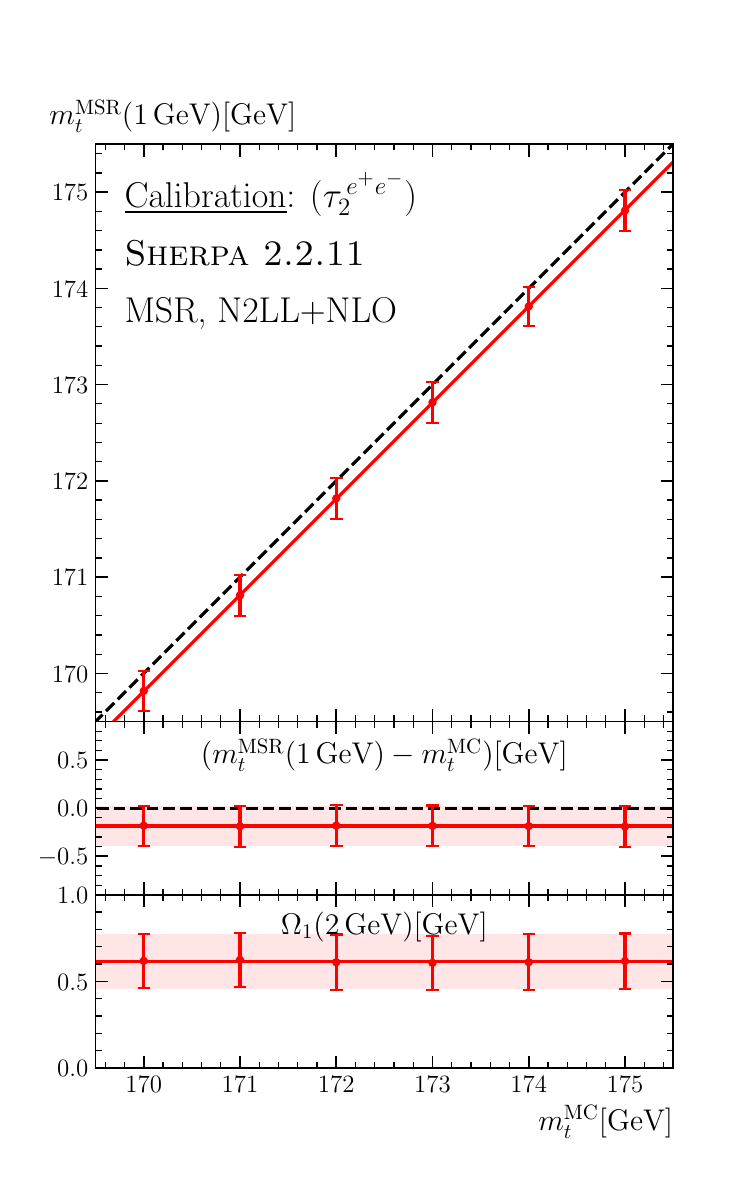}
	}
	\caption{Dependence of the fit result for the MSR mass $m_t^{\rm MSR}(R=1\,\mbox{GeV})$ and $\Omega_{1}(\SI{2}{\GeV})$ on the input $m_t^\mathrm{MC}$ for \pythia, \herwig and \sherpa. The fit results for $\Omega_1$ are $m_t^\mathrm{MC}$-independent. \label{fig:summaryallmtMC}}
\end{figure}

\begin{figure}[t]
	\makebox[\textwidth]{
		\hspace*{-0.15in}
		\includegraphics[width=0.35\textwidth]{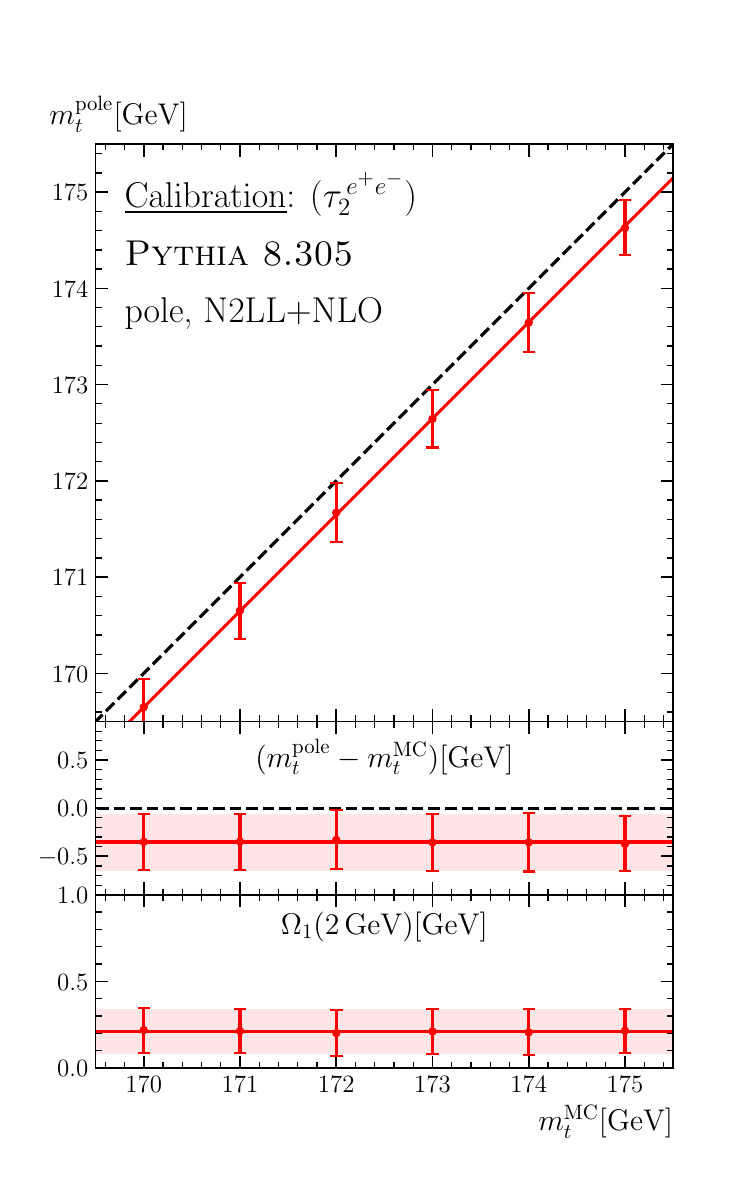}
		\hspace*{-0.3in}
		\includegraphics[width=0.35\textwidth]{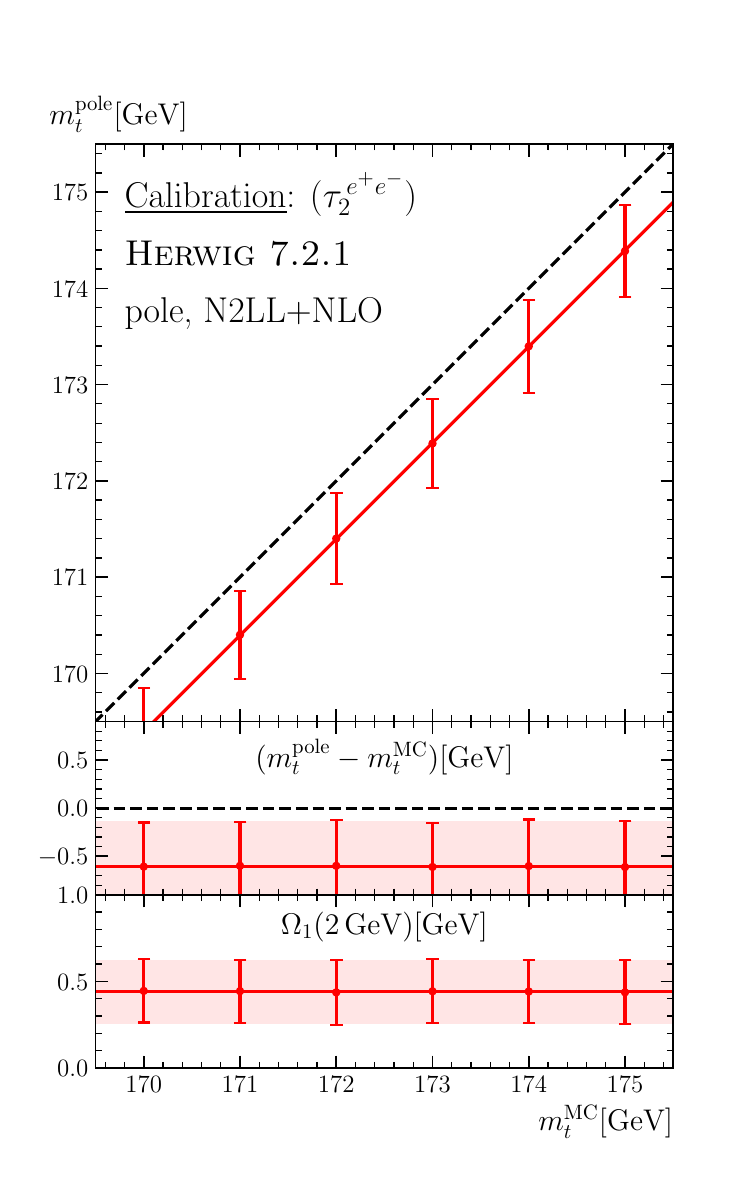}
		\hspace*{-0.3in}
		\includegraphics[width=0.35\textwidth]{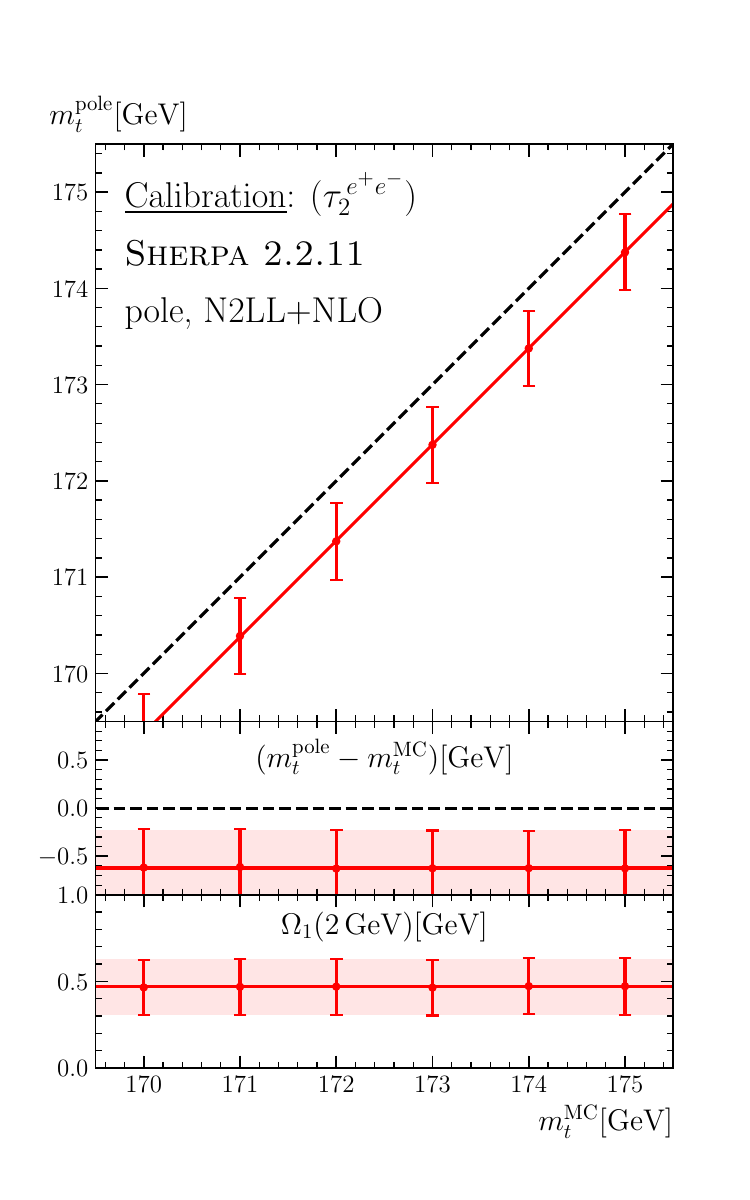}
	}
	\caption{Dependence of the fit result for the pole mass $m_t^{\rm pole}$ and $\Omega_{1}(\SI{2}{\GeV})$ on the input $m_t^\mathrm{MC}$ for \pythia, \herwig and \sherpa. \label{fig:summaryallmtMCpole}}
\end{figure}

\begin{figure}[t]
	\centering
	\begin{subfigure}{\textwidth}
		\includegraphics[width=1.01\textwidth]{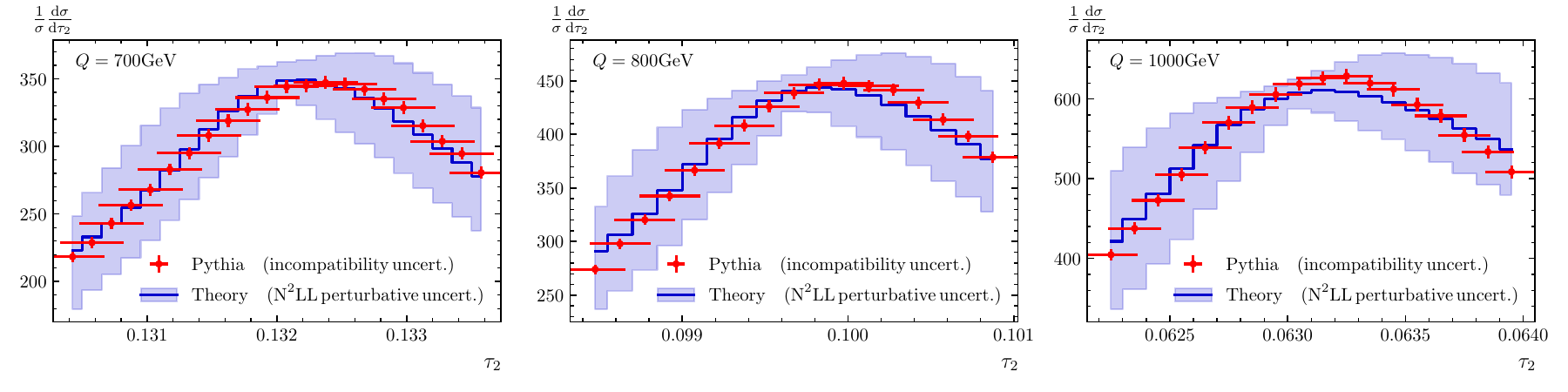}
		\vspace*{-0.40in}
	\end{subfigure}
	\begin{subfigure}{\textwidth}
		\includegraphics[width=\textwidth]{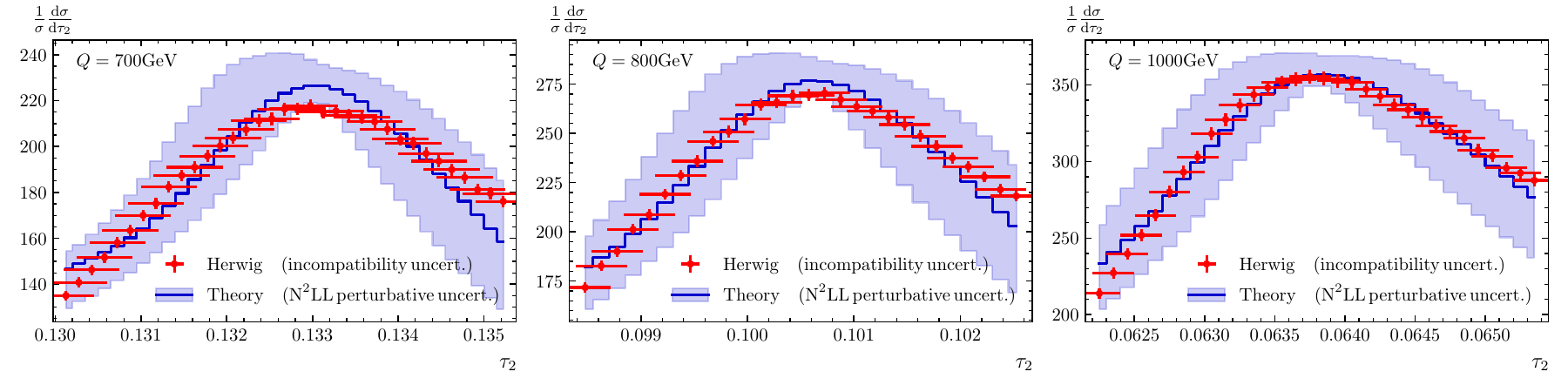}
		\vspace*{-0.40in}
	\end{subfigure}
	\begin{subfigure}{\textwidth}
		\includegraphics[width=\textwidth]{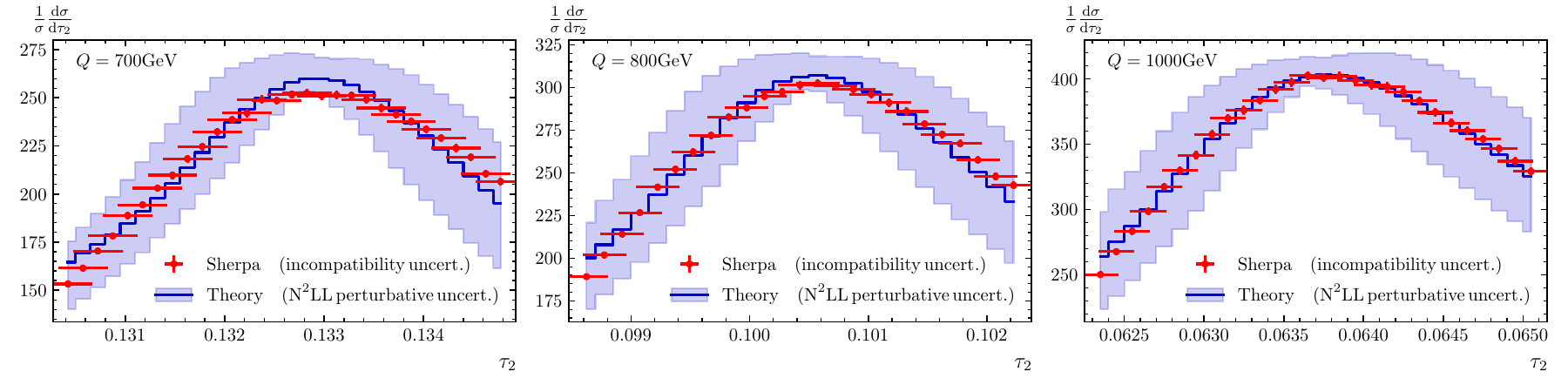}
	\end{subfigure}
	\caption{Comparison of the \pythia, \herwig and \sherpa 2-jettiness distributions (red dots) to the best-fit N$^2$LL$+$NLO theoretical predictions in the MSR scheme for gap scheme~2. The blue band shows the perturbative uncertainty from a random scan over 501 profile functions (and $\xi$ parameters), where the fit parameters are those from the $\chi^2$ analysis using the default profile (central blue line). Vertical error bars on the MC points (which are quite small) are obtained by a global rescaling of the \pythia statistical uncertainties such that the average $\chi_\mathrm{min}^2/\mathrm{dof}$ equals $1$. Horizontal error bars are related to the N\textsuperscript{2}LL incompatibility uncertainty of the fitted MSR mass. }
	\label{fig:summarybestfit2jettiness}
\end{figure}

\section{Conclusions}
\label{sec:conclusions}

In this article we have updated and generalized the Monte Carlo (MC) top quark mass calibration framework of Ref.~\cite{Butenschoen:2016lpz} that was based on the 2-jettiness distribution for boosted top pair production in $e^+e^-$ annihilation and applied to relate the \pythia~8.205 top quark mass parameter $m_t^{\rm MC}$ to top quark masses in unambiguously defined renormalization schemes. The calibration approach uses binned hadron-level distributions generated by the MC for a given  $m_t^{\rm MC}$ and N$^2$LL$+$NLO factorized and resummed hadron-level theory predictions with ${\cal O}(\Lambda_{\rm QCD})$ renormalon subtractions to obtain from $\chi^2$ fits the top quark mass in the MSR  $m_t^{\rm MSR}(R=1\,\mbox{GeV})$ and pole $m_t^{\rm pole}$ schemes together with the first moment $\Omega_1$ of the shape function describing the non-perturbative effects related to large-angle soft radiation. The results are relevant since the current most precise direct measurements determine $m_t^\mathrm{MC}$ of the MC simulations used for the experimental analyses.

We have generalized the original framework of Ref.~\cite{Butenschoen:2016lpz}, which is based on a bHQET factorization formula ---\,matched to SCET and full QCD\,--- in several ways: (i)~including two more shape variables, namely the sum of (squared) hemisphere jet masses $\tau_s$ and the newly designed modified jet mass $\tau_m$, and (ii)~accounting for two additional gap subtraction schemes that remove the ${\cal O}(\Lambda_{\rm QCD})$ renormalon effects coming from large-angle soft radiation. The treatment of different gap subtraction schemes requires a more general fit procedure for the parameters of shape function describing the hadronization corrections. The inclusion of two more shape observables revealed the importance of carefully treating $(m_t/Q)^2$ power corrections already in the singular bHQET factorization formula to achieve observable-independent results. Furthermore, we have updated the calibration framework to use standard file and event record formats, and presented all theoretical ingredients in great detail, which was missing in Ref.~\cite{Butenschoen:2016lpz} due to lack of space.

We applied the updated calibration framework to \pythia~8.305, \herwig~7.2 and \sherpa~2.2.11. For \pythia~8.305, the calibration results are fully consistent (within uncertainties) with the results of the original calibration  of Ref.~\cite{Butenschoen:2016lpz} based on \pythia~8.205 and exhibit an increase of the best fit $m_t^{\rm MSR}(R=1\,\mbox{GeV})$ and $m_t^{\rm pole}$ values of about $150$\,MeV. Using the calibration at N$^2$LL$+$NLO, the theoretical uncertainty in  \mbox{$m_t^{\rm MSR}(R=1\,\mbox{GeV})$} is around $200$\,MeV for all three generators, while for $m_t^{\rm pole}$ it is generator dependent and varies between $350$ and $600$\,MeV. The probably most instructive result of our analysis is that, even though \pythia~8.305, \herwig~7.2 and \sherpa~2.2.11 with their standard tunes produce resonance shape distributions that are visibly different as far as the peak position and shape are concerned, the interpretation of their top quark mass parameters $m_t^{\rm MC}$ agree with each other within $200$\,MeV. We find from the fit results for $\Omega_1$, which are $m_t^{\rm MC}$-independent, that the differences are associated to the different hadronization modeling used by the generators.  

While the calibration framework presented in this article provides concrete numerical relations between $m_t^{\rm MC}$ and the top quark mass in well-defined renormalization schemes, it is not capable of testing the physical aspects of the interplay between the parton-level description and the hadronization modeling contained in the MCs. These two components are usually blended together in state-of-the-art MCs within the tuning procedure where the shower cut is treated as a tuned parameter. The next important conceptual step towards a better understanding in the interpretation of the MC top quark mass parameter $m_t^{\rm MC}$ is to carefully study the hadronization models. This shall be addressed in future work. In this context, the calibration framework presented in this article will play an important numerical diagnostic tool.

\section{Acknowledgments}
\label{sec:acknowledgements}

We acknowledge support by the FWF Austrian Science Fund under the Project No.~P32383-N27 and under the FWF Doctoral Program ``Particles and Interactions'' No.~W1252-N27, the Spanish MECD Grants Nos.\ PID2019-105439GB-C22 and PID2022-141910NB-I00., the EU STRONG-2020 project under Program No.\ H2020-INFRAIA-2018-1, Grant Agreement No.\ 824093 and the COST Action No.\ CA16201 PARTICLEFACE. The work of B.D. has been partially funded by the Helmholtz Association Grant W2/W3-116 and the Deutsche Forschungsgemeinschaft (DFG, German Research Foundation) - 491245950. We thank Simon Pl\"atzer for collaboration in early stages of this project. We are grateful the Erwin-Schr\"odinger International Institute for Mathematics and Physics for partial support during the Programme ``Quantum Field Theory at the Frontiers of the Strong Interactions'', July 31 - September 1, 2023, where this article has been finalized.

\appendix

\section{Fixed-Order NLO QCD Results}
\label{app:FONLOQCD}

\subsection{Notation and Tree-Level Results}
The full QCD, NLO fixed-order calculation for the different event-shape distributions in $e^+e^-$ annihilation to a stable, massive quark-antiquark pair is required for the treatment of the $m_t^2/Q^2$ power corrections and to obtain the QCD non-singular contributions mandatory for a full N$^2$LL$+$NLO prediction. The result for a generic even shape variable $\tau$ can be written in the form \mbox{($\mhat_t=m_t/Q$)}
\begin{align}
\label{eq:QCDdistribution}
\frac{1}{\sigma_0^C} \dv{\sigma^C_\mathrm{QCD}}{\tau} ={}& R_0^C(\mhat_t) \biggl\{ \delta(\tau-\taumin) + \frac{C_F\alpha_s}{4 \pi}   A_\tau^C(\mhat_t) \delta(\tau-\taumin)  \\
&+  \frac{C_F\alpha_s}{4 \pi}  B_\mathrm{plus}(\mhat_t) \left[\frac{1}{\tau-\taumin}\right]_+ \biggr\} + \frac{C_F\alpha_s}{4 \pi} F_\tau^\mathrm{NS,C}(\tau, \mhat_t)  + \mathcal{O}\bigl(\alpha_s^2\bigr),\nonumber
\end{align}
where $C$ stands for either the vector (V) or axial-vector (A) current induced massive quark-antiquark production. The quark mass $m_t$ is defined in the pole renormalization scheme. The minimal (and tree-level) event shape values $\taumin$ are 
\begin{align}
\tau_{2,{\rm min}} =\,& 1 - v = 1-\sqrt{1-4\mhat_t^2} & \qquad \mbox{(2-jettiness)}\,,\\
\tau_{s,{\rm min}} =\,& 2\mhat_t^2  & \qquad \mbox{(jet mass sum, sJM)}\,, \nonumber\\
\tau_{m,{\rm min}} =\,& 2\mhat_t^2+2\mhat_t^4 & \qquad \mbox{(modified jet mass, mJM)}\,.\nonumber
\end{align}

All distributive terms for $\tau\to\taumin$ are encoded in the coefficients $A_\tau^C(\mhat_t)$ and $B_\mathrm{plus}(\mhat_t)$ such that the functions $F_\tau^\mathrm{NS,C}(\tau, \mhat_t)$ are non-singular, which means that they are integrable at $\tau=\taumin$. The terms $\sigma_0^C$ stand for the Born cross section for massless quark production:
\begin{align}
\label{eq:bornXSection}
\sigma^V_0={}&\frac{N_c}{3}\frac{4\pi\alphaem^2}{Q^2}\left[Q^2_q+\frac{v^2_q(v^2_e+a_e^2)}{(1-\mhat^2_Z)^2+\Bigl(\frac{\Gamma_Z}{m_Z}\Bigr)^{\!2}}-\frac{2Q_qv_ev_q(1-\mhat^2_Z)}{(1-\mhat_Z^2)^2+\Bigl(\frac{\Gamma_Z}{m_Z}\Bigr)^{\!2}}\right]\!,\\
\sigma^A_0={}&\frac{N_c}{3}\frac{4\pi\alphaem^2}{Q^2}\left[\frac{a^2_q(v^2_e+a_e^2)}{(1-\mhat^2_Z)^2+\Bigl(\frac{\Gamma_Z}{m_Z}\Bigr)^{\!2}}\right]\!,\nonumber
\end{align}
with $N_c$ the number of colors, $\alphaem$ the electromagnetic coupling, $Q_q$ the quark electric charge, $\hat{m}_Z=m_Z/Q$ the reduced $Z$-boson mass, $\Gamma_Z$ the finite width of the $Z$ boson, and $v_i=(T_3^i-2Q_i\sin^2{\theta_W})/\sin(2\theta_W)$ and $a_i=T_3^i/\sin(2\theta_W)$ the vector and axial-vector couplings of the electron or quark to the $Z$ boson, respectively. The coefficients $R_0^C(\mhat_t)$ show the quark mass dependence of the tree-level total cross section and read
\begin{align}
R_0^V(\mhat_t)={}&\frac{(3-v^2)v}{2}={(1+2\mhat_t^2)v=1-6\mhat_t^4 + \mathcal{O}(\mhat_t^6)},\\
R^A_0(\mhat_t)={}&v^3=(1-4\mhat_t^2)v=1-6\mhat_t^2 + \mathcal{O}(\mhat_t^4)\,.\nonumber
\end{align}

\subsection{NLO Results}

A full generic analytic method to determine the NLO fixed-order corrections to massive quark event-shape distribution was developed in Ref.~\cite{Lepenik:2019jjk} and earlier calculations were already provided in Refs.~\cite{Dehnadi:2016snl,Preisser:2018yfv}. Here we use the notation of Ref.~\cite{Lepenik:2019jjk} to write down the results, where also the ingredients needed for the computation can be found.

The NLO delta function coefficients read
\begin{align}
\label{eq:sigmaNLO}
A_\tau^V&(\mhat_t) = \frac{4}{v}  \biggl\{ (1 - 2\mhat_t^2) \biggl[ \li2\biggl( -\frac{v(1+v)}{2\mhat_t^2} \biggr) - 3 \li2\biggl( \frac{v(1 - v)}{2 \mhat_t^2} \biggr) + 2 \ln^2(\mhat_t) + \pi^2 \\
& -  2\ln^2\biggl( \frac{ 1 + v }{2} \biggr) \biggr] + 2v [ \ln(\mhat_t) - 1 ] - 2I_\tau(\mhat_t)   \biggr\}  + \frac{4}{(1 + 2\mhat_t^2)v} ( 4 + v^2 - 16\mhat_t^4) L_v \nonumber\\
A_\tau^A&(\mhat_t) = \frac{4}{v} \biggl\{  (4 + v^2) L_v + 2v [ \ln(\mhat_t) -1 ] - 2I_\tau(\mhat_t) + ( 1 - 2\mhat_t^2 ) \nonumber\\
& \times  \biggl[ \li2\biggl( - \frac{ v( 1 - v ) }{ 2\mhat_t^2 } \biggr) - 3\li2\biggl( \frac{ v ( 1 - v )}{ 2\mhat_t^2 } \biggr) + \pi^2 + 2 \ln^2(\mhat_t) - 2 \ln^2\biggl( \frac{ 1 + v }{2} \biggr) \biggr] \biggr\},\nonumber
\end{align}
with 
\begin{equation}
L_v\equiv \ln\biggl(\frac{1+v}{2\mhat_t}\biggr)\,,
\end{equation}
and the only event-shape dependent contribution is encoded in the term $I_\tau(\mhat_t)$. For sJM it has the form
\begin{align}
I_{\tau_s}(\mhat_t) =&{} \frac{1}{24} \biggl\{ \!\pi^2(v^2 + 1)-12 (v^2 + 1) \!\li2\!\biggl(\frac{v+1}{2}\biggr) 
+ 6 \ln(1 - v)\! \biggl[(v^2 + 1) \!\ln\biggl(\frac{4}{1 - v}\biggr) - 4v\biggr]\nonumber\\
&  - 6\{v[v(2+\ln^2 2) + 2 - 4 \ln 2] + \ln^2 2\} %\nonumber\\ &
   + 6 (v^2 - 1) \ln\biggl(\frac{1 - v}{1 + v}\biggr) \!\biggr\} , %\nonumber
\end{align}
and agrees with the case of the heavy jet mass distribution already given in Ref.~\cite{Lepenik:2019jjk}.
The results for 2-jettiness and mJM read
\begin{align}
I_{\tau_2}(\mhat_t) &\,= I_{\tau_s}(\mhat_t) - \ln(v) [ (1 + v^2) L_v - v]\,,\\
I_{\tau_m}(\mhat_t) &\,= I_{\tau_s}(\mhat_t) + \ln(1+2\mhat_t^2)[ (1 + v^2) L_v - v]\,.\nonumber
\end{align}
The coefficient of the plus distribution is universal for any event shape distribution and whether we consider vector or axial-vector current induced quark pair production: 
\begin{equation}
B_\mathrm{plus}(\mhat_t)=\frac{8}{v}\,[(1+v^2)L_v-v]=-8[1+2\ln(\mhat_t)]
	-16\mhat_t^2+\mathcal{O}\bigl(\mhat_t^4\bigr).
\end{equation}
This fact motivates factoring out the tree-level mass correction terms $R_0^C(\mhat_t)$ in Eq.~(\ref{eq:QCDdistribution}). Our treatment of power corrections concerning the overall factor $R_0^C(\mhat_t)$ in Sec.~\ref{sec:absorbconcept} is based on the assumption that this universality is not accidental and also valid for the singular QCD corrections beyond NLO, which are assumed to be event-shape independent as well.  

The integrable functions $F^{\mathrm{NS},C}_\tau(\tau, \mhat_t)$ can be obtained computing the quark-antiquark plus gluon phase space for a given event shape value $\tau>\taumin$ in four dimensions, see Eq.~(4.16) of Ref.~\cite{Lepenik:2019jjk}. The result for the full distribution for $\tau>\taumin$, which is referred to as $F^C_\tau(\tau, \mhat_t)$, receives contributions where either only the quark, only the antiquark or only the gluon are populating one of the two hemispheres. We call these phase space regions quark (qu), antiquark and gluon (gl) regions, and we find that the quark and antiquark region results are identical. The result for the $F^C_\tau(\tau, \mhat_t)$ can then be written in the form 
\begin{equation}
F^C_\tau(\tau, \mhat_t) ={} F^C_{\tau,\quark}(\tau,\mhat_t) + F^C_{\tau,\gluon}(\tau, \mhat_t)\,.
\end{equation}
The expressions for the integrable functions  $F^{\mathrm{NS},C}_\tau(\tau, \mhat_t)$ are then obtained by subtracting the singular contributions proportional to $1/(\tau-\taumin)$ shown in
 Eq.~(\ref{eq:QCDdistribution}):
\begin{equation}
\label{eq:fns}
F^{\mathrm{NS},C}_\tau(\tau, \mhat_t) = F_\tau^C(\tau, \mhat_t) -   \frac{R_0^C(\mhat_t) B_\mathrm{plus}(\mhat_t) }{ \tau - \taumin } \,.
\end{equation}
For 2-jettiness the results read~\cite{Dehnadi:2016snl}  
\begin{align}
F^V_{\tau_2,\quark}&(\tau,\mhat_t)=\frac{2 t_\tau }{(\zq -1) \zq ^2 (\xi -t_\tau^2-4 \mhat_t^2)}\{4 (\zq -1) \zq ^2 \tanh ^{-1}(1-2 \zq )\\
&\times(4 \mhat_t^2 \xi -8 \mhat_t^4+(\tau -2) \tau +2)-(2 \zq -1) [4 \mhat_t^2 (2 \zq  ((\xi -1) \zq +2-\xi)-1)\nonumber\\
&+8 \mhat_t^4 \zq +(\zq -1) ((\tau -2) \tau +4 (\xi -1) \zq +2-2 \xi)]\}\,,\nonumber\\
F^A_{\tau_2,\quark}&(\tau,\mhat_t)=\frac{2 t_\tau}{(\zq -1) \zq ^2 (\xi -t_\tau^2-4 \mhat_t^2)}\{(2 \zq -1) [2 \mhat_t^2 ((\tau -2) \tau \nonumber\\
&+4 \zq ^2 (4 \xi -(\tau -2) \tau -4)+\zq  (3 (\tau -2) \tau +10-14 \xi )+4-2 \xi)\nonumber\\
&+8 \mhat_t^4 (\zq  (5-4 \zq )+1)-(\zq -1) ((\tau -2) \tau +4 (\xi -1) \zq +2-2 \xi)]\nonumber\\
&+4 (\zq -1) \zq ^2 \tanh ^{-1}(1-2 \zq ) [2 \mhat_t^2 ((\tau -2) \tau +2-6 \xi )+24 \mhat_t^4+(\tau -2) \tau +2]\}\,,\nonumber\\
F^V_{\tau_2,\gluon}&(\tau,\mhat_t)=\frac{4}{t_\tau}\biggl\{[2-(2-t_\tau)t_\tau-4\mhat_t^2 t_\tau-8\mhat_t^4]\ln\biggl(\frac{1}{\zg }-1\biggr)\nonumber\\
&-\frac{(1-2\zg )[(1-\zg )\zg t_\tau^2+2\mhat_t^2+4\mhat_t^4]}{(1-\zg )\zg } \biggr\},\nonumber\\
F^A_{\tau_2,\gluon}&(\tau,\mhat_t)=\frac{4}{t_\tau}\biggl\{[2-2t_\tau+t_\tau^2+2\mhat_t^2(t_\tau^2+4t_\tau-6)+16\mhat_t^4]\ln\biggl(\frac{1}{\zg }-1\biggr)\nonumber\\
&-\frac{(1-2\zg )[t_\tau^2(1-\zg )\zg +2\mhat_t^2-8\mhat_t^4]}{(1-\zg )\zg }\biggr\},\nonumber
\end{align}
with $t_\tau\equiv1-\tau$, $\xi\equiv\sqrt{t_\tau^2+4\mhat_t^2}$, $r\equiv\sqrt{1 - 3 \mhat_t^2}$ and \\
\begin{align}
\zq={}&\begin{cases}
(1+\tau-\xi)/2 &  \qquad \quad \qquad \quad \qquad \qquad\;\;\,1-v<\tau\le\mhat_t/(1-\mhat_t)\\
(1-\xi)/t_\tau &  \qquad \quad \qquad \quad \quad \;\;\mhat_t/(1-\mhat_t)\le\tau \le(5-4r)/3
\end{cases}\,,
\\
\zg={}&\begin{cases}
[1-\sqrt{1-4\mhat_t^2/\tau}]/2 & \qquad \quad\;\; 4\mhat_t^2<\tau\le\mhat_t/(1-\mhat_t)\nonumber\\
[1-\sqrt{(1-\tau)^2+4\mhat_t^2}]/(1-\tau) &\mhat_t/(1-\mhat_t)\le\tau\le(5-4r)/3
\end{cases}\,,
\end{align}
Note that $\tau_{2,{\rm max}}=(5-4r)/3$ is the maximal 2-jettiness value at NLO.
We also mention that the $F^C_{\tau_2,\gluon}(\tau,\mhat_t)$ coincide with those of the heavy jet mass (HJM) $\rho$ distribution that have been already calculated in Ref.~\cite{Lepenik:2019jjk}, i.e.\ $F^C_{\tau_2,\gluon}(\tau,\mhat_t)=F^C_{\rho,\gluon}(\tau,\mhat_t)$.

The sJM and mJM expressions for $F^{C}_{\tau_s}(\tau_s, \mhat_t)$ and $F^{C}_{\tau_m}(\tau_m, \mhat_t)$  can be written in terms of the heavy jet mass (hJM) result $F^{C}_{\rho, \quark}(\rho, \mhat_t)$ since at NLO they are related by simple bijective (quark- and gluon-region dependent) mappings. 
For sJM the result reads
\begin{align}
		F^C_{\tau_{s}, \quark}(\tau_s, \mhat_t)=
		 F^C_{\rho, \quark}(\rho=\tau_s-\mhat_t^2, \mhat_t)\,,\qquad
		F^C_{\tau_{s}, \gluon}(\tau_s, \mhat_t)=
		F^{C}_{\rho, \gluon}(\tau=\tau_s, \mhat_t)\,,
\end{align}
where the hJM results take the following form~\cite{Lepenik:2019jjk}:
\begin{align}\label{eq:FHJM}
		F^V_{\rho,\quark}(\rho,\mhat_t)={}&\frac{(2-4z)[\rho(1-z)(\rho-4z)-2\mhat_t^2(\rho+2(1-2\rho)z^2+3\rho z)+\mhat_t^4(1-z-8z^2)]}{(1-z)z^2(\rho-\mhat_t^2)}\nonumber\\
		&+4\biggl(\rho-2-5\mhat_t^2+2\frac{1-4\mhat_t^4}{\rho-\mhat_t^2}\biggr)\ln(\frac{1-z}{z})\,,\nonumber\\
		F^A_{\rho,\quark}(\rho,\mhat_t)={}&4[4-8\mhat_t^2(2+\rho)+8\mhat_t^4]+\frac{2}{\rho-\mhat_t^2}
		\biggl\{\frac{(1+2\mhat_t^2)(\rho-\mhat_t^2)^2}{z^2}+\frac{4(1-4\mhat_t^2)\mhat_t^2}{1-z}\nonumber\\
		&+2[2-(2-\rho)\rho+2\mhat_t^2(\rho(\rho+3)-5)+\mhat_t^4(9-4\rho)+2\mhat_t^6]\ln\biggl(\frac{1-z}{z}\biggr)\nonumber\\
				&-2\frac{\rho(2+\rho)-2\mhat_t^2\rho(5+\rho)+\mhat_t^4(1+4\rho)-2\mhat_t^6}{z}\biggr\}\,,
\end{align}
with $t_\rho\equiv1+\rho-\mhat_t^2$, $\xi_\rho\equiv\sqrt{t_\rho^2-4\rho}$, $r\equiv\sqrt{1 - 3 \mhat_t^2}$ and
\begin{align}
		z={}&\begin{cases}
			(t_\rho-\xi_\rho)/2 & \qquad\qquad\,\mhat_t^2<\rho\le\frac{\mhat_t(1-\mhat_t-\mhat_t^2)}{1-\mhat_t}\\
			(t_\rho-1)/\sqrt{(1-\rho)^2-2\mhat_t^2(1+\rho)+\mhat_t^4} & \frac{\mhat_t(1-\mhat_t-\mhat_t^2)}{1-\mhat_t}\le\rho\le\frac{2r-1}{3}+\mhat_t^2
		\end{cases}\,.
\end{align}
At NLO we have $\tau_{s,{\rm max}}=\tau_{s,{\rm gl, max}}=(5-4r)/3$ which agrees with the maximal 2-jettiness value and $\tau_{s,{\rm qu, max}}=(2r-1)/3+2\mhat_t^2$.

For mJM, since $\tau_m = \tau_s + \tau_s^2/2$, the results involve an additional Jacobian factor \mbox{$\dv{\tau_s}{\tau_m}=(1+2\tau_m)^{-1/2}$}, so that the results read
\begin{align}
F^C_{\tau_{m}, \quark}(\tau_m, \mhat_t)={}& (1+2\tau_m)^{-\frac{1}{2}}F^C_{\rho, \quark}(\rho=\sqrt{1+2\tau_m}-1-\mhat_t^2, \mhat_t)\,,\\
F^C_{\tau_{m}, \gluon}(\tau_m, \mhat_t)={}& (1+2\tau_m)^{-\frac{1}{2}} F^{C}_{\tau, \gluon}(\tau=\sqrt{1+2\tau_m}-1, \mhat_t)\,.\nonumber
\end{align}

\section{Evolutions}
\label{eq:evolutions}
\subsection{Evolution Factors and Anomalous Dimensions}\label{app:evolfact}
We follow the notation and convention of Ref.~\cite{Bachu:2020nqn}, except that our $\Gamma^\mathrm{cusp}$ has $C_F$ absorbed.
The RGE evolution factors read~\cite{Fleming:2007xt}
\begin{align}
	\label{eq:evolfact}
U_{H_Q}(Q,\mu_0,\mu_1)={}&e^{K_{H_Q}}\biggl(\frac{\mu_0}{Q}\biggr)^{\!\omega_{H_Q}}\,,\\
U_{v}(\rho,\mu_1,\mu_0)={}&e^{K_v}\varrho^{-\omega_v}\,,\nonumber\\
U_F(t,\mu_1,\mu_0)={}&\frac{e^{K_F}(e^{\gamma_E})^{\omega_F}}{\Gamma(-\omega_F)}\plusFunc{0,\omega_F}{\mu_0}{t}\,,\nonumber
\end{align}
where $F\in\{B_\tau,S_\tau\}$ and $\mathcal{L}_{0,\omega_F}^{\mu}$ is the fractional plus distribution defined in App.~\ref{app:distributions}. They are the solutions to the renormalization group equations
\begin{align}\label{eq:factrge}
\mu\dv{\mu}H_Q(Q,\mu)={}&\biggl[\Gamma_H[\alpha_s]\ln\biggl(\frac{\mu}{Q}\biggr)+\gamma_{H_Q}[\alpha_s]\biggr]H_Q(Q,\mu)\,,\\
\mu\dv{\mu}\mathcal{J}_\nu(\tau,\mu)={}&\left(\Gamma_v[\alpha_s]\ln({\varrho^{-1}})+\gamma_v[\alpha_s]\right)\mathcal{J}_\nu(\tau,\mu)\,,\nonumber\\
\mu\dv{\mu}F(t,\mu)={}&\int_{-\infty}^{+\infty}\dd{t'}\gamma_F(t-t',\mu)F(t',\mu)\,,\nonumber\\
\gamma_F(t-t',\mu)={}&-\!\Gamma_F[\alpha_s]\plusFunc{0}{\mu_F}{t-t'}+\gamma_F[\alpha_s]\delta(t-t')\,,\nonumber
\end{align}
with $\mathcal{J}_\nu$ defined as the squared bHQET current $B_\tau\otimes S_\tau$, such that the running of this combination is not a convolution anymore. The terms $\Gamma_F$ and $\gamma_F$ are the cusp and non-cusp anomalous dimensions, respectively. Note, that the implementation of power corrections in the measurement function discussed in Sec.~\ref{sec:absorbconcept} rescales the boost factor between the soft and ultra-collinear momenta. As a consequence, the instances of $\varrho$ in the formulae above have to be replaced by $r_s \varrho$ to obtain a consistent running, that is independent of the starting scale of $U_\nu$.

The evolution kernels are given by [$\alpha_i\equiv\alpha_s(\mu_i)$]:
\begin{align}\label{eq:evolKernel}
\omega(\Gamma;\mu_1,\mu_0)={}&\int_{\alpha_0}^{\alpha_1}\frac{\dd{\alpha}}{\beta(\alpha)}\Gamma[\alpha]\,,\\
K(\Gamma,\gamma,j;\mu_1,\mu_0)-\omega\left(\gamma;\mu_1,\mu_0\right)={}&j\!\int_{\alpha_0}^{\alpha_1}
\frac{\dd{\alpha}}{\beta(\alpha)}\Gamma[\alpha]\int_{\alpha_0}^{\alpha}\frac{\dd{\alpha'}}{\beta(\alpha')}\,,\nonumber
\end{align}
where $j$ is the mass dimension of the variable in the logarithm of the cusp piece. Given our notation in Eq.~(\ref{eq:factrge}) all $j=1$, except for $\mathcal{J}_\nu$ for which $j=0$. The results at \nnll read~\cite{Abbate:2010xh}
\begin{align}
\omega&^\mathrm{N\textsuperscript{3}LL}(\Gamma;\mu_1,\mu_0)=-\frac{\Gamma_0}{2\beta_0}\biggl\{\ln{r}+\frac{\alpha_0}{4\pi}
\biggl(\frac{\Gamma_1}{\Gamma_0}-\frac{\beta_1}{\beta_0}\biggr)(r-1)\\
&+\frac{1}{2}\biggl(\frac{\alpha_0}{4\pi}\biggr)^2\biggl(\frac{\beta_1^2}{\beta_0^2}-\frac{\beta_2}{\beta_0}+\frac{\Gamma_2}{\Gamma_0}-\frac{\Gamma_1\beta_1}{\Gamma_0\beta_0}\biggr)(r^2-1)\nonumber\\
&+\frac{1}{3}\biggl(\frac{\alpha_0}{4\pi}\biggr)^3\biggl[\frac{\Gamma_3}{\Gamma_0}-\frac{\beta_3}{\beta_0}+\frac{\Gamma_1}{\Gamma_0}\biggl(\frac{\beta_1^2}{\beta_0^2}-\frac{\beta_2}{\beta_0}\biggr)-\frac{\beta_1}{\beta_0}\biggl(\frac{\beta_1^2}{\beta_0^2}-2\frac{\beta_2}{\beta_0}+\frac{\Gamma_2}{\Gamma_0}\biggr)\biggr](r^3-1)\biggr\},\nonumber\\
K&^{\text{\nnll}}(\Gamma,\gamma,j;\mu_1,\mu_0)=\frac{j\Gamma_0}{4\beta_0^2}\biggl\{\frac{4\pi}{r\alpha_0}(r\ln{r}+1-r)+\biggl(\frac{\Gamma_1}{\Gamma_0}-\frac{\beta_1}{\beta_0}\biggr)(r-1-\ln{r}) \nonumber \\
&-\frac{\beta_1}{2\beta_0}\ln^2{r}+\frac{\alpha_0}{4\pi}\biggl[\biggl(\frac{\Gamma_1\beta_1}{\Gamma_0\beta_0}-\frac{\beta_1^2}{\beta_0^2}\biggr)(r-1-r\ln{r})-B_2\ln{r}\nonumber\\
&+\biggl(\frac{\Gamma_2}{\Gamma_0}-\frac{\Gamma_1\beta_1}{\Gamma_0\beta_0}+B_2\biggr)\frac{r^2-1}{2}+\biggl(\frac{\Gamma_1\beta_1}{\Gamma_0\beta_0}-\frac{\Gamma_2}{\Gamma_0}\biggr)(r-1)\biggr]\biggr\}+\omega^{\mathrm{NLL}}(\gamma;\mu_1,\mu_0)\,,\nonumber
\end{align}
where $r=\alpha_1/\alpha_0$ depends on the 4-loop running coupling and $B_2=\beta_1^2/\beta_0^2-\beta_2/\beta_0$. 
The QCD beta function and the cusp and non-cusp anomalous dimensions are given by the series
\begin{equation}
\dv{\alpha_s(\mu)}{\ln{\mu}}=\beta[\alpha_s]=-2\alpha_s\sum_{n=0}^{\infty}\beta_n\left(\frac{\alpha_s}{4\pi}\right)^{n+1},\qquad\Gamma[\alpha_s] = \sum_{n=0}^\infty \Gamma_n\left(\frac{\alpha_s}{4\pi}\right)^{n+1},
\end{equation}
where $\Gamma$ stands for either $\Gamma_F$, $\gamma_F$ or the QCD cusp anomalous dimension $\Gamma^\mathrm{cusp}$. 
The cusp anomalous dimensions are proportional to $\Gamma^\mathrm{cusp}$~\cite{Fleming:2007xt,Hoang:2015vua,Moch:2004pa,Moch:2005id,Jain:2008gb,Becher:2006mr}
\begin{align}
\Gamma_{B_\tau}[\alpha_s]={}&\Gamma_\nu[\alpha_s]=-\Gamma_{S_\tau}[\alpha_s]=4\Gamma^{\mathrm{cusp}(5)}[\alpha_s]\,,\\
\Gamma_{H_Q}[\alpha_s]={}&-4\Gamma^{\mathrm{cusp}(6)}[\alpha_s]\,,\nonumber
\end{align}
with the universal cusp anomalous dimension coefficients given by
\begin{align}
\label{eq:cuspAnomdim}
\Gamma^\mathrm{cusp}_0={}&\frac{16}{3}\,,\qquad\Gamma_1^\mathrm{cusp}=\frac{1072}{9}-\frac{16}{3}\pi^2-\frac{160}{27}n_f\,,\\
\Gamma_2^\mathrm{cusp}={}&1960-\frac{2144}{9}\pi^2+\frac{176}{15}\pi^4+352\zeta_3+
\biggl(\frac{320}{27}\pi^2-\frac{5104}{27}-\frac{832}{9}\zeta_3\biggr)n_f-\frac{64}{81}n_f^2\,.\nonumber
\end{align}
Consistency in the running gives the relation
\begin{equation}
\gamma_\nu[\alpha_s]=\gamma_{B_\tau}[\alpha_s]+\gamma_{S_\tau}[\alpha_s]\,,
\end{equation}
and the expressions for the non-cusp anomalous dimensions read 
\begin{align}
\gamma^{H_Q}_0&=-16\,,& \gamma^{H_Q}_1&=-\frac{7976}{27}-\frac{136}{9}\pi^2+\frac{736}{3}\zeta_3+\biggl(\frac{1040}{81}+\frac{16}{9}\pi^2\biggr)n_f,\\
\gamma^{B_\tau}_0&=\frac{32}{3},&\gamma^{B_\tau}_1&=\frac{11168}{27}-\frac{184}{9}\pi^2-160\zeta_3+
\biggl(-\frac{1856}{81}+\frac{16}{27}\pi^2\biggr)n_f\,,\nonumber\\
\gamma^{S_\tau}_0&=0\,,&\gamma^{S_\tau}_1&=-\frac{6464}{27}+\frac{88}{9}\pi^2+224\zeta_3+
\biggl(\frac{896}{81}-\frac{16}{27}\pi^2\biggr)n_f\,\nonumber.
\end{align}
The beta function coefficients are given by~\cite{Chetyrkin:2000yt}
\begin{align}
	\label{eq:betacoeff}
\beta_0={}&11-\frac{2}{3}n_f,\qquad \beta_1=102-\frac{38}{3}n_f,\qquad\beta_2=\frac{2857}{2}-\frac{5033}{18}n_f+\frac{325}{54}n_f^2\,,\\
\beta_3={}&\frac{149753}{6}+3564\,\zeta_3-\biggl(\frac{1078361}{162}+\frac{6508}{27}\zeta_3\biggr)n_f+\!\biggl(\frac{50065}{162}+\frac{6472}{81}\zeta_3\biggr)n_f^2+\frac{1093}{729}n_f^3\,.\nonumber
\end{align}
Note that the flavor number $n_f$ is either 5 or 6 depending on whether the quantity refers to scales above or below the top quark mass.

\subsection[{$R$-evolution}]{$\mathbf R$-evolution}
\label{app:Revo}

The MSR mass and soft gap parameter $R$-RGEs can be determined from the fact that $m_t^\mathrm{pole}=m_t^\mathrm{MSR}(R)+\delta m_t(R)$ and $\Delta=\overline{\Delta}(R,R)+\bar{\delta}(R,R)$ are scale independent. Given a perturbative series of the form
\begin{equation}
f(R)=\text{const.}-R\sum_{n=1}^\infty\alphapi{R}^n f_n\,,
\end{equation}
the $R$-RGE can be written as
\begin{equation}
\dv{f(R)}{\ln{R}}=R\dv{f(R)}{R}=-R\sum_{n=0}^{\infty}\gamma_n^{f,R}\alphapi{R}^{n+1}\,,
\end{equation}
with the following anomalous dimension coefficients
\begin{align}
\gamma_0^{f,R}={}&f_1\,, \\
\gamma_n^{f,R}={}&f_{n+1}-2\sum_{j=0}^{n-1}(n-j)\beta_jf_{n-j} \qquad (n\ge1)\,.\nonumber
\end{align}
The solution for the evolution is therefore
\begin{equation}
f(R_1)-f(R_0)=-\sum_{n=0}^\infty\gamma_n^{f,R}\int_{R_0}^{R_1}\dd{R}\alphapi{R}^{n+1}.
\end{equation}
For $f(R)=m_t^\mathrm{MSR}(R)$ we have $f_n=a_n^{\mathrm{MSR}}(n_\ell)$ as defined in Eq.~(\ref{eq:pMSR}) and given by~\cite{Hoang:2017suc}
\begin{align}
a_1^{\mathrm{MSR}}(n_\ell)={} & 5.33333\,,  \\
\quad a_2^{\mathrm{MSR}}(n_\ell)={} & 213.437 - 16.6619 \,n_\ell\,,\nonumber \\
a_3^{\mathrm{MSR}}(n_\ell)={} &  12075. -1707.35 \,n_\ell + 41.7722 \,n_\ell^2\,,\nonumber
\end{align}
where we exclusively have $n_\ell=5$.
For $f(R)=\overline{\Delta}(R,R)$ we use $f_n= d_i(R,R)$ as defined in Eq.~(\ref{eq:gapdef}). These in turn depend on the coefficients in the exponent of the position-space soft function as defined in Eq.~(\ref{eq:softFourier}), which can be generated by~\cite{Bachu:2020nqn}
\begin{equation}\label{eq:soft_coeff}
s_{mn}=s^{[0]}_{mn}[\beta]+s^{[1]}_{mn}[\beta] + s^{[2]}_{mn}[\beta]\,,
\end{equation}
where each term follows a recursion relation for $m>1$ and $1\le n-k\le m-1$
\begin{equation}
s^{[k]}_{mn}[\beta]=\frac{2}{n}\sum_{i=n-k}^{m-1}i\, s_{i(n-1)}[\beta]\beta_{m-i-1}\,,
\end{equation}
and the starting values (with $m\ge 1$) read
\begin{equation}
s^{[0]}_{m0}[\beta]= s_{m0}\,,\qquad s^{[1]}_{m1}[\beta]= \gamma^{S_\tau}_{m-1}\,,\qquad s^{[2]}_{m2}[\beta]= \frac{1}{2} \Gamma^{S_\tau}_{m-1}\,.
\end{equation}
The anomalous dimensions are listed in the previous section and the relevant non-logarithmic terms are given by~\cite{Fleming:2007xt,Monni:2011gb} 
\begin{align}
s_{10} =
13.1595 \,,\qquad
s_{20}=
 -225.996 + 28.9270 \,n_\ell\,.
\end{align}
The gap subtraction series coefficients as defined in Eq.~(\ref{eq:gapdef}) for gap 1 are then given by
\begin{align}
d_1^{(1)}( R_s,\mu_S)&=-18.9981L_R\,,\\
d_2^{(1)}( R_s,\mu_S)&=-43.9543 -131.242L_R -145.652L_R^2\,,\nonumber
\end{align}
where $L_R=\ln(\mu_S/R_s)$. For gap 2 they read
\begin{align}
d_1^{(2)}( R_s,\mu_S)&=-3.9363\,,\\
d_2^{(2)}( R_s,\mu_S)&=-94.8742-60.3566L_R\,.\nonumber
\end{align}
For gap 3 they take the form
\begin{align}
d_1^{(3)}( R_s,\mu_S)&=-8.35669\,,\\
d_2^{(3)}( R_s,\mu_S)&=-72.4431-128.136L_R\,.\nonumber
\end{align}
The gap $R$-evolution is therefore
\begin{equation}
\bar{\Delta}(R_1,R_1)-\bar{\Delta}(R_0,R_0)=-\sum_{n=0}^\infty\gamma_n^{\Delta}\int_{R_0}^{R_1}\dd{R}\alphapi{R}^{n+1},
\end{equation}
where the $R$-anomalous dimensions for the 3 gap subtraction schemes are given by
\begin{align}
\{\gamma_i^\text{gap 1}\}_{0\le i\le 1}&=\{0, -43.9543\},\\
\{\gamma_i^\text{gap 2}\}_{0\le i\le 1}&=\{-3.9363, -34.5176\},\nonumber\\
\{\gamma_i^\text{gap 3}\}_{0\le i\le 1}&=\{-8.35669, 55.6927\}.\nonumber
\end{align}
Gap 2 and 3 are $\mu$ independent, but gap 1 inherits a non-trivial $\mu$-anomalous dimension from the soft function and hence requires an additional $\mu$ evolution. This $\mu$-RGE reads~\cite{Dehnadi:2016snl}
\begin{equation}
\mu \dv{\mu} \bar{\Delta}^{(1)}(R,\mu)=-\mu\dv{\mu}\bar{\delta}^{(1)}(R,\mu)=2\,R\,e^{\gamma_E}\,\Gamma^\mathrm{cusp}[\alpha_s]\,,
\end{equation}
which follows from the gap definition Eq.~({\ref{eq:gap1defshort}}) and the soft function RGE in position space, and where the cusp anomalous dimension coefficients are given in Eq.~(\ref{eq:cuspAnomdim}). The solution is
\begin{equation}
\bar{\Delta}^{(1)}(R,\mu)-\bar{\Delta}^{(1)}(R,\mu_0)=2\,R\,e^{\gamma_E}\omega(\Gamma^\mathrm{cusp},\mu,\mu_0)\,,
\end{equation}
with the evolution kernel as defined in Eq.~(\ref{eq:evolKernel}).

\section{Distributions}\label{app:distributions}
The plus function with a fractional exponent $1+\omega$ and $\omega<1$ is defined by~\cite{Fleming:2007xt}
\begin{equation}
\biggl[\frac{\Theta(x)}{(x)^{1+\omega}}\biggr]_+\equiv\lim_{\beta \rightarrow 0}\biggl[\frac{\theta(x-\beta)}{(x)^{1+\omega}}-\delta(x-\beta)\frac{\beta^{-\omega}}{\omega}\biggr].
\end{equation}
Expanding this equation for small $\omega$ defines  plus distributions for positive integer $n$,
\begin{equation}
\biggl[\frac{\Theta(x)\ln^n{\!(x)}}{x}\biggr]_+\equiv\lim_{\beta \rightarrow 0}\biggl[\frac{\theta(x-\beta)\ln^n{\!(x)}}{x}-\delta(x-\beta)\frac{\ln^{n+1}{\!(x)}}{n+1}\biggr].
\end{equation}
Integrating plus distributions with a test function $f(x)$ gives
\begin{equation}
\int_0^\Delta \dd{x} \left[\frac{\Theta(x)\ln^n{\!(x)}}{x}\right]_+ f(x)=\int_0^\Delta \dd{x} \frac{f(x)-f(0)}{x}\ln^n{\!(x)}+f(0)\frac{\ln^{n+1}{\!(\Delta)}}{n+1}\,.
\end{equation}
Plus distributions appear in the jet and soft function and their evolutions. We use the following shorthand notation for them,
\begin{align}\label{eq:plusdist}
\plusFunc{0,\omega}{\mu}{\ell}\equiv%{}&
\frac{1}{\mu^j}\biggl[\frac{\Theta(\ell)}{(\ell/\mu^j)^{1+\omega}}\biggr]_+ \,,\qquad %\nonumber \\ 
\plusFunc{n}{\mu}{\ell}\equiv%{}&
\frac{1}{\mu^j}\biggl[\frac{\Theta(\ell)\ln^n(\ell/\mu^j)}{\ell/\mu^j}\biggr]_+ \,,
\end{align}
where the exponent $j$ is the mass dimension of the variable $\ell$. In the case of a dimensionless argument we will also use the notation
\begin{equation}
\biggl[\frac{1}{e}\biggr]_+\equiv\biggl[\frac{\Theta(e)}{e}\biggr]_+\,.
\end{equation}
The rescaling identity for plus distribution arguments reads 
\begin{equation}\label{eq:plusresc}
\kappa\biggl[\frac{\theta(x)\ln^n(\kappa x)}{\kappa x}\biggr]_+=\frac{\ln^{n+1}(\kappa)}{n+1}\delta(x)+\sum^n_{k=0}\frac{n!}{(n-k)!k!}\ln^{n-k}(\kappa)\biggl[\frac{\theta(x)\ln^k(x)}{x}\biggr]_+\,.
\end{equation}

\section{MC Simulation Settings}
\label{sec:MCsettings}

In the following subsections we will give all the relevant MC settings that are sufficient to describe the process $e^+e^-\to t\bar t$. Any other standard instructions/settings (random seeds) that might be necessary for the operation of the MC, which do not change the statistical population of the final state, can be found in the respective manuals or example input files and they have been left out below.  
\subsection{PYTHIA}
The following flags were set in our \pythia \textsc{8.305} \texttt{main program}. We kept the default Monash 2013 tune,
\begin{verbatim}
Tune:ee = 7
\end{verbatim}
We select the process $e^+e^-\rightarrow t\bar{t}$ and turn off initial state radiation (ISR). The center of mass energy is set to $\texttt{Q}=Q/\textrm{GeV}$,
\begin{verbatim}
Top:ffbar2ttbar(s:gmZ) = on
Beams:idA = 11
Beams:idB = -11
PDF:lepton = off
Beams:eCM = Q
\end{verbatim}
The top quark mass is set to $\texttt{m}=m_t^\mathrm{MC}/\mathrm{GeV}$ and the top width was fixed to \SI{1.4}{\GeV},
\begin{verbatim}
6:m0 = m
6:mWidth = 1.4
6:doForceWidth = true
\end{verbatim}

\subsection{HERWIG}
The following instructions were given in the \herwig 7.2.1 \texttt{input file}.
\begin{verbatim}
read snippets/EECollider.in
\end{verbatim}
We use an internal LO $e^+e^-\rightarrow t\bar{t}$ matrix element and turn off initial state radiation (ISR). We have to explicitly turn on off-shell production of the top quarks. The center of mass energy is set to $\texttt{Q}=Q/\textrm{GeV}$,
\begin{verbatim}
cd /Herwig/MatrixElements
insert SubProcess:MatrixElements 0 MEee2gZ2qq
set MEee2gZ2qq:MinimumFlavour 6
set MEee2gZ2qq:MaximumFlavour 6
set MEee2gZ2qq:TopMassOption OffShell
set /Herwig/Particles/e-:PDF /Herwig/Partons/NoPDF
set /Herwig/Particles/e+:PDF /Herwig/Partons/NoPDF
cd /Herwig/Generators
set EventGenerator:EventHandler:LuminosityFunction:Energy Q*GeV
\end{verbatim}
The top quark mass is set to $\texttt{m}=m_t^\mathrm{MC}/\mathrm{GeV}$ and the top width was fixed to \SI{1.4}{\GeV},
\begin{verbatim}
cd /Herwig/Particles
set t:NominalMass m*GeV
set tbar:NominalMass m*GeV
set t:Width 1.4
set t:Width_generator:Initialize Yes
set t:Mass_generator:Initialize Yes
set tbar:Width 1.4
set tbar:Width_generator:Initialize Yes
set tbar:Mass_generator:Initialize Yes
set /Herwig/Decays/Top:Initialize Yes
\end{verbatim}

\subsection{SHERPA}
The following parameters were set in the \sherpa 2.2.11 \texttt{Run.dat}. We set the c.m.\ energy $\texttt{Q}=Q/\textrm{GeV}$ and we specify $e^+e^-$ beams without ISR. The top quark mass is set to \mbox{$\texttt{m}=m_t^\mathrm{MC}/\mathrm{GeV}$}. The decay of hard process final state particles has to be explicitly turned on. The top quark decay width was fixed to \SI{1.4}{\GeV},
\begin{verbatim}
(run){
BEAM_1  11; BEAM_ENERGY_1 Q/2;
BEAM_2 -11; BEAM_ENERGY_2 Q/2;
PDF_LIBRARY None;

MASS[6] m;

HARD_DECAYS 1;
WIDTH[6] 0;
HDH_WIDTH[6,24,5]=1.4;
HDH_WIDTH[-6,-24,-5]=1.4;
}(run)
\end{verbatim}
We select the process $e^+e^-\rightarrow t\bar{t}$ 
\begin{verbatim}
(processes){
Process 11 -11 -> 6 -6;
Order (*,2);
End process;
}(processes)
\end{verbatim}

\bibliography{sources}
\bibliographystyle{JHEP}

\end{document}